\documentclass{aa}  
\usepackage{graphicx}
\usepackage{array}
\usepackage{lscape}                                % Tables in landscape
\usepackage{natbib}
\usepackage{longtable}
\usepackage{color}
\usepackage{mathrsfs} 
\usepackage{subcaption}
\usepackage{etoolbox}
\usepackage[utf8]{inputenc}
\usepackage{booktabs}

\usepackage{txfonts}
\usepackage{xcolor}
\usepackage[breaklinks=true]{hyperref} %% to avoid \citeads line fills
\bibpunct{(}{)}{;}{a}{}{,} %% natbib format for A&A and ApJ
\bibpunct{(}{)}{;}{a}{}{,}
\newcommand{\kms} {km\,s$^{-1}$}
\newcommand{\vsini} {$v$\,sin\,$i$}

\newcommand{\Teff} {$T_{\mathrm{eff}}$}
\newcommand{\grav} {$\log\,{g}$}

\newcommand{\fwhb}{\textit{FW3414(H$\beta$)}}

\begin{document} 

   \title{The IACOB project \thanks{Spectra of all suggested standard and reference stars can be downloaded from the webpage \url{https://astroplus.ua.es/mkbtypestds} }
}

   \subtitle{XII. A new grid of northern standards for the spectral \\ classification of B-type stars}

   \author{I. Negueruela\inst{1,2}, S.~Sim\'on-D\'iaz\inst{3,4}, A. de Burgos\inst{3,4}, A. Casasbuenas\inst{4,5}, P.~G.~Beck\inst{3,4}} 

\institute{
Departamento de F\'{\i}sica Aplicada, Facultad de Ciencias, Universidad de Alicante, Carretera de San Vicente s/n, E03690, San Vicente del Raspeig, Spain
\and
Instituto Universitario de Investigaci\'on Inform\'atica, Universidad de Alicante, San Vicente del Raspeig, Spain
\and
Instituto de Astrof\'{\i}sica de Canarias, E-38200 La Laguna, Tenerife, Spain              
\and
Universidad de La Laguna, Dpto. Astrof\'{\i}sica, E-38206 La Laguna, Tenerife, Spain
\and
Nordic Optical Telescope, Rambla Jos\'{e} Ana Fern\'{a}ndez P\'{e}rez 7, ES-38711 Bre\~{n}a Baja, Spain 
}

\offprints{ignacio.negueruela@ua.es}

\date{Submitted/Accepted}

\titlerunning{B-type star classification}

\authorrunning{I. Negueruela et al.}
% \abstract{}{}{}{}{} 
% 5 {} token are mandatory
 
  \abstract
  % context heading (optional)
  % {} leave it empty if necessary  
   {With the advent of large spectroscopic surveys, automated stellar parameter determination has become commonplace. Nevertheless, spectral classification still offers a quick and useful alternative for obtaining parameter estimates for large samples of spectra of varying quality.}
  % aims heading (mandatory)
   {We present a new atlas of stellar spectra covering the B-type range, with the intention of providing detailed classification criteria valid for modern spectra and improving the grid of reliable standards. This new grid will be used in future works to provide classification criteria beyond the classical classification range and addressing, in particular,  the use of \textit{Gaia}/RVS spectra.}
  % methods heading (mandatory)
   {We analysed historical standards by means of multiple high-resolution spectra, marking out problematic cases and complementing the grid with a new set of reliable comparators. We then elaborated on a new set of classification criteria based on high-quality $R=4\,000$ spectra.}
  % results heading (mandatory)
   {Our new classification grid is much thicker than any previous set of standards, presenting a high degree of self-consistency. Although it is based entirely on morphological criteria, the grid demonstrates a much better correlation with different physical parameters.}
  % conclusions heading (optional), leave it empty if necessary 
   {The new grid can be used to study classification criteria in other spectral ranges, providing a valuable tool for the study of B-type stars that covers a very wide range of temperatures, luminosities, and stellar masses. The very process of classification also offers valuable insights into stellar evolution.}


   \maketitle
%
%-------------------------------------------------------------------

\section{Introduction}

For over a century, spectral classification has served as a fundamental tool of astrophysics. The development of continuously improving stellar models \citep[see the latest report by  IAU Commission~36 in][]{Puls2016} has now rendered the determination of actual stellar parameters from
spectroscopy, such as the effective temperature ($T_{\mathrm{eff}}$) or the surface gravity ($\log\,g$),  a straightforward procedure. With the advent of powerful (at least partly) automated schemes for these tasks \citep[e.g.][]{Mokiem2005,SimonDiaz2011,Allende2015,Tabernero2019,Bestenlehner2023}, it is now possible to derive physical properties for very large samples with a moderate time investment. Spectral classification, however, still has an important role to play \citep[cf.][]{keenan87,garrison03}. Leaving aside practical difficulties in the derivation of parameters from stellar models, for instance, the availability of an appropriate grid or the need for moderate signal-to-noise ratio (S/N) values, spectral classification does not depend on models. From its inception \citep{morgan37,morgan38}, the process of spectral type and luminosity class assignment was conceived as purely relaying on morphological features in the spectra. In consequence, it is free of the underlying dependence on stellar models that characterises any temperature scale. As per \citet{garrison03}, the appearance of a stellar spectrum is a fundamental quantity and the classification must be autonomous, namely,\ independent of any theoretical interpretations.

Because of this independence, spectral classification is suitable for the characterisation of large numbers of spectra that cannot be used to derive accurate stellar parameters. Examples of such spectra for hot stars are provided by the LAMOST survey of the northern sky \citep{luo15,luo22}, whose spectral resolution only allows for rough estimations of \Teff\ and log\,$g$. There are also the many obscured objects that will be targeted by the WEAVE-SCIP strand \citep{jin24}, for which the blue section of the spectrum will be affected by heavy extinction, resulting in low S/N or simply no counts in the region where the main diagnostic lines for stellar parameter determination are located. 

An even more pertinent example is the collection of \textit{Gaia} RVS spectra of hot stars \citep{blomme23}. The low number of lines and species that can be seen in the RVS spectra of O and B-type stars and the lack of explicit $T_{\mathrm{eff}}$ diagnostics do not permit accurate spectroscopic analysis. This has been recognised for long \citep[e.g.][]{mt99}, with the implicit implication that little quantitative information can be obtained from RVS spectra for these stars, beyond a radial velocity \citep[and even this only when tailored templates and procedures are considered; see, e.g.][]{blomme23}. In contrast, there have been several attempts to obtain valuable qualitative information from spectra of OB stars in this spectral range. A summary of previous work can be found in appendix~A of \citet{negueruela10Wd1}, where the prospects of spectral classification are explored, based on existing stellar libraries. The conclusion of this exploration is that approximate classification can be obtained even with the restricted spectral range given by the RVS, although a final assessment requires a deeper dedicated study, based on high-quality spectra of standard stars.
 
Our long-term aim is to perform such an investigation, by using \'{e}chelle spectra of a high S/N and resolution, so that the classical classification region \citep[i.e. the blue part of the optical spectrum where photographic plates were most sensitive; e.g.][]{lesh68,walborn90,lennon90,sota11,sota14} and the RVS region (8470\,--\,8740\,\AA) are observed simultaneously. Today, such a set of spectra is accessible as part of the collection associated with the IACOB project \citep[see][and Sect.~\ref{sec:obs} below]{simondiaz2020} and we intend to use them to explore the true possibilities presented by spectra of OB stars outside the classical classification region. In the present work, we redefine the grid of B-type standards by using high-quality modern digital spectra. In the following papers, we present the spectral atlas and classification criteria for the RVS region, while exploring the use of quantitative measurements as an alternative to classification and deriving spectroscopic parameters for the whole B-type standard grid, as was recently done for the O-type grid by \citet{holgado18}.

A new assessment of criteria for the classification of B-type stars has been deemed necessary because strong inconsistencies in the relation between luminosity class and intrinsic magnitude, already noticed by \citet{jaschek98}, were found when preparing the list of stars to observe. Moreover, there has been no systematic study of the whole B-type range since the advent of digital spectroscopy. The pioneering work of \citet{walborn90} was centred on O-type stars and touched on B-type stars only in passing, while the atlas of \citet{lennon92} included only supergiants. Although the work of N.~R.~Walborn and collaborators has updated and strengthened the O-type standard grid, fully revised in \citet{sota11}, no similar effort has been carried out for B-type stars. The only modern guide to spectral classification of B-type stars is the collection of online notes by E.~Mamajek\footnote{\url{https://github.com/emamajek/SpectralType}} used to build the calibration presented in \citet{pm13}. Even worse, this calibration of intrinsic colours and $T_{\mathrm{eff}}$ is the only modern attempt at characterising the B-type stars, a most striking situation when we consider that the B spectral types contain the vast majority of intermediate-mass stars and, given the shape of the initial mass function (IMF), the bulk of supernova progenitors. Consequently, our second main long-term objective is obtaining a robust characterisation of physical parameters for the different B types.

In this first paper, we present the spectral library in Section~\ref{sec:obs}.We proceed to discuss the strengths and limitations of the classification process in Section~\ref{sec:class}. The core of the paper is formed by Section~\ref{sec:crit}, where classification criteria suited to modern spectra are discussed, and Appendix~\ref{app:grid}, where the new grid of standards is defined. Finally, we discuss the validity of the new system in Section~\ref{sec:discuss}. The other appendices provide abundant ancillary information on the selection of standards and the choice of criteria.

\section{Observations}
\label{sec:obs}

For the sake of homogeneity, all spectra considered in this and forthcoming papers were obtained with the High Efficiency and Resolution Mercator Echelle Spectrograph (HERMES), operated at the 1.2~m Mercator Telescope (La Palma, Spain). HERMES offers a resolving power $R= 85\,000$, and a spectral coverage from 377 to 900~nm, although there are some small gaps beyond 850~nm \citep{raskin14}. 
Data were homogeneously reduced using version 4.0 of the HermesDRS\footnote{\url{http://www.mercator.iac.es/instruments/hermes/hermesdrs.php}} automated data reduction pipeline offered at the telescope. A complete set of bias, flat, and arc frames was obtained each night and used to this aim. For wavelength calibration, we used a combination of a thorium-argon lamp equipped with a red-blocking filter to cut off otherwise saturated argon lines and a neon lamp for additional lines in the near infrared. 
%The arc images typically have a S/R in the $\sim80$\,--\,120 range. 

The HermesDRS pipeline provides wavelength-calibrated, blaze-corrected, order-merged spectra. We then used our own programmes developed in IDL, as well as the {\tt pyIACOB} tool\footnote{{\tt pyIACOB} Github repository: \href{https://github.com/Abelink23/pyIACOB}{https://github.com/Abelink23/pyIACOB}} \citep[see][]{deBurgos23}, to normalise the spectra, remove unwanted cosmic rays and other cosmetic defects, and correct all the spectra for heliocentric velocity.

We gathered a minimum of three HERMES spectra for an initial sample of about 200 B-type stars (plus a handful of early A-type stars). In all cases we reached a S/N in the $\sim$4500~\AA\ region of at least 100 with an exposure time from a few minutes up to 15 minutes, depending on the magnitude of the star. This initial sample was intended to include all the stars that have been used as standards for spectral classification in the B0\,--\,A0 star domain by different authors throughout time, with special attention to the original sources listed in Section~\ref{sec:basic} and the obvious limitation of observability from La Palma. After close examination, many of these stars had to be removed from the list of standards for a variety of reasons. For example, they may appear as double-lined spectroscopic binaries (SB2) in modern spectra, or they may have entered a Be phase at some point. The main standards that have been considered inadequate, as well as objects that have received different classifications in reference publications, are discussed individually in Appendix~\ref{app:gone}. Secondary standards that have been rejected or have received a new classification are mentioned in the main text. Once a standard had been ruled out, we looked for an appropriate replacement and observed it as well. After completion of the process, we are left with a list of 157 reference stars, which is presented in Table~\ref{bigtable}. For each star, Table~\ref{bigtable} lists (1) its final spectral classification; (2) whether the star has been identified as a single-line spectroscopic binary (SB1) based on the compiled spectra; (3) its projected rotational velocity \citep[\vsini, derived following the methodologogy presented in][]{ssimon14}; and (4) some other quantities of interest, introduced in Sect.~\ref{quantities}.

Following the discussion in Sect.~\ref{class_approach}, for the purposes of this paper, we downgraded the resolution of all the spectra to an equivalent resolving power of $R$\,$\sim$\,4\,000. To this aim, we used a tailored procedure included within the {\tt pyIACOB} tool that convolves the original spectra with a Gaussian kernel with $\sigma\,=\,\frac{\lambda_{0}}{2\sqrt{2\ln2} R}$, where $\lambda_{0}$ is the average wavelength of the spectrum and $R$ is the desired resolving power. %spectral resolution. %Previous to this, we used {\tt pyIACOB} to correct the spectra from radial velocity. \ind{Esto no lo hemos dicho ya antes?}

\begin{table*}
\caption{Grid of primary classification standards that are kept after the  revision process and retain their classification -- if intermediate types are not interpolated. The anchor standards of \citet{garrison94} are shown in bold. Standards marked with an asterisk present complications which are discussed in the respective subsections. \label{tab:daggers}}
\centering
\begin{tabular}{l c c c c c c c }
\hline\hline
\noalign{\smallskip}
& V & IV & III & II & Ib & Iab & Ia \\
\noalign{\smallskip}
\hline
\noalign{\smallskip}
B0 & $\tau$ Sco & $ $ &$ $ & $ $&$ $&$ $& \textbf{$\epsilon$ Ori}\\
\noalign{\smallskip}
B0.5& & & & & & &$\kappa$ Ori\\
\noalign{\smallskip}
B1& $\omega^{1}$ Sco& & & & $\zeta$~Per& &\\
\noalign{\smallskip}
B2&$\beta^2$~Sco$^{*}$ & $\gamma$~Peg, $\zeta$~Cas& $\gamma$~Ori & &\textbf{9~Cep}&&\textbf{$\chi^{2}$~Ori}\\
\noalign{\smallskip}
B2.5& $\sigma$~Sgr & &&&&&\\
\noalign{\smallskip}
B3&29~Per, \textbf{$\eta$~UMa}, \textbf{$\eta$~Aur}& $\iota$~Her&&&&&\textbf{$o^2$~CMa}\\
\noalign{\smallskip}
%B4&\multirow{2}{*}{$\rho$~Aur}&&\\
%\noalign{\smallskip}
B5&$\rho$~Aur$^{*}$&&&&67~Oph$^{*}$ & &\textbf{$\eta$~CMa}\\
\noalign{\smallskip}
B6&& 19~Tau&&&&&\\
\noalign{\smallskip}
B7&HD~21071$^{*}$&&$\beta$~Tau&&&\\
\noalign{\smallskip}
B8&18~Tau& &27~Tau$^{*}$&&&&\textbf{$\beta$~Ori}\\
\noalign{\smallskip}
B9&& $\alpha$~Del&&&&\\
%&\multicolumn{2}{c}{Exposure times (s)}\\
%Filter & Long times & Short times \\
%\hline
%$U$ & 900 & 250\\
%$B$ & 200 & 60 \\
%$V$ & 40& 10 \\
\noalign{\smallskip}
\hline
\end{tabular}
\end{table*}

\section{Classification}
\label{sec:class}

\subsection{Basic principles}
\label{sec:basic}
 
The modern spectral classification system starts with the work of \citet{mkk}, who elaborated on the classification of \citet{hd}. The system is defined by a set of standard stars, fully presented in the seminal paper by \citet[from now on, JM53]{jm53}, where it is tied to the $UBV$ photometric system. Spectral classification is achieved by comparison to the standard stars, if possible, observed with the same instrumentation at the same resolution. The classification is intended to be independent of intrinsic physical properties \citep{morgan37}, but at the same time it is expected to provide some information on them. For further historical details and an overview of the classification process, we refer to \citet{graybook}.
\defcitealias{jm53}{JM53}

The original set of standard stars in \citetalias{jm53} was intended as a collection of representative objects for each subtype. They were observed on photographic plates with dispersions of 100\,\AA /mm at H$\delta$ (resolution of 2\,\AA). Along the years, works by W.~W.~Morgan and collaborators, focussed on stars in young open clusters, led to a revision of the standards. This was first advanced by \citet{garrison67} and then by \citet{lesh68}, who presented detailed classification criteria and introduced the possibility of distinguishing half subtypes, making extensive use of the new spectral types B2.5 and B4. A number of complementary standards, only visible from the southern hemisphere, was presented by \citet{garrison77}.

The consequence of this process was the production of a revised list of 'primary' standards, the 'revised MK system', published by \citet[from now on MK73]{mk73} for the stars earlier than the Sun.\defcitealias{mk73}{MK73} To differentiate among the new types and the old \citetalias{jm53} types, Morgan marked them with a $\dagger$ sign, and thus these came to be known as 'dagger' standards. According to \citetalias{mk73}, [t]hese are fundamental reference points for the revised MK system;  any future changes in their values will effectively alter the MK system itself. The grid of standards for early-type stars is complemented with a number of classification criteria, presented in \citet{morgan78}. The new system is sometimes referred to as MK78. A good description of the MK78 system is given in \citet[from now on, MK78]{keenan85}, where a list of primary standards (similar, but not identical to that of \citetalias{mk73}) is given. The set of standards is given in Table~\ref{tab:daggers}, where objects that (for any reason) have been discontinued as standards or have been reassigned to a different type are not included. Many of these reassignments are  commented on later in this work (see, in particular, Appendix~\ref{app:gone}).
\defcitealias{keenan85}{MK78}

As an evolution of this system, \citet{garrison94} proposed the development of a hierarchy of standards, in which objects whose classification had remained untouched after successive modifications could be considered 'anchor points' of the system, while other primary standards, representative of each type, would fill the gaps between them. The number of anchor points is quite small, and they therefore represent the system itself, as [t]he MK Process ... is a methodology that uses the objects themselves as standards \citep{garrison94}.

The philosophy behind the MK system can be summarised by quoting directly from \citetalias{mk73}: The MK system is a phenomenology of spectral lines, blends, and bands, based on a general progression of color index (abscissa) and luminosity (ordinate) ... defined by an array of standard stars, located on the two-dimensional (2D) spectral type-versus-luminosity diagram. A number of points listed as characteristics of the MK system in \citetalias{mk73} are of particular relevance for this project:
\begin{itemize}
\item Such a system is defined by the spectra of an array of standards stars.
\item Such a system is autonomous and self-consistent with regard to the various lines, bands, and blends in the ordinary photographic region.
\item When types are determined from other spectral regions, they must be collated with results from the ordinary photographic region.
\end{itemize} 

As a practical corollary of these basic principles, we may expect stars classified with the same spectral type and luminosity class to have roughly similar $T_{\mathrm{eff}}$ and \grav, but certainly \textit{not} identical, properties. This is due to two main reasons:
\begin{enumerate}
\item Spectral classification, like most classification systems, translates a continuum of (in this case, physical) properties into a number of categories. In this sense, a given spectral classification may be understood as a "box" in the parameter space of physical properties (see, e.g.\ \citealt{simon14cal}, for the mapping of physical properties on to spectral types of O dwarfs).
\item Physical properties other than $T_{\mathrm{eff}}$ and $\log\, g$ (for instance, stellar abundances or mass loss in the form of a radiative wind) affect the appearance of spectra that still have to be mapped on to a 2D grid.
\end{enumerate}

These effects guarantee that the resulting classification will never be a true representation of an underlying theoretical Hertzsprung-Russell diagram (HRD). On the other hand, the system is much more robust and self-consistent, because it is based solely on morphological criteria, that is, criteria that can be derived directly from the description of the spectra themselves \citep{walborn79}. Of course, this fundamental dependence on morphology implies that classification criteria may vary if the quality of the spectra varies. Higher spectral resolution or higher S/N may allow for a finer classification. However, if the system is robust, the grid of standards should still be self-consistent, even though some new standards will have to be added to account for the finer classification.

As a prime example, \citet{walborn71} studied the spectroscopic characteristics of OB stars by using spectrograms of dispersion 63\,\AA\,mm$^{-1}$, i.e.\ better resolution ($\sim1.2$\,\AA) than used in the \citet{mk73} system. As a result, he felt forced to introduce three new interpolated spectral types: O9.7 (originally defined only for supergiants), B0.2, and B0.7. The finer grid led to the reshuffling of some standard stars in the B0\,--\,B1 range. Moreover, as the \citetalias{keenan85} system was being developed in parallel, some standards were assigned slightly different spectral types in both systems, which has led to some degree of confusion among the community  ever since.

A major change in the way that the classification criteria were applied came with the introduction of digital detectors. The process of comparing widened spectrograms on photographic plates to those of standards became a much more direct comparison of 1D traces. \citet{walborn90} transposed the classification criteria for OB stars and B-type supergiants to digital spectrograms with resolution 1.5\,\AA. Shortly afterwards, \citet{lennon92} published the first CCD spectroscopic atlas of B-type supergiants, at a dispersion of 0.4\,\AA/pixel. Much more recently, \citet{sota11} published a new atlas of CCD spectra of O-type stars at $R\sim2\,500$ and high S/N \citep[see also][]{sota14}. Their work extends into the B-type range until the spectral type B0.5. To our knowledge, no similar work has been performed for the B-type range.

\subsection{Known issues relative to the classification of B stars}
\label{sec:issues}

Astronomers who wish to use the MK process as a tool to study B-type stars face a number of complications. An initial difficulty stems from the fact that the MK process is based on comparison with the grid of standards observed with the same (or, at lest, very similar) instrumentation. With modern digital spectra, this difficulty disappears, as it is possible to use relatively simple tools to modify existing spectra of the standards and adapt them to the resolution or typical S/N of the programme spectra. Apart from the atlases of OB stars \citep{walborn90,sota11}, several general purpose stellar atlases contain large numbers of B-type stars, for instance the MILES\footnote{\url{http://miles.iac.es/}} library at low resolution \citep{miles} or the Indo-US library\footnote{\url{https://noirlab.edu/science/observing-noirlab/observing-kitt-peak/telescope-and-instrument-documentation/cflib}} at intermediate resolution \citep{indo-us}. These libraries, however, have some limitations, as the number of MK standards is not high, and the spectral types provided for many stars are not accurate. In addition, while it is possible to degrade the spectra to a lower resolution, they cannot be used as good comparators for spectra taken at higher resolution. Our new atlas, on the other hand, consists of high-resolution spectra of high quality, which have been convolved to classification resolution (see Sect.~\ref{class_approach}), reaching very high S/N.

Other complications are more intimately related to the MK process itself. Firstly, the existence of several versions of the standard grid, in which some stars have slightly different classifications, leads to confusion. This difficulty is easy to overcome by sticking to a given system \citepalias[e.g.][for low resolution spectra]{keenan85}. Second, improvements in the classification have led to changes in the spectral types of some standards. Again, a consistent classification requires adhesion to a given version of the system. Third, some standards have been found not to be the best choice to represent a class because of unexpected developments. For example, some stars that present broad lines at a given resolution are found to be SB2 when observed at higher resolution. As a consequence, even if they are not detected as SB2 at lower resolution, their line profiles are likely to change with time, rendering them inadequate standards (an example is the B1\,III primary standard $o$~Per; see Appendix~\ref{oper}). Another example is the \citetalias{mk73} B7\,III primary standard $\eta$~Tau (Alcyone = HD~23630) that entered an extended Be-shell phase in 1973, and had to be abandoned.

\begin{figure}[ht]
\resizebox{\columnwidth}{!}{\includegraphics{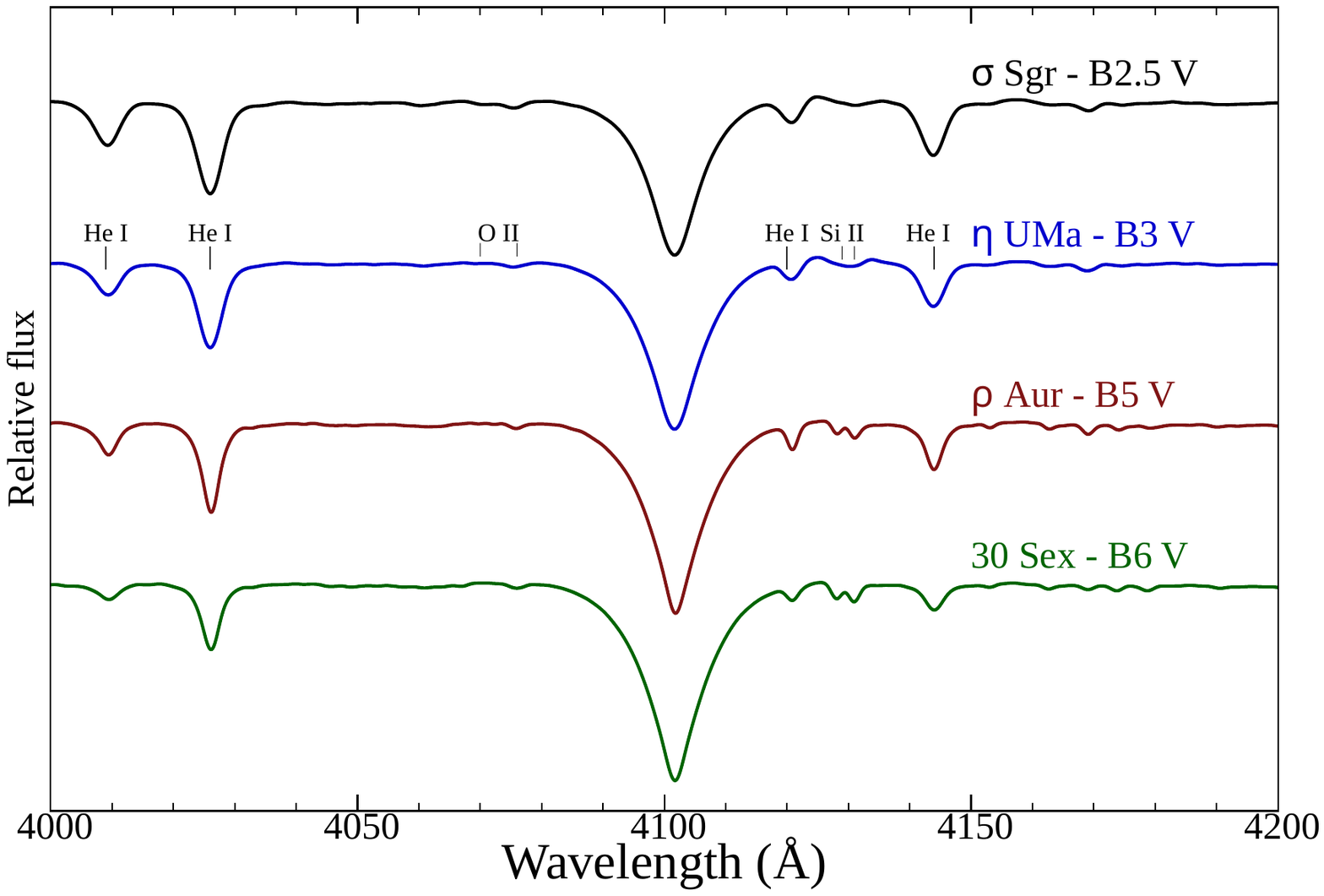}}
\centering
\caption{Sequence of mid-B dwarf MK standards in the region around H$\delta$ showing the metallic and \ion{He}{i} lines used for classification.\label{fig:crit_lesh} } 
\end{figure}

Two more serious concerns arise from the historical development of the MK process. Firstly, despite all the improvements to the system, some of the standards still do not seem truly representative of the class, in the sense that they do not even provide an approximate map to the underlying HRD. This is the main motivation for the present work and  is addressed below. Finally, with the application of the classification process to digital spectra of ever-increasing quality, the classification criteria that were used to build the grid have lost their significance. 

To illustrate this point, we use the classification criteria for mid-B dwarfs given by \citet{lesh68}. These are the criteria used to build the standard grid for the \citetalias{mk73} system. Quoting directly from \citet{lesh68}, these criteria are:
\begin{itemize}
\item \textbf{B2.5\,V:} \ion{Si}{ii}~4129\,\AA\ present, \ion{O}{ii}~4072\,\AA\ absent.
\item \textbf{B3\,V:} \ion{He}{i}~4121\,\AA\ $<$ \ion{Si}{ii}~4129\,\AA\ $<$ \ion{He}{i}~4144\,\AA, \ion{He}{i}~4009\,\AA\ moderately weak.
\item \textbf{B5\,V:} \ion{Si}{ii}~4129\,\AA\ $\sim$ \ion{He}{i}~4144\,\AA, but \ion{He}{i}~4121\,\AA\ still present, \ion{He}{i}~4009\,\AA\ very weak.
\item \textbf{B6\,V:} \ion{Si}{ii}~4129\,\AA\, $>$ \ion{He}{i}~4144\,\AA, \ion{He}{i}~4121\,\AA\ very weak.
\end{itemize}

Figure~\ref{fig:crit_lesh} shows modern, high S/N spectra at $R\,=\,4\,000$ of the primary standards defining these types in the spectral range of interest. A cursory look at Fig.~\ref{fig:crit_lesh} shows that none of the criteria can be applied. The standards form a smooth sequence of decreasing temperature and the diagnostic lines still behave smoothly in the sense defined by the criteria, but the discrete steps that were meaningful in the low-resolution plate spectra have lost most of their relevance. This is, however, not a major complication. The system is robust precisely because it relies solely on the standard grid. We simply need to find new criteria to define the spectral types.

Nevertheless, criteria based on line ratios are very strongly dependent not only on spectral resolution but also on stellar rotation, since some lines are intrinsically broader than others \citep[cf.][where illustrative effects in the O-type range are discussed]{Markova2011}. A prime example is the ratio of the \ion{He}{i}~4471\,\AA\ line to the neighbouring \ion{Mg}{ii}~4481\,\AA\ doublet, a very important diagnostic for mid- and late-B stars. The \ion{He}{i} line is subject to considerable Stark broadening, while the \ion{Mg}{ii} feature is intrinsically narrow. As a consequence, a high rotational velocity, which will broaden both lines, will affect the depth of the \ion{Mg}{ii} line (and hence its perceived intensity) much more strongly. To counter this effect, \citet{garrison_g94} proposed the introduction of standards for both low and high rotational velocity in the B7\,--\,B9 range.

\begin{figure*}[!t]
\resizebox{\columnwidth}{!}{\includegraphics{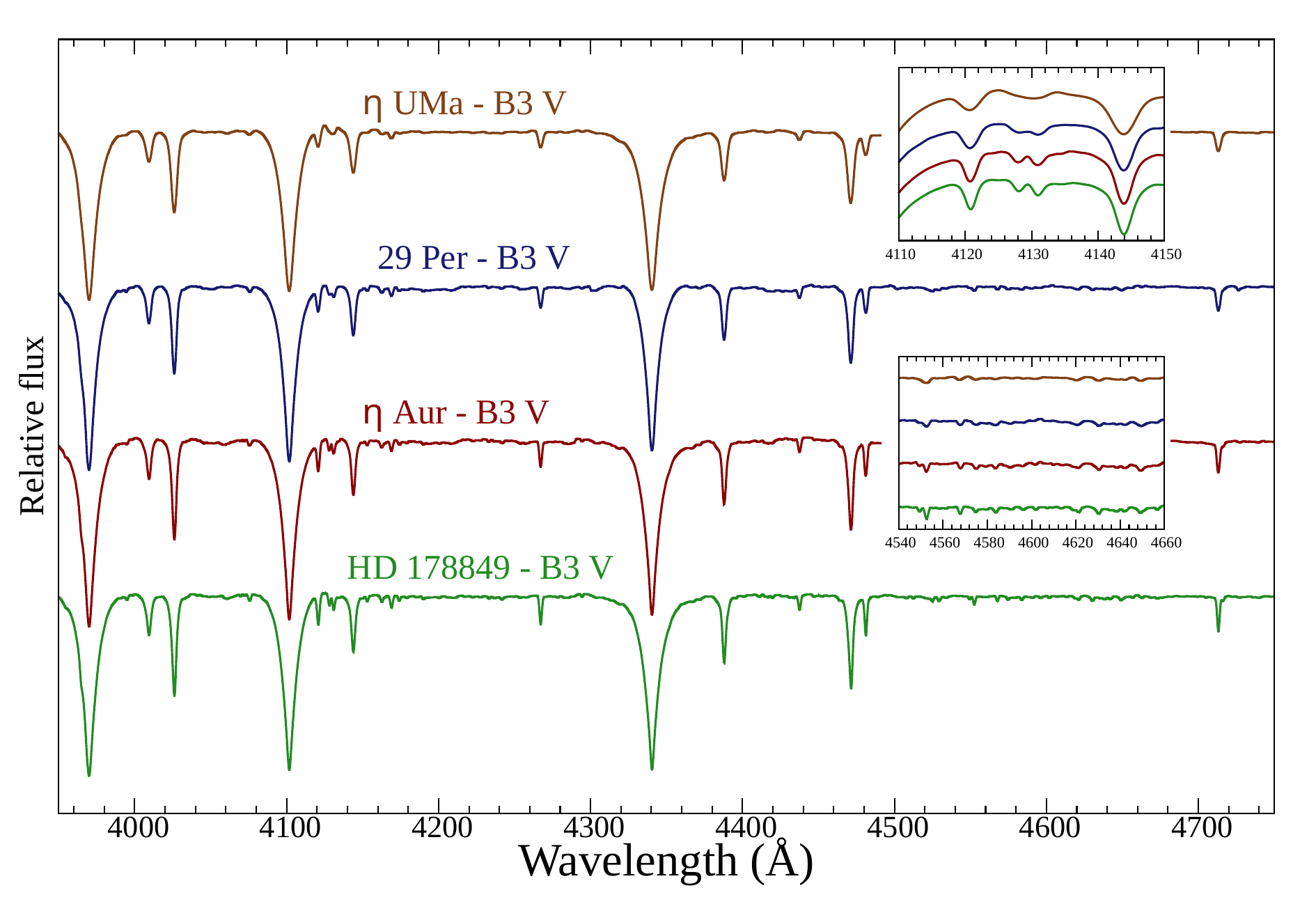}}
\resizebox{\columnwidth}{!}{\includegraphics{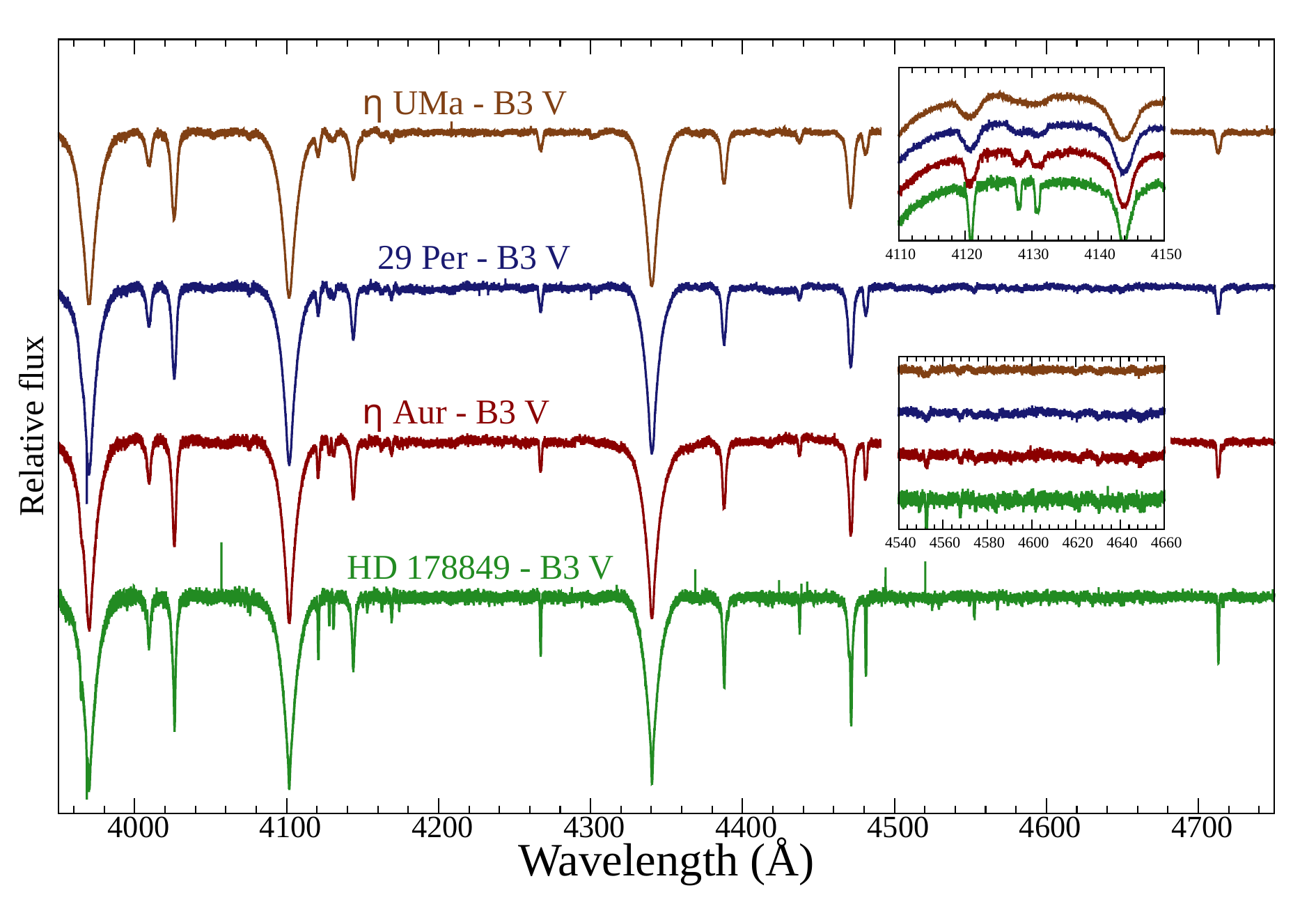}}
\centering
\caption{Effect of rotational velocity (left panel, increasing from bottom to top) on spectral features at classification resolution. All the stars shown are B3\,V standards. At the very high S/N of these spectra, the star with the slowest rotation ($v\,\sin\,i\approx35\:\mathrm{km}\,\mathrm{s}^{-1}$) still displays, although very weakly,  many of the metallic lines visible at earlier types (or higher luminosities), such as the \ion{Si}{iii} triplet at $\lambda\lambda$~4553, 4568, 4575. The other stars have rotational velocities of $\approx100$, $\approx125$ and $\approx160\:\mathrm{km}\,\mathrm{s}^{-1}$, respectively. We note the effect on the \ion{Si}{ii}~4129\,\AA\ doublet and the apparent ratio between \ion{He}{i}~4471\,\AA\ and \ion{Mg}{ii}~4481\,\AA. Line identifications after \citet{kilian91}. The insets show the details of the metallic lines. {Right panel shows the same details at the original ($R\approx85\,000$) resolution, to appreciate the full extent of the effects. We note the clearly separated \ion{Si}{ii} doublet at 4129\,\AA\ and the triplet structure in \ion{He}{i}~4471\,\AA\ in the spectrum of HD~178849.}\label{fig:btreses} } 
\end{figure*}

\begin{figure}
\resizebox{\columnwidth}{!}{\includegraphics{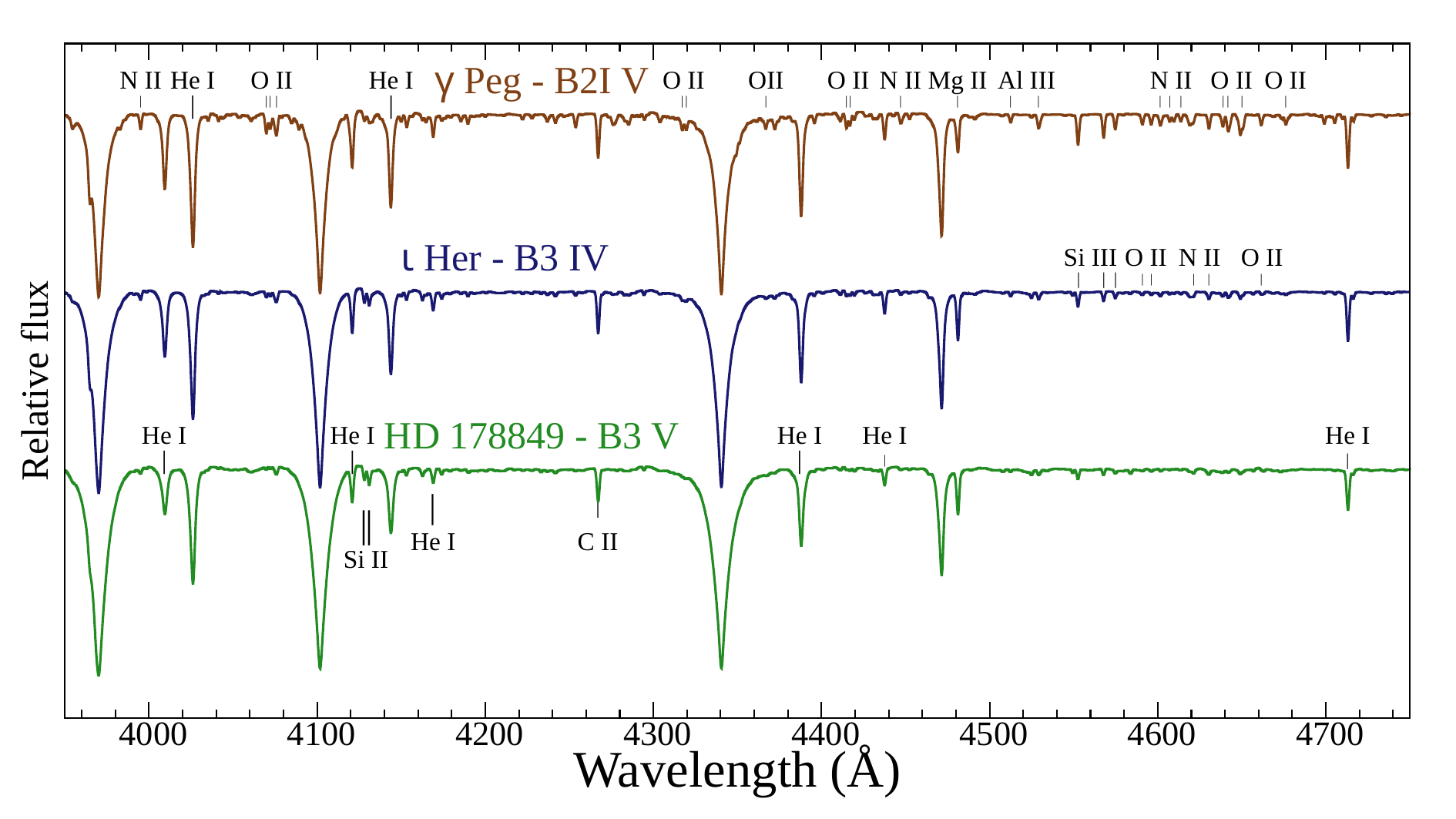}}
\centering
\caption{A comparison of the slowly rotating B3\,V standard HD~178849 with standard stars of similar spectral type and luminosity class IV. Line identifications follow \citet{kilian91}. \label{fig:rotcheats} } 
\end{figure}

In fact, the effects of rotational velocity on spectral classification are much stronger than generally assumed. As an example, in Figure~\ref{fig:btreses}, we show four B3\,V standard stars, the three primary standards listed in Table~\ref{tab:daggers} and the \citetalias{jm53} standard HD~178849, which has a low rotational velocity ($v\,\sin\,i\approx35\:\mathrm{km}\,\mathrm{s}^{-1}$). Rotational velocity affects very strongly the perceived ratio between \ion{He}{i}~4471\,\AA\ and \ion{Mg}{ii}~4481\,\AA\ -- the effect is more clearly seen when comparing the spectra of $\eta$~UMa ($v\,\sin\,i\approx160\:\mathrm{km}\,\mathrm{s}^{-1}$) and HD~178849 --, but its most obvious impact is on the detectability of features. Because the lines are very narrow, it is possible to see in the spectrum of HD~178849 a large number of weak features that can at best only be guessed in the other B3\,V standards. The \ion{Si}{iii} triplet at 4553, 4568, 4575\,\AA\ is an obvious example (see insets to Fig.~\ref{fig:rotcheats}), but there are many other weak features corresponding to \ion{O}{ii}, \ion{N}{ii}, and other ionised metals.

The very large difference in the overall aspect of slowly and fast rotating stars has historically led to significant confusion. In Figure~\ref{fig:rotcheats}, the narrow-lined B3\,V standard HD~178849 is compared to two nearby dagger standards, $\iota$~Her (B3\,IV) and $\gamma$~Peg (B2\,IV). The overall resemblance between HD~178849 and $\iota$~Her is probably closer than between HD~178849 and the other B3\,V standards. If we were tempted to think that HD~178849 is not a B3\,V star, given its dissimilarity to 29~Per and $\eta$~UMa, we are unlikely to assign it an earlier spectral type, despite the presence of many metallic lines seen in earlier types, because the main temperature discriminator, the ratio between the \ion{Si}{ii}~4128\,\AA\ doublet and the \ion{Si}{iii} lines, reverses between B2 and B3 (compare to the spectrum of $\gamma$~Peg). We would with high probability think that it has a higher luminosity, grouping it together with $\iota$~Her. In fact, as we explain below, the luminosity classification of non-supergiant mid- and late-B stars has historically reflected the rotational velocity more closely than the stellar luminosity, leading to many of the inconsistencies that have motivated this work. Further examples of the effect of rotation on the overall aspect of a spectrum can be found in Figures~\ref{fig:two1p5v_stds} and~\ref{fig:two2p5v_stds}, while changes in individual lines are explored in Figures~\ref{fig:wingsb8} and~\ref{fig:wingsb7}.

When accurate distances for nearby stars became available, thanks to the \textit{Hipparcos} catalogue, \citet{jaschek98} analysed all the early-type standards with errors in parallaxes, guaranteeing errors in bolometric magnitude $\leq 0.3$~mag. They found that there was a very significant dispersion at a given type and luminosity. Although giants and dwarfs had statistically different absolute magnitudes, the separation between luminosity classes V and III was not evident. Furthermore, a number of stars had very strongly deviant magnitudes for their luminosity class. The solution proposed by \citet{jaschek98} was simply dropping these extreme cases from the list of standards. Here, instead, we analyse the reason for this discrepancy and introduce changes in the grid of standards whenever necessary.

\subsection{Our approach}
\label{class_approach}

When attempting to redefine the standard grid, a number of decisions have to be made. As mentioned in Sect.~\ref{sec:obs}, our observational data consists of high quality spectra covering the whole optical range (3900\,\AA\,--\,9\,200\,\AA) at $R\,=\,85\,000$. The final purpose of this project is investigating the use of features over the whole range to classify B-type stars, with an emphasis on the RVS spectral range. However, for the definition of the standard grid, we choose to use the classical classification region, not only for ease of comparison with previous work, but also because this is the part of the spectrum where B-type stars are richer in features, and the creators of the system insisted on using it as a reference for other spectral regions. Traditionally, this classification region extends from 3950\,\AA\ to 4750\,\AA, corresponding to the range of sensitivity of the photographic plates. With modern detectors, sensitivity is, on the other hand, higher to longer wavelengths, and the classification region has been extended to include H$\beta$, which is a powerful diagnostic for stellar winds or disk emission (e.g. \citealt{sota11}, or the GOSC catalogue; \citealt{maiz16}). In most figures, we have only shown the classical 3950\,--\,4750\,\AA\ range (broader-range spectra are shown in some spectral sequences in the Appendices), but we have consistently used the spectra up to 5100\,\AA\ for classification.

A key feature of our approach is the attempt to keep as many of the primary standards as possible in order to fully respect the philosophy of the MK73 system. As mentioned, there are many reasons why a given standard may have to be discarded, and indeed a number of dagger standards have been removed from the grid. The motivation for this decision is presented case by case in Appendix~\ref{app:gone}, while the list of surviving primary standards is given in Table~\ref{tab:daggers}. Importantly, all the anchor points of \citet{garrison94} are kept, although $\upsilon$~Ori has been moved to O9.7 and thus falls outside the table (see Appendix~\ref{upsori}).

\subsubsection{Choice of resolution}
\label{sec:resol}

High resolution is not needed for an accurate classification of B-type stars, since the density of spectral features is not high and the line profiles are generally broadened by a number of processes, among which rotation and macroturbulence are dominant in most cases \citep[e.g.][]{simon14rot, simondiaz17}. For digital spectra, a high S/N is probably more important to obtain an accurate classification than resolution. Reaching a high S/N is not difficult when the grid of standards is observed, but may be more challenging for the stars whose classification must be accomplished. For this reason, we decided to use a resolution as low as it may seem sensible. On the other hand,  to try to fight the limitations found in the \citetalias{keenan85} system, a higher resolution than the original $\sim2$\,\AA\ will be needed. 

Our compromise involves performing the classification at $R\sim$\,4\,000. This is comparable to the resolution used by \citet{walborn71} to investigate the stability of the reference frame with respect to resolution, and allows for most modern spectrographs on professional telescopes to secure moderately high S/N with modest exposure times on relatively faint targets. Moreover, as discussed in Sect.~\ref{sec:issues}, the higher the spectral resolution used, the more critical the relative effect of \vsini\ on the strength of the various key diagnostic lines becomes. Thus, ideally, lower values of $R$ imply less dependence on \vsini\ effects. On the contrary, lower values of $R$ imply a higher difficulty in detecting faint lines, especially in fast rotators. Following the decisions taken by \citet{Markova2011} for the case of O-type stars, and after performing some tests ourselves on B-type spectra, we confirm that $R\sim$\,4\,000 is also the best compromise to account for these effects.

\begin{figure}
\resizebox{\columnwidth}{!}{\includegraphics{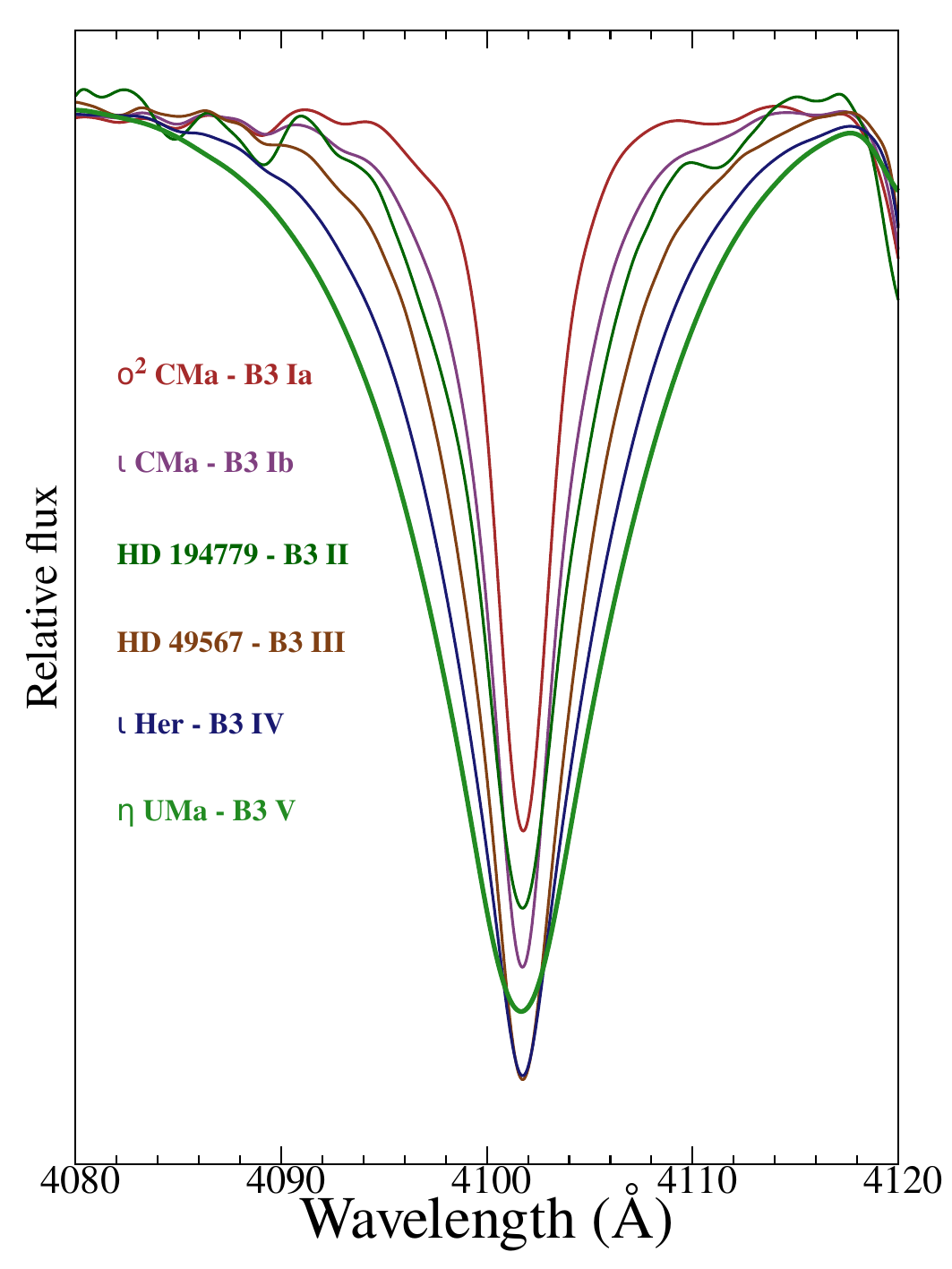}}
\centering
\caption{H$\delta$ line for a number of B3 stars illustrates our criterion for assigning luminosity class. We note that the B3\,II star is much closer to the supergiants than to the giant standard. The only fast rotator in the set is the  B3\,V standard $\eta$~UMa. Its profile is drawn with a thicker line, as high rotational velocity modifies its shape so importantly that it crosses the lines for higher luminosity standards. } 
\label{fig:wingsb3}
\end{figure}

\subsubsection{A word of caution about broad-lined stars}
\label{sec:rotbroad}

Our choice of resolution ($R$\,=\,4000) corresponds to a velocity resolution of \vsini\,$\approx$\,75~\kms. For stars with significantly smaller projected rotational velocity, the profiles of narrow lines should be dominated by instrumental broadening, blurring the difference between stars of diverse rotational velocities. For stars rotating faster than this velocity, projected rotational velocity becomes the main factor broadening the lines. When choosing standard stars, one is thus faced with a dichotomy. On the one hand, slow rotators will have sharper, well-marked features that can be more easily identified. However, many stars, especially dwarfs, subgiants and giants (i.e. luminosity classes V, IV, and III) can have much higher projected rotational velocities, easily reaching 300\,--\,350~\kms\ \citep[e.g.][]{Abt2002}. The estimated average of \vsini\ for B0\,--\,B9-type stars with these luminosity classes ranges between $\approx$75 and 150~\kms. The classifier must be aware of the differences introduced by this higher velocity when comparing such objects to the standards. Contrarily, a fast-rotating star is not a good standard, because many features are washed out or blended due to rotational broadening.
 
To counter this effect, we have (whenever possible) chosen a minimum of two standards for each spectral type, one with a rotational velocity below our resolution, and a moderately fast rotator. In total, our revised grid of B-type standard and reference stars comprises $\sim$130 objects with \vsini\,$\lesssim$\,70\,\kms, plus some 30 additional targets, between dwarfs, subgiants, and giants of diverse spectral types, with a \vsini\ in the range $\approx$\,100\,--\,300\,\kms\ (see Table~\ref{bigtable}). In a future work, we will explore whether quantitative measurements, such as equivalent widths, provide an objective way of establishing spectral types. For this paper, we have limited our pursuit to a purely morphological classification. Nevertheless, with current technology, it is easy and not time-consuming to generate grids of standard stars by artificially spinning up slowly rotating stars with tailored software. Such a procedure is preferably carried out by convolving our original spectra (at $R=85\,000$) with appropriate broadening functions.

\subsubsection{Luminosity classification}
\label{sec:lumis}

At lower resolutions, the overall profiles and strengths of Balmer and \ion{He}{i} lines have been used to classify mid- and late-B stars \citep[see][]{graybook}. As discussed in Sect.~\ref{sec:issues}, inconsistencies are found when the luminosity class of the standards is compared to their actual intrinsic brightness. For this reason, we have started out from a well-known observational fact: the effective gravity of stars determines to a large degree the appearance of Balmer lines in the B star range. Stellar models show beyond any reasonable doubt that the width of Balmer lines, and in particular the extension of their wings, correlates very strongly with effective gravity \citep[see also][and Sect.~\ref{quantities} for a posteriori confirmation]{deBurgos23}. This has always been considered a prime luminosity classification criterion for A-type stars \citep{graybook}. Exploration of the standard grid leads us to propose the following basic classification criterion: a star belongs to a given luminosity class (e.g. III) if the wings of a given Balmer line (e.g. H$\delta$) fit comfortably inside the profile of the same Balmer line in a star of the next higher luminosity class (IV, in this case) at our classification resolution ($R\,=\,4\,000$). This criterion is nicely illustrated in Fig.~\ref{fig:wingsb3} for the H$\delta$ line in stars of spectral type B3 (see Fig.~\ref{fig:wingsb8} for spectral type B8), but it should also be valid for the three Balmer lines in the classification region, for H$\beta$ (see Fig.~\ref{fig:wingsb7}) and for H$\zeta$ (for successive Balmer lines, the surrounding continuum may not be sufficiently well defined to apply the criterion). The global effect can be appreciated, for example, in Fig.~\ref{fig:B8seq}. Of course, a correct application of this criterion requires careful, homogeneous normalisation of the spectra, which is not always easy to achieve. Moreover, this criterion cannot be used effectively at low resolution. We have explored the effect of degrading resolution on its application, and the criterion can be used, at high S/N, down to resolving powers of 1\,500\,--\,2\,000. Nevertheless, this limitation does not impede classification. Once the standard grid is firmly set by using this procedure, classification can resort back to the traditional methodology of seeking the highest similarity with one of the standard spectra.

 \begin{figure*}[t!]
\resizebox{\columnwidth}{!}{\includegraphics{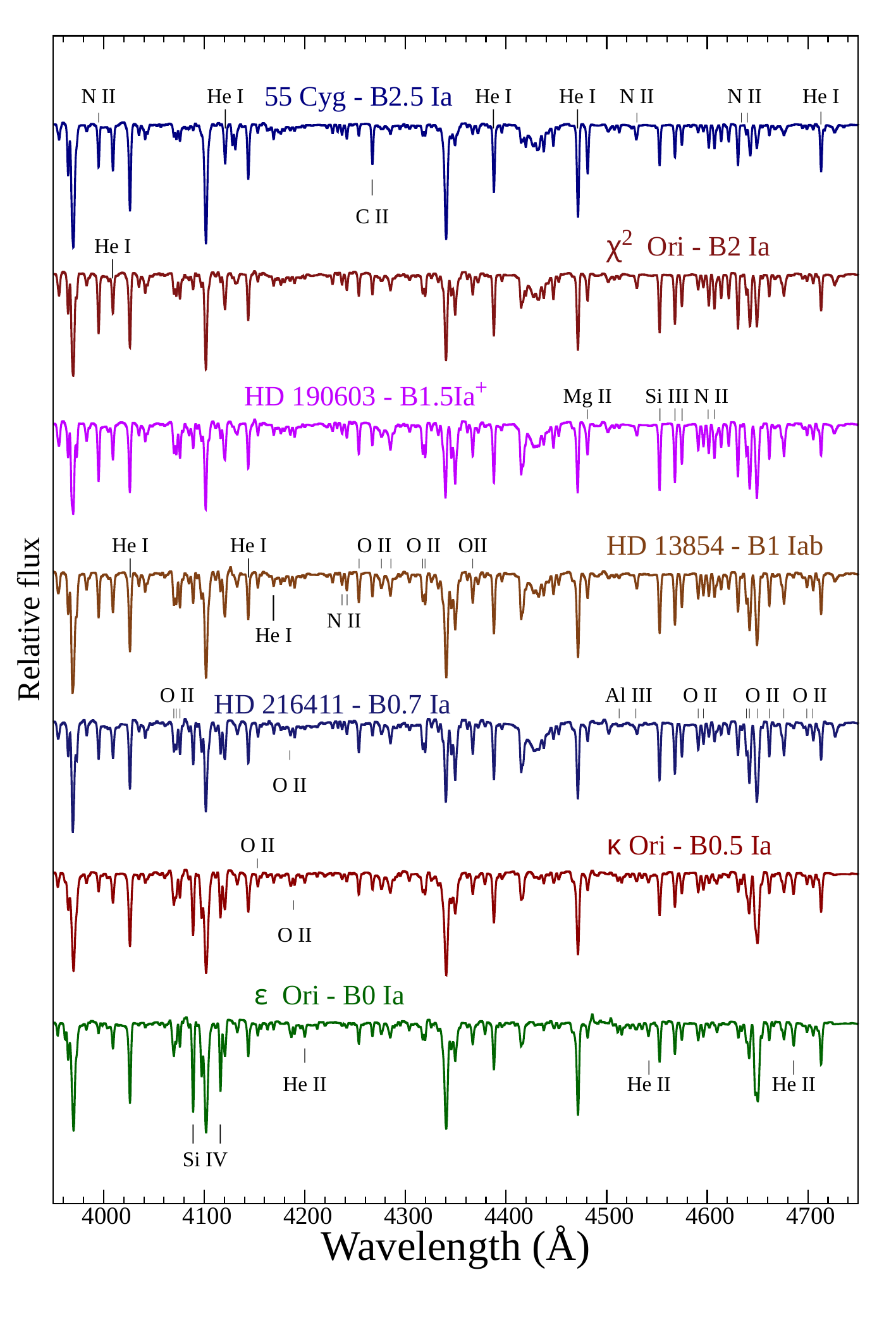}}
\resizebox{\columnwidth}{!}{\includegraphics{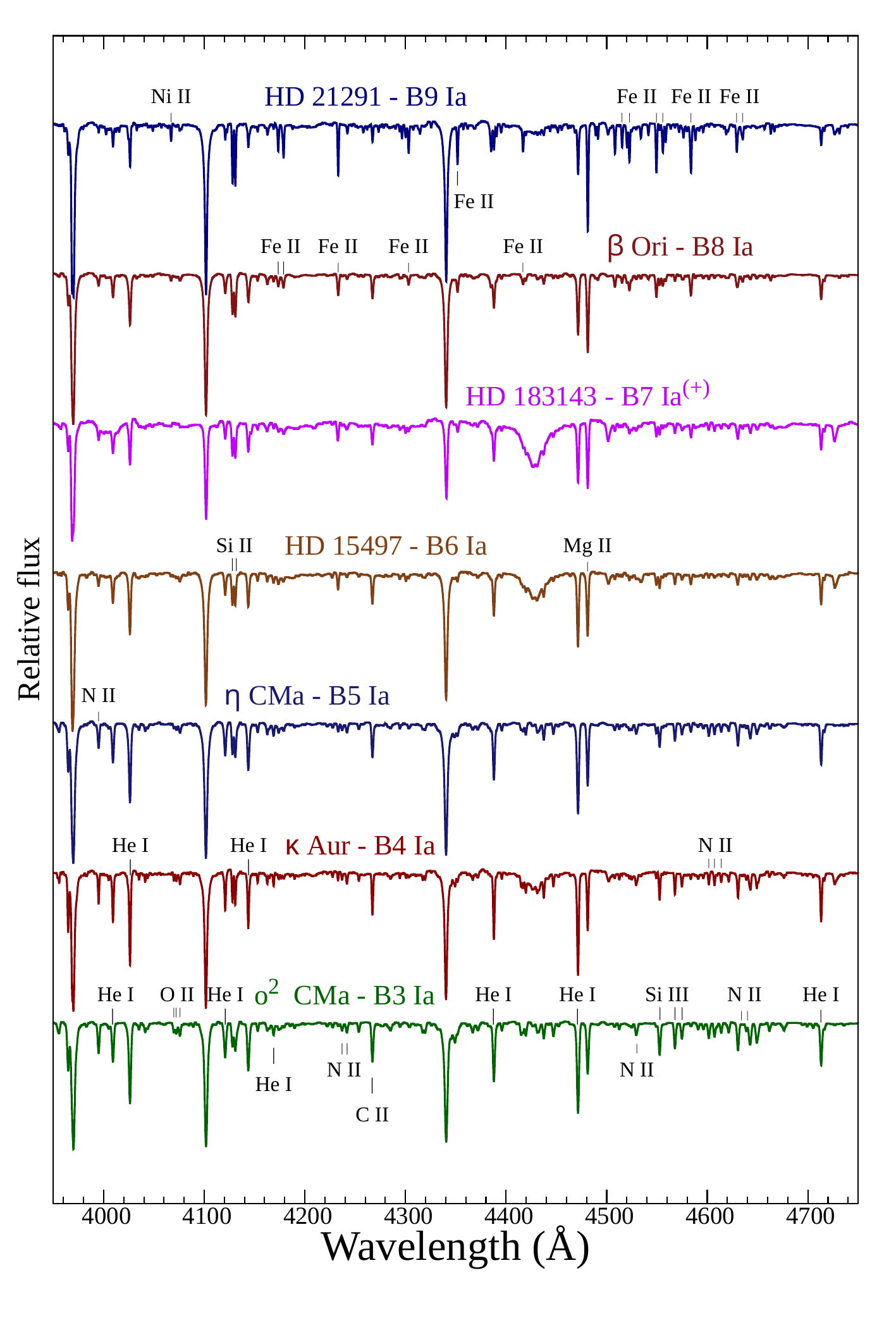}}
\centering
\caption{Spectral sequence for luminous B-type supergiants. Left panel shows the evolution of the \ion{Si}{iii}/\ion{Si}{iv} ratio, the growth of the \ion{O}{ii} spectrum to a maximum at B1, and its subsequent decrease, and the rise of the \ion{N}{ii} spectrum to a maximum at B2. Right panel illustrates the smooth growth of the \ion{Si}{ii} doublet and \ion{Mg}{ii}~4881\,\AA\ with respect to the neighbouring \ion{He}{i} lines. \ion{Si}{iii}, \ion{O}{ii} and \ion{N}{ii} are essentially gone by B6, but the \ion{Fe}{ii} spectrum is already growing. We note that some of these spectra, unlike those of the main sequence stars, contain deep diffuse interstellar bands, which are more prominent in the spectrum of HD~183143, where the broad band at $\sim4430$\,\AA\ is the strongest feature. Most of Garrison's anchor standards for the MK system (cf.~Table~\ref{tab:daggers}) are included in this sequence. \label{fig:bsgs} } 
\end{figure*}

\begin{figure*}[t!]
\resizebox{\columnwidth}{!}{\includegraphics{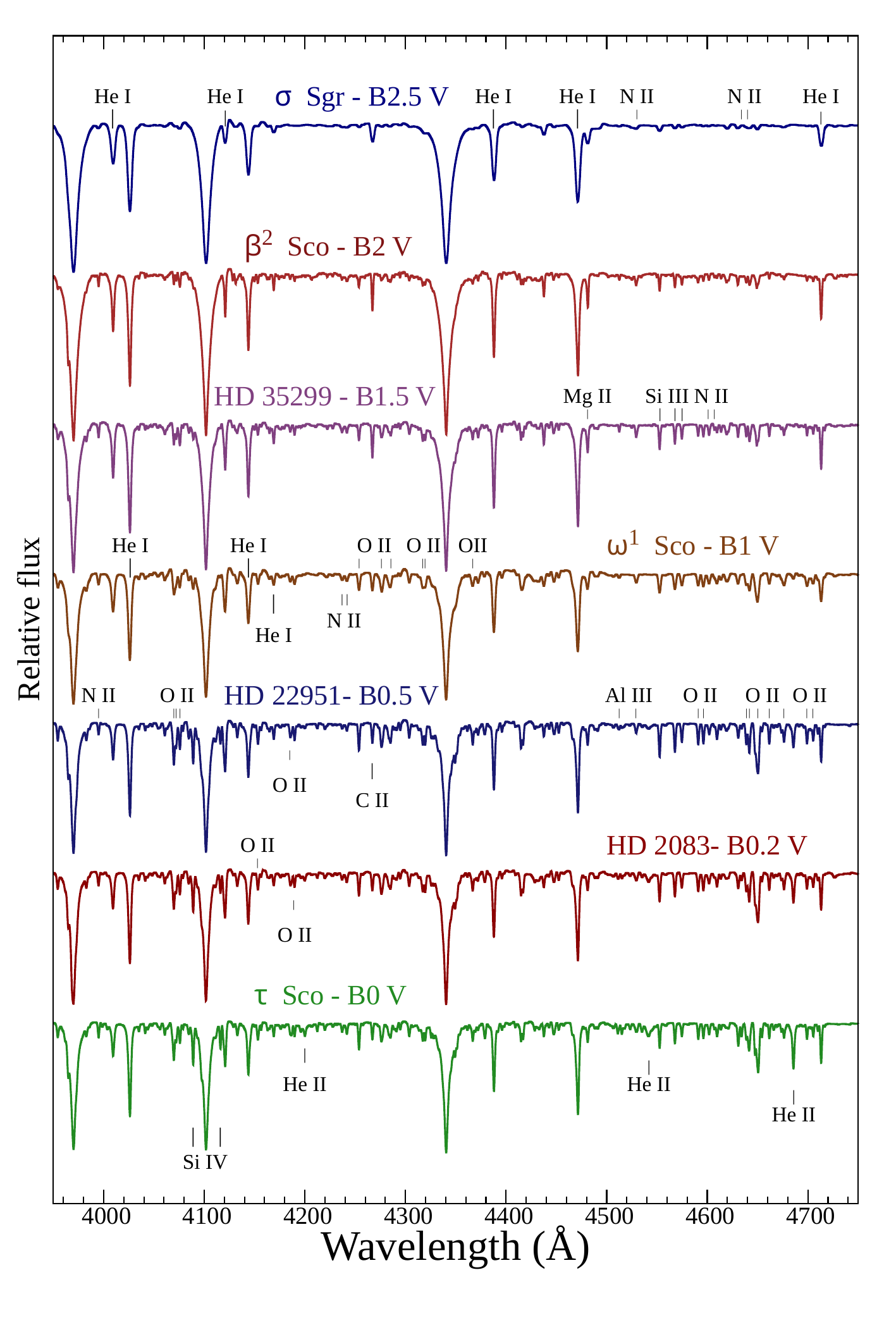}}
\resizebox{\columnwidth}{!}{\includegraphics{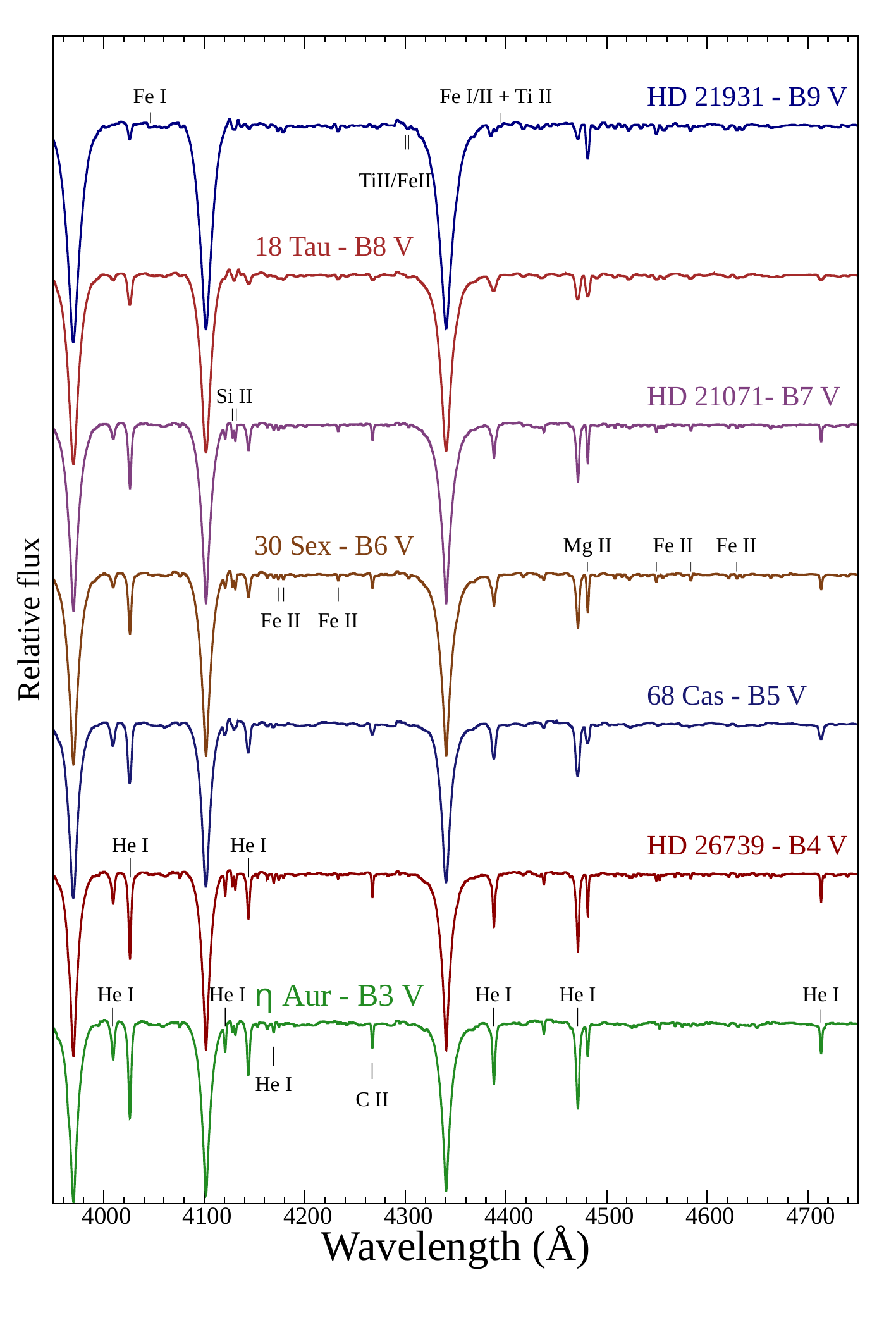}}
\centering
\caption{Spectral sequence for B-type dwarfs. Stars in the left panel present numerous \ion{O}{ii} lines (some of which are mixed with \ion{C}{iii} features in objects earlier than B1), together with \ion{Si}{ii/iii/iv} features that provide a powerful temperature discriminator. Stars in the right panel have simpler spectra. Objects later than B5 present some \ion{Fe}{ii} lines, but they tend to rotate fast and all weak features become very shallow and broad, as in the B8\,V standard 18~Tau, a member of the Pleiades.}
\label{fig:bms}
\end{figure*} 

\section{Classification criteria}\label{sec:crit}

Historically, B-type stars were defined as those with \ion{He}{i}, but no \ion{He}{ii} lines in old photographic spectra. This is still generally true with modern digital spectra, but \ion{He}{ii}~4686\AA\ can be seen as late as B0.7, while \ion{He}{i} lines can be found in the spectra of early-A supergiants. An overview of B-type spectra can be obtained by looking at Figures~\ref{fig:bsgs} and~\ref{fig:bms}, which display luminous supergiant and dwarf standards, respectively. In dwarfs,  the strength of the Balmer lines increases monotonically with spectral type, reaching a maximum close to A2 (Fig.~\ref{fig:bms}). However, it is easy to see, by comparing Figures~\ref{fig:bsgs} and~\ref{fig:bms}, that the strength of the Balmer lines diminishes with increasing luminosity, with all the Ia supergiants having H lines of comparable strength. Another global effect seen in Fig.~\ref{fig:bms} is the evolution of the \ion{He}{i} spectrum, which grows moderately with spectral type among the early types, reaches a maximum at B2 and then gradually weakens until its disappearance around A0. The dependence is well behaved. In fact, for B-type stars with normal He abundances, it is possible to determine moderately accurate stellar parameters from the properties of H and He lines alone \citep[e.g.][]{castro12}. Nevertheless, B-type stars contain many other features suitable for classification, and a very useful description of general criteria for the spectral classification of B-type stars based on modern spectra can be found in \citet{graybook}.

The main novelty of our methodology lies in the use of the width of the wings of Balmer lines as the principal luminosity classification criterion. Once a luminosity class scale has been established with this simple criterion, we just need to search for consistency across the standard grid, by relating the spectral type to the traditional criteria. This procedure, which is fully developed in Appendix~\ref{app:grid}, leads to some reassignments of spectral type and the definition of a new list of standards, still anchored on the set of fundamental MK standards listed in Table~\ref{tab:daggers}.  Our new luminosity criterion proves highly effective for mid- and late-B stars of low and moderate luminosity, particularly addressing the inconsistencies observed in traditional standards within this region. Nevertheless, given its primarily comparative nature, some caution is needed when its edges are considered. The key elements that we have taken into account are as follows. 

\begin{itemize}

\item {At early types: }  For spectral types earlier than B2, the Balmer lines start to lose their sensitivity to luminosity (see e.g. Fig.~\ref{fig:B1seq}), which is almost gone by B0. {Although the wing-width criterion can still be used for a rough estimation of the luminosity of early B stars, it is preferable to rely on classical criteria based on} the ratios of Si lines to their \ion{He}{i} neighbours. Clear rules have been put forward by \citet{walborn90} and \citet{sota11}. 

This choice leads to a natural separation of the B-type stars in two groups. For stars earlier than B2, we list the most important spectral type and luminosity criteria in Table~\ref{tab:earlyB}, noting that in the B0\,--\,B0.5 range we have simply adopted the standard stars and criteria of \citet{sota11}, whenever possible. The traditional criteria are, of course, rather sensitive to metallicity, as they involve comparison of an $\alpha$ element to He. This is, however, an intrinsic characteristic of the system: the spectral type criteria for mid-B stars also involve comparison between lines of Si (and Mg) and \ion{He}{i}, as these are the only features seen in their spectra. The grid and accompanying classification criteria presented here assume a standard solar neighbourhood composition \citep{nieva12}. For a comprehensive discussion of the requirements for extension to a distinctly lower metallicity, we refer the reader to \citet{evans03} and \citet{evans04}.

\item {Close to the ZAMS: }  As we come to late-B stars (which have $\sim3\:$M$_{\sun}$) stellar lifetimes are sufficiently long for evolutionary effects to be observable on the main sequence. A B8\,V star in the open cluster NGC~2516 \citep{abt69} has already lived a significant part of its main sequence lifetime, while a B8\,V star in a young cluster, such as NGC~2264 \citep{morgan65} can be considered to lie on the Zero Age Main Sequence (ZAMS). Because of the very strong dependence of the Balmer line width on gravity for stars close to A0 \citep[see][chapter 5]{graybook}, in the range between B9 and A2, it seems possible to distinguish between ZAMS stars and normal main sequence stars. In view of this, \citet{morgan78} split spectral types B9\,V and B9.5\,V into luminosity classes Va and Vb, with Vb representing stars with very wide lines. We are not adopting this division here, but are aware of the possible dependence of spectral features with stellar age and the danger {this entrails for a system based on direct comparison of line widths}.

For this reason, we have tried to tie our set of standards to stars in open clusters. In particular, for the early-B type dwarfs, we have selected, when possible, standards among a set of B dwarfs in the Orion star-forming region with well-established spectral classification and accurate physical parameters determined from atmospheric analysis \citep{simon10}. For mid- and late-B stars, we have used confirmed members of three northern open clusters whose parameters have been determined with very high accuracy by the TGAS release of \textit{Gaia} data \citep{leeuwen17}, namely, the $\alpha$~Persei cluster, IC~4665, and the Pleiades. The latter was already employed as an anchor by \citet{mk73}, who used several stars in the Pleiades as dagger standards, and benefited from the fact that stars of luminosity classes V, IV, and III can be found in this cluster. The other two are younger clusters, with the main sequence  extending up to B3\,V, which can provide examples of ZAMS stars. Details of these three clusters and an exhaustive description of their stellar content are presented in Appendix~\ref{app:clusters}.

\begin{sidewaystable*}
\caption{Main standard stars and classification criteria for early-B stars. Criteria in italics apply to all luminosity classes. \label{tab:earlyB}}
\centering
\begin{tabular}{l|c|c|c|c|c|c|c|c|}
\hline
\hline
\noalign{\smallskip}
 & \large{V}& \large{IV} & \large{III}& \large{II} & \large{Ib} & \large{Iab}& \large{Ia} & \large{Ia$^{+}$}\\
 \noalign{\smallskip}
\hline
\noalign{\smallskip}
\Large{B0}   & {\small  {$\tau$ Sco}}     &                 & {\small  {HD 48434}}      &                 & {\small  {HD~205196}}           &  & {\small  {$\varepsilon$ Ori}} & \\
  &                     &  \multicolumn{3}{c|}{\ion{Si}{iv}~4116/\ion{He}{i}~4121 increases} &   &    &   Very strong \ion{Si}{iv}~4089 & \\
  & \textit{All \ion{He}{ii} lines present}  & \multicolumn{3}{c|}{\ion{He}{ii}~4686 weaker compared  to}  & {\small  { HD~164402}} & & Very strong \ion{C}{iii}~4650 &         \\
  &see main text for criteria against O-type & \multicolumn{3}{c|}{other \ion{He}{ii} or \ion{He}{i}~4713}  & & & Weak \ion{He}{ii}~4686 & \\
\hline
\noalign{\smallskip}
\Large{B0.2} &     {\small  {  HD 2083 }}        & {\small  { $\phi^{1}$ Ori}} &    {\small  {HD 6675 }}          &    &            {\small  {HD~16808}}             &{\small  {69~Cyg}} & {\small  { HD~171012}} & \\              
  & & \multicolumn{3}{c|}{As in B0} & & & As in B0 & \\
  & \textit{\ion{He}{ii}~4200 hardly seen or absent} &&&&&&  & \\
\hline
\noalign{\smallskip}
\Large{B0.5} & {\small HD 36960}     &             &     {\small  { 1 Cas }}           & {\small  {HD~193007}}               & {\small  {26~Cep}}          &                & {\small  { $\kappa$ Ori}} &      \\
 &    &              &               &               & {\small  {HD~192422}}        &             & {\small  {HD~194839}} &      \\
& \textit{\ion{He}{ii}~4541 hardly seen or absent} &&&&&&  & \\
& \textit{\ion{Si}{iii}~4553 comparable to \ion{Si}{iv}~4089} &\multicolumn{6}{c|}{increasing \ion{Si}{iv}~4089/\ion{He}{i}~4144 and \ion{Si}{iii}~4553/\ion{He}{i}~4387}  & \\
\hline
\noalign{\smallskip}
\Large{B0.7}   & {\small  {HD 37042}},{\small  {HD 201795}}                        & {\small  {$\xi^{1}$~CMa}}  &       {\small  {HD~13969}}                  & {\small  {HD~193076}}                 &      {\small  {HD 190066}}    & & {\small  {HD~216411}} &\\                       
     &    &                        &      &                         &               {\small  {HD~14053}}    &            &               &          \\
          & \textit{\ion{He}{ii}~4686 barely noticeable}    &  \multicolumn{6}{c|}{\ion{Si}{iii}~4553 somewhat stronger than \ion{Si}{iv}~4089}            &                     \\
          & \textit{\ion{C}{iii}/\ion{O}{ii}~4650 weaker compared to other \ion{O}{ii}}&  \multicolumn{6}{c|}{\textit{Same luminosity criteria as in B0.5}}            &                     \\          
\hline
\noalign{\smallskip}
\Large{B1} &      {\small  {HD~24131}}, {\small  {$\omega^{1}$~Sco}}      &                        &    \multicolumn{2}{c|}{{\small  {$\beta$~CMa (B1\,II--III)}}}      &       {\small  {$\zeta$~Per}}                 &   {\small  {HD~13854}}               &            &  {\small  {HD~169454}}                        \\
     &    &                        &                       &                         &                   &             &                 &        \\
     &    \textit{\ion{He}{ii}~4686 not seen} &                        &                       &                         &                   &             &                 &        \\
          & \multicolumn{3}{c|}{\ion{Si}{iv}~4089 hardly seen, many \ion{O}{ii} lines}   & \multicolumn{3}{c|}{\textit{\ion{Si}{iii}~4553/\ion{He}{i}~4387, 4713 main luminosity criterion}}                         &                              &        \\
\hline
\noalign{\smallskip}
\Large{B1.5}   &  {\small  {HD 35299}},{\small  {HD 215191}} &                        & {\small  {12 Lac}}    &            {\small  {$\varepsilon$ CMa}}&   {\small  {HD~13841}}               & {\small  {HD~5551}}   &       {\small  {HD~14956}}  & {\small  {HD~190603}}                \\
     &   \multicolumn{3}{c|}{\ion{Si}{iv}~4089 not seen}                          &                   &                       &                   &     & \\
     &                      &                        &  &                       &    {\small  {HD~193183}}                &                  \multicolumn{3}{c|}{\ion{Si}{ii}~4128 appears}        \\
  &\textit{\ion{C}{ii}~4267 stronger than neighbour \ion{O}{ii}}   &                      &                        &  &                              &     \multicolumn{3}{c|}{Strong \ion{N}{ii} lines}       \\     
\hline
\end{tabular}
\end{sidewaystable*}

\item {Among supergiants: }  When we come to the most luminous supergiants, our criterion may not be fully reliable, because the Balmer lines become so narrow that their profiles can be dominated by stellar wind effects associated with heavy mass loss. There are large morphological differences between the Balmer lines of stars classified as Ia, even at a given spectral type. A cursory look at Fig.~\ref{fig:bsgs} shows that some stars (notably HD~216411, $\chi^2$~Ori and HD~183143) have much weaker Balmer lines than neighbouring standards, more similar to those of the B1.5\,Ia$^+$ hypergiant HD~190603. Mass loss in blue supergiants is related to luminosity, but also has a strong dependency on temperature. The H$\alpha$ line is almost always in emission in Ia supergiants, but we find a wide range of H$\beta$ profiles, from strong almost-symmetric absorption lines to very weak, peculiar features with emission components \citep[see][for a description of different types of profiles found in H$\beta$]{deBurgos23}. For this reason, it is preferable not to use H$\beta$ in the classification of supergiants, even if we extend the classification spectra to profit from the many metallic lines between $\approx4900$\,\AA\ and $\approx5100$\,\AA. In Section~\ref{sec:hyper}, we  come back to the defining characteristics of hypergiants, but all of them present emission profiles in H$\beta$ \citep{clark12_MW}, testifying to a very heavy mass loss. As can be seen in Table~\ref{tab:daggers}, the system is mostly defined in terms of the Ia standards. There are only two Ib standards, and no standards of class II or Iab are given. Although there are many reasons that can explain this situation, it presents a complication as it is not a good idea to try to define Iab starting from Ia, given the diversity of line strengths in the latter. {As we show in Sect.~\ref{sec:iab}, luminosity class Iab is ill-defined in many cases, and so the problem is likely inherent to the system and not to our choice of luminosity criteria. Contrarily, except for the earliest (<B1) supergiants, our criterion provides the best luminosity sorting.}
\end{itemize}

\begin{sidewaystable*}
\caption{Main standard stars and classification criteria for mid- and late-B stars. Criteria in italics apply to all luminosity classes. \label{tab:mainB}}
\centering
\begin{tabular}{l|c|c|c|c|c|c|c|}
\hline
\hline
\noalign{\smallskip}
 & \large{V}& \large{IV} & \large{III}& \large{II} & \large{Ib} & \large{Iab}& \large{Ia} \\
 \noalign{\smallskip}
\hline
\noalign{\smallskip}
\Large{B2}   & {\small  {$\beta^2$ Sco}}                      & {\small  {$\gamma$~Peg}}      &{\small  {$\gamma$~Ori}}           & {\small  {  HD 31327 }}   & {\small  {9~Cep}} & &{\small  {$\chi^2$ Ori}} \\
  & Very weak metallic lines                    &  \multicolumn{3}{c|}{\ion{Si}{ii}~4129 starts to be seen} &   &    &   Very strong \ion{N}{ii} lines \\
  & \textit{\ion{He}{i}~4009 strong compared to \ion{He}{i}~4026}  &   &  &  \multicolumn{4}{c|}{\ion{N}{ii} lines comparable  in strength to \ion{O}{ii}}       \\
  &\textit{\ion{C}{ii}~4267 is strong } & && & & & \\
\hline
\noalign{\smallskip}
\Large{B2.5} &     {\small  {  22 Sco }}        &  &    {\small  {$\pi^2$~Cyg}}          &    &            {\small  {3~Gem}}             & & {\small  {55~Cyg}} \\              
  &Even weaker metallic lines& \multicolumn{6}{c|}{\ion{N}{ii} stronger than \ion{O}{ii}}  \\
  & \textit{\ion{Mg}{ii}~4481 becoming strong} &&&&&&  \\
\hline
\noalign{\smallskip}
\Large{B3} & {\small  {$\eta$~Aur}}    &    {\small  {$\iota$~Her}}          &     {\small  {HD~21483}}           & {\small  {HD~194779}}               &  {\small  {$\iota$~CMa}}         &                & {\small  { $o^2$ CMa}}  \\
& \textit{\ion{Mg}{ii}~4481 clearly seen even in fast rotators} &\multicolumn{4}{c|}{\textit{\ion{Si}{ii}~4129\,$\gtrsim$\,\ion{Si}{iii}~4553}}&&  \\
&\ion{Si}{ii}~4129 becoming strong &&& \multicolumn{4}{c|}{\textit{\ion{Si}{iii}~4553 weaker than some \ion{N}{ii} lines}} \\
\hline
\noalign{\smallskip}
\Large{B5}   & {\small  {HD 4142}}                        &   &       {\small  {HD~170682}}                  & {\small  {HD~191243}}                 &      {\small  {HD 9311}}    &  {\small  {HD 7902}} & {\small  {$\eta$~CMa}}\\                       
     & \textit{\ion{Si}{ii}~4129 comparable to \ion{He}{i}~4121 }   &                        &      &                         &                 &            &                \\
          & \textit{\ion{He}{i}~4009 weaker compared to \ion{He}{i}~4026}    &  \multicolumn{6}{c|}{\ion{Mg}{ii}~4481 growing in strength with luminosity}            \\
          & \textit{\ion{C}{ii}~4267 becoming weaker} &  \multicolumn{5}{c|}{\textit{}}            &   \ion{Fe}{ii} very weakly present                  \\          
\hline
\noalign{\smallskip}
\Large{B6/7} &      {\small  {HD~21071}}      &                        &    {\small  {$\beta$~Tau}}      &      & & {\small  {13~Cep}}                 &   {\small  {HD~199478}}         \\
     &    &                        &                       &                         &                   &             &                 \\
     &    \textit{\ion{He}{i}~4009, 4388 decline strongly}  &  \textit{\ion{Si}{ii}~4129 comparable or}                      &                     \multicolumn{5}{c|}{\textit{Same trends as in B5}}            \\
          & \textit{with respect to other \ion{He}{i} lines}   & \textit{stronger than \ion{He}{i}~4144}                        &     &&&&                      \\
\hline
\noalign{\smallskip}
\Large{B8}   &  {\small  {21 Tau}} &     {\small  {$\pi$ Cet}}                    & {\small  {$\tau$ And}}     &           &   {\small  {53~Cas}}               & {\small  {HD~14542}}   &       {\small  {$\beta$~Ori}} \\
     &   \multicolumn{3}{c|}{\ion{Mg}{ii}~4481 comparable}                          & &&&  \\
     &   \multicolumn{3}{c|}{to \ion{He}{i}~4471}                     &                      &       \multicolumn{3}{c|}{Strong \ion{Fe}{ii} }        \\
  &  \textit{\ion{He}{i}~4009, 4388 very weak}   &                      &                        &  &                  \multicolumn{3}{c|}{\ion{Mg}{ii} still growing }        \\     
  \hline
\noalign{\smallskip}
\Large{B9}   &  {\small  {134~Tau}} &        {\small  {($\alpha$~Del)}}                & {\small  {12 Cas}}    &            {\small  {HD~21661 (II--III)}}&   {\small  {HD~35600}}               & {\small  {$\sigma$~Cyg}}   &       {\small  {HD~21291}}               \\
     &  \multicolumn{3}{c|}{\ion{He}{i}~4009, 4388 not seen}                         &                   &                       &                   &      \\
     &  \textit{\ion{Mg}{ii}~4481 much stronger than \ion{He}{i}~4471}                    &        &                &      &                        \multicolumn{3}{c|}{\textit{Weak \ion{He}{i}  }}                \\
  &\textit{Metallic lines even in fast rotators}   &                                             &  &                              &     \multicolumn{3}{c|}{\textit{Very strong \ion{Si}{ii} and \ion{Fe}{ii}}}      \\     
\hline
\end{tabular}
\end{sidewaystable*}

{For stars of spectral type B2 and later, the wings of Balmer lines on their own can be used to determine luminosity. At spectral types B2, B2.5 and B3 (except for dwarfs), the ratio of \ion{Si}{iii}~4553\,\AA\ to \ion{He}{i} lines can still be used as a luminosity criterion, and we do not find any inconsistencies between the two approaches. For later-type stars, in whose spectra we may find very few metallic lines, a number of general criteria can be used for classification. These criteria are:}

\begin{itemize}
\item At a given spectral type, the \ion{He}{i} lines become narrower and deeper with increasing luminosity. Combined with the narrowing of Balmer lines and presence of weak metallic lines, this results in a very large difference in general appearance between supergiants and dwarfs, with the latter dominated by a few broad Balmer lines, while the former display many narrow lines (see, for example, Figures~\ref{fig:B5seq}, \ref{fig:b7seq}
and~\ref{fig:B8seq}).
\item As we move to later spectral types, the \ion{He}{i} spectrum becomes weaker. Not all lines weaken at the same pace, and thus pairs such as \ion{He}{i}~4471\,\AA/\ion{He}{i}~4387\,\AA\ or \ion{He}{i}~4009\,\AA/\ion{He}{i}~4026\,\AA\ are sensitive to spectral type. The latter is particularly useful, as the two lines lie close together. Likewise, the combination of growing Balmer and decreasing \ion{He}{i} strength provides a quick assessment of the spectral type of non-supergiant stars.
\item The ratios \ion{Mg}{ii}~4481\,\AA/\ion{He}{i}~4471\,\AA\ and \ion{Si}{ii}~4129\,\AA/\ion{He}{i}~4144\,\AA\ are responsive to luminosity within a given spectral type, but also to temperature, increasing with growing spectral type and decreasing luminosity class.  In consequence, it is preferable to determine the luminosity class first from the width of the Balmer lines and then the spectral type from the strength of the \ion{He}{i} and metallic lines, assuming no chemical peculiarities.
\end{itemize}

{A number of other, more specific criteria, relevant to the definition of individual types, is given in Table~\ref{tab:mainB}. A further complication comes from chemical peculiarities. Stars with non-standard He surface abundances are found in relatively small numbers \citep{cidale07}. Stars with weak He are found at mid and late spectral types, while He-strong stars concentrate around spectral type B2 \citep{walborn83}, in many cases in connection with large-scale magnetic fields \citep{schultz19}. Among stars later than B7, we find a moderate fraction of chemically peculiar Bp stars. These objects are naturally connected to the Ap stars. A very large fraction of A-type stars are chemically peculiar \citep{graybook}, to the point that it is difficult to define what a ``normal'' A-type star is. \citet{abt95} found that two of the main standards for the spectral type, Vega (A0\,V; one of the anchor points of \citealt{garrison94}) and $\alpha$~Dra (A0\,III) show mild chemical peculiarities. Moreover, these peculiarities seem to be intimately related to rotational velocity, with Ap (and also the Am stars, which tend to occur in binaries) having much lower rotational velocities (on average) than ``normal'' early-A stars \citep{abt95}. This behaviour is almost certain to extend to the late-B stars. Moreover, based on the distribution of rotational velocities among a large sample of A0\,--\,A1 stars of luminosity classes IV and V and their \textit{Hipparcos} parallaxes, \citet{abt00} speculated that the separation between luminosity classes V and IV may not be good enough on photographic spectra. In view of these complications, although the effects of Stark broadening guarantee that we can effectively separate well the luminosity class of late-B stars, traditional standards may not be a safe guide.  Criteria for chemically peculiar stars and references for further study are given in \citet{graybook}.}

%\section{A revised grid of B-type standards}\label{sec:typedescription}

%Armed with our well-defined luminosity criteria, we can now explore -- and redefine when convenient -- the standard grid. For illustration, we show the whole B sequence for luminous supergiants in Fig.~\ref{fig:bsgs} and the whole B sequence for dwarfs in Fig.~\ref{fig:bms}. Both figures can be used as reference for the characteristics given for each subtype in the corresponding subsection of Appendix~\ref{app:grid}. 
%%%Other spectral sequences illustrating the evolution of classification criteria with spectral type and luminosity class are given in Appendix~\ref{app:sequences}. 
%In the main text, we only include the spectral subtypes that are in general use and we consider necessary for a fine grid (although we note some caveats for B0.7 and B6/7). Other interpolated subtypes which have been widely used and may be convenient in some cases (B4, B8.5 and B9.5) are discussed in full in Appendix~\ref{app:others}.

\section{Discussion}
\label{sec:discuss}

\subsection{A revised grid of B-type standards}
\label{sec:typedescription}
{Armed with a high-quality and homogeneous spectroscopic dataset and our well-defined luminosity criteria, we have explored -- and redefined when convenient -- the standard grid by requiring consistency, whenever possible, with the traditional criteria for spectral type classification. In Appendix~\ref{app:grid}, we detail the defining characteristics of the main spectral subtypes, list primary and secondary standards used in literature, and select appropriate replacements for those that are discarded, as well as new standards to fill in the gaps. Some primary standards have been disused, and we discuss the motivations leading to their exclusion in Appendix~\ref{app:gone}. Some intermediate spectral types that have been used in previous works are not considered fully justified within the present framework. These include B6 -- which is discussed together with B7, as some of the main standards are perfectly interchangeable --, B4 -- which is useful, but would require a complex reassignment of primary standards --, B8.5 and B9.5, both of which are felt to be unnecessary. Their characteristics and representative stars are presented in Appendix~\ref{app:others}.}

{Table~\ref{bigtable} contains all the stars selected as useful standards and other reference stars that illustrate the spectral types, but cannot be considered standards, and provides basic observational data on them. A selection of the best B-type standards, covering as many subtypes as we have been able to, is given in Table~\ref{tab:std_sqr}. Since instrumental broadening should dominate profiles for stars with low projected rotational velocity, in Table~\ref{tab:std_sqr} we highlight stars with \vsini$\lesssim$70~\kms (cf. Sect.~\ref{sec:crit}) for luminosity classes III\,--\,V (this is not an important issue for supergiants). }

{For illustration, we show the whole B sequence for luminous supergiants and dwarfs in Figs.~\ref{fig:bsgs} and \ref{fig:bms}, respectively. Both figures can be used as reference for the characteristics given for each subtype in the corresponding subsection of Appendix~\ref{app:grid}. }
%Other spectral sequences illustrating the evolution of classification criteria with spectral type and luminosity class are given in Appendix~\ref{app:sequences}. 
%In the main text, 

{With the definition of a new grid of standards (Table~\ref{tab:std_sqr} and Appendix~\ref{app:grid}) and the new set of classification criteria adequate for modern CCD spectrographs (Sect.~\ref{sec:crit} and Tables \ref{tab:earlyB} and \ref{tab:mainB}), most of the major inconsistencies that plagued B-type classification must have been removed. There are a few minor standing issues, which  we comment on below (Sect.~\ref{issues}), before we proceed to check the consistency of the new grid against other indicators of stellar parameters (Sect.~\ref{quantities}).}

\begin{table*}
\caption{Main standards within the new grid$^{1}$. Among the non-supergiants, objects with low projected rotational velocity ($<70\:$\kms) are in italics.\label{tab:std_sqr}}
\centering
\begin{tabular}{l|c|c|c|c|c|c|c|c|}
\hline
\hline
\noalign{\smallskip}
 & \large{V}& \large{IV} & \large{III}& \large{II} & \large{Ib} & \large{Iab}& \large{Ia} & \large{Ia$^{+}$}\\
 \noalign{\smallskip}
\hline
\noalign{\smallskip}
\Large{B0}   & {\small  \textit{$\tau$ Sco}}     &                 & {\small  \textit{HD 48434}}      &                 & {\small  {HD~205196}}           &  & {\small  {$\varepsilon$ Ori}} & \\
 \hline
\noalign{\smallskip}
\Large{B0.2} &     {\small  \textit{  HD 2083 }}        & {\small  \textit{ $\phi^{1}$ Ori}} &    {\small  \textit{HD 6675 }}          &    &            {\small  {HD~16808}}             &{\small  {69~Cyg}} & {\small  { HD~171012}} & \\              
\hline
\noalign{\smallskip}
\Large{B0.5} & {\small  \textit{ HD 36960} }    &             &     {\small  \textit{ 1 Cas }}           & {\small  {HD~193007}}               & {\small  {26~Cep}}          &                & {\small  { $\kappa$ Ori}} &      \\
 &    &              &               &               & {\small  {HD~192422}}        &             & {\small  {HD~194839}} &      \\
\hline
\noalign{\smallskip}
\Large{B0.7}   & {\small  \textit{HD 37042}},                        & {\small  \textit{$\xi^{1}$~CMa}}  &       {\small  \textit{HD~13969}}                  & {\small  {HD~193076}}                 &      {\small  {HD 190066}}    & & {\small  {HD~216411}} &\\   &{\small  \textit{HD 201795}}        &                        &  {\small  {HD~14053}}     &                         &                  &            &               &          \\    
\hline
\noalign{\smallskip}
\Large{B1} &      {\small  {HD~24131}},       &                        &    \multicolumn{2}{c|}{{\small  \textit{$\beta$~CMa (B1\,II--III)}}}      &       {\small  {$\zeta$~Per}}                 &   {\small  {HD~13854}}               &            &  {\small  {HD~169454}}                   \\
&{\small  {$\omega^{1}$~Sco}}        &                        &      &                         &                  &            &               &          \\ 
\hline
\noalign{\smallskip}
\Large{B1.5}   &  {\small  \textit{HD 35299}} &                        & {\small  \textit{12 Lac}}    &            {\small  {$\varepsilon$ CMa}}&   {\small  {HD~13841}}               & {\small  {HD~5551}}   &       {\small  {HD~14956}}  & {\small  {HD~190603}}                \\
&{\small  {HD 215191}}        &                        &      &                         &      {\small  {HD~193183}}             &           &               &          \\ 
\hline
\noalign{\smallskip}
\Large{B2}   & {\small  \textit{HD 36629}}  & {\small  \textit{$\gamma$ Peg}} & {\small  \textit{$\gamma$ Ori}}   &            {\small  {HD 31327}}&   {\small  {9 Cep}}               &  &       {\small  {$\chi^{2}$~Ori}}  &               \\
&{\small  \textit{$\beta^{2}$ Sco}}       &  {\small  \textit{$\zeta$ Cas}}                      &  {\small  \textit{$\pi^4$ Ori}}      &                         &                  &          &  {\small  {HD 14143}}             &          \\ 
\hline
\noalign{\smallskip}
\Large{B2.5}   &  {\small  {$22$ Sco}} &                        & {\small  \textit{$\pi^2$ Cyg}}    &           &    {\small  {3 Gem}}             & {\small  {}}   &       {\small  {55 Cyg}}&               \\
&{\small  {$\sigma$~Sgr}}      &                        &      &                         &                  &          &             &          \\ 
\hline
\noalign{\smallskip}
\Large{B3}   &  {\small  {$\eta$ Aur}} &    {\small  \textit{$\iota$ Her}}                    & {\small  \textit{HD 49567}}  &            {\small  {HD 194779}}&   {\small  {$\iota$ CMa}}               &   &       {\small  {$o^{2}$ CMa}}   & {\small  {}}  \\
&  {\small  \textit{HD 178849}}   &   {\small  {$\xi$ Cas}}                      &   {\small  {HD 21483}}   &                           &                  &          &    {\small  {HD 14134}}         &          \\ 
\hline
\noalign{\smallskip}
\Large{B5}   &  {\small  \textit{$\tau$ Her}} & &{\small  \textit{HD~211924}}                       & {\small  {HD~191243}}    &            {\small  {HD 9311}}&   {\small  {HD 7902}}               & {\small  {$\eta$ CMa}}   &                 \\
& {\small  {HD 4142}}    &                        &   {\small  {HD 170682}}    &                       &                  &          &             &          \\ 
\hline
\noalign{\smallskip}
\Large{B7}   & {\small  \textit{3 Vul}} &         {\small  {19 Tau} (B6\,IV)}                & {\small  \textit{HD 1279}}    &           &               & {\small  {13 Cep}}   &       {\small  {HD 199478}}  & {\small  ({HD 183143})}                \\
&  {\small  {16 Tau}}   &                       &  {\small  \textit{$\beta$ Tau}}    &                          &                  &          &          &          \\ 
\hline
\noalign{\smallskip}
\Large{B8}   &  {\small  \textit{HD 171301}}& {\small  \textit{$\pi$ Cet}}&  {\small  \textit{21 Aql}}   &            {\small  {}}&   {\small  {53 Cas}}               & {\small  {HD 14542}}   &       {\small  {$\beta$~Ori}} & {\small  {}}                \\
& {\small  {21 Tau}}     &  {\small  {HD 3240}}                      & {\small  {$\tau$ And}}      &                         &                 &          &          &          \\ 
\hline
\noalign{\smallskip}
\Large{B9}   &  {\small  \textit{134 Tau}}&       {\small ({$\alpha$ Del})}                  & {\small  {32 Peg}}   &            {\small  \textit{HD 21661} (II--III)}&   {\small  {HD 35600}}               &{\small  {$\sigma$ Cyg}}   &       {\small  {HD 223960}} &               \\
& {\small  {HD 21931}}      &                      &  {\small  {12 Cas}}    &                         &        &          &     {\small  {HD 21291}}     &          \\ 
\hline
\end{tabular}
\tablefoot{$^{1}$  An online version of this table linked to downloadable spectra is available at \url{https://astroplus.ua.es/mkbtypestds}.}
\end{table*}
 
\subsection{Remaining issues}\label{issues}

The new selection of standard stars resolves many of the inconsistencies mentioned by \citet{jaschek98}. This does not mean that the system is fully self-consistent, even if it is much more self-consistent than previous versions. There are a number of questions, both methodological and physical, that still demand special attention when addressing them. We now list the most important ones.

\subsubsection{The earliest B-type stars}
\label{sec:earliest}

Stars of spectral types B0\,--\,B0.5 were included in the work by \citet{sota11} on the classification of O-type stars. We have chosen to respect their choice of standards, even though some of them are certainly problematic. The B0\,V standard, $\tau$~Sco is known to be a magnetic star with extremely slow rotation, while the B0.2\,V standard HD~2083 is an SB2 when seen at high resolution. Beyond this practical complication, we feel that this spectral range needs to be re-examined with great care. Although \citet{walborn71} introduced the interpolated spectral types B0.2 and B0.7 to eliminate inconsistencies that were observed when trying to place stars within the grid, some of such inconsistencies seem to persist when they are used. Many evolved stars in this range still present absolute magnitudes at odds with their luminosity class. An example at hand is HD~23675, whose spectrum is almost indistinguishable from that of the B0.2\,III standard. Its membership in the open cluster NGC~1444 implies an absolute magnitude close to those of supergiants. Conversely, HD~205196, which has been consistently given as a B0\,Ib standard since \citet{morgan50}, has an absolute magnitude typical of a giant \citep[cf.][confirmed by its \textit{Gaia} distance]{humphreys78}. 

The reason for these discrepancies is unclear. Unfortunately, in this range of temperatures, Balmer lines are not so responsive to luminosity as in later spectral types. Although they can still be used for a basic distinction between dwarfs, giants, and supergiants, the criterion used at later types cannot be applied. In some cases, the discrepancy may be due to hidden binarity. Possible examples are stars such as HD~217490, which was classified B0\,II by \citet{walborn71} but displays an absolute magnitude typical of luminous supergiants\footnote{Interestingly, this object was classified B0.5\,Ia by \citet{morgan55}, although our spectrum seems compatible with Walborn's classification. Two unresolved components mimicking a lower luminosity are a possible explanation for this discrepancy}. This, however, is unlikely to explain all the cases, as many stars with discrepant luminosity do not show any evidence of binarity at high resolution. There may be an inconsistency intrinsic to the system: the main spectral type indicator is the ratio of \ion{Si}{iii} to \ion{Si}{iv}, while the luminosity class is mostly determined by the ratio of the same Si lines to neighbouring \ion{He}{i}~lines. However, the change in the \ion{Si}{iii}/\ion{Si}{iv} ratio is so large between B0 and B1 that even with interpolated subtypes, the range of ratios falling within a given subtype may be too broad to allow for a consistent luminosity classification. Examples of giants and luminous supergiants covering the whole spectral range can be found in Figures~\ref{fig:b0b1iii} and~\ref{fig:b0b1Ia}, respectively, where the very large changes in the depth of the \ion{Si}{iv} line at 4089\,\AA\ between consecutive subtypes can be appreciated.

\begin{figure*}
\resizebox{\textwidth}{!}{\includegraphics{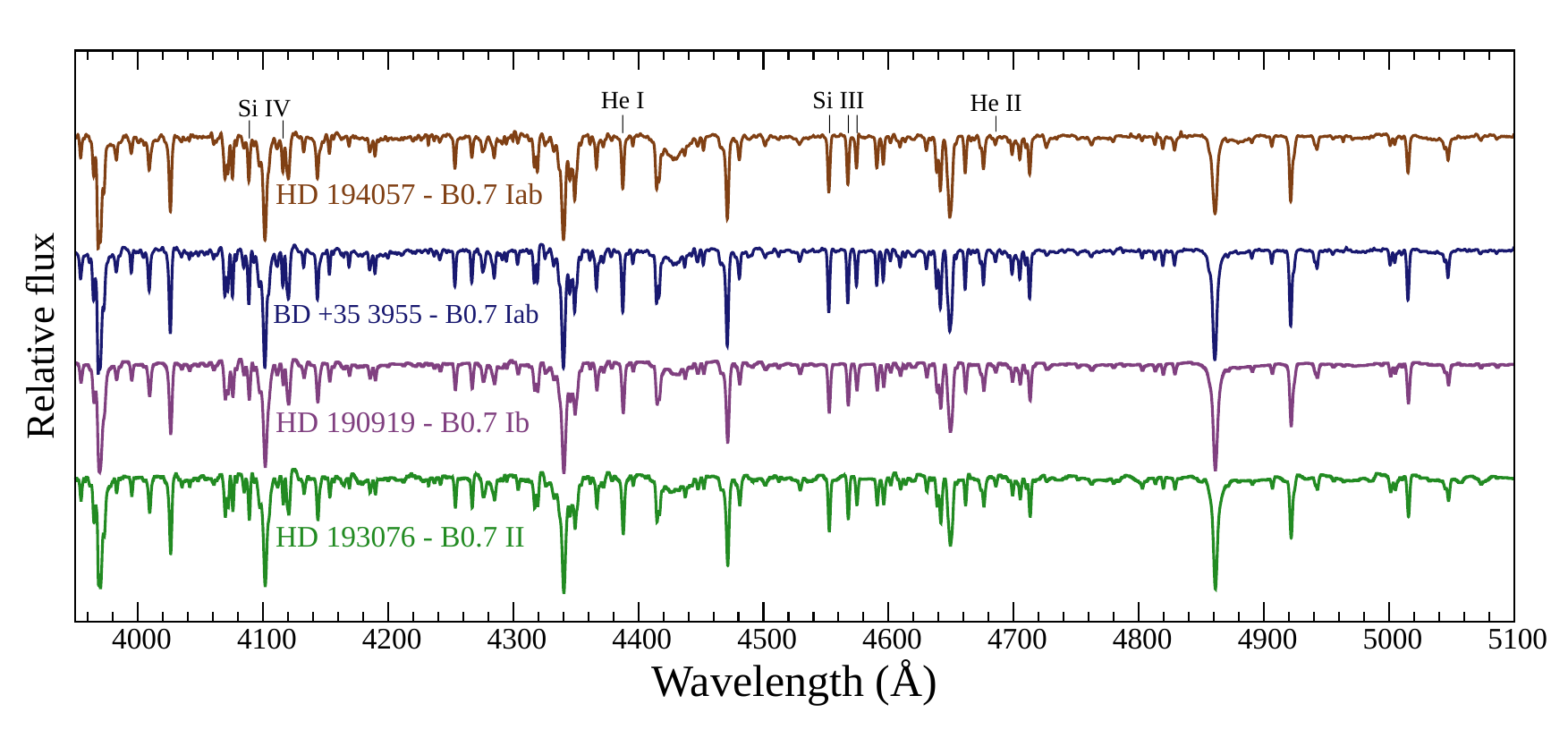}}
\centering
\caption{Four stars classified as B0.7 supergiants by \citet{walborn71}, with their original luminosity classes.  Inspection of the region around H$\delta$, with application of the \ion{Si}{iv}~4116\,\AA/\ion{He}{i}~4144\,\AA\ criterion, supports these classifications. However, the primary luminosity criterion \ion{Si}{iii}~4553\,\AA/\ion{He}{i}~4387\,\AA\ suggests that the three bottom stars all have approximately the same luminosity. The higher luminosity of HD~194057 is also reflected in the much weaker Balmer lines and stronger \ion{O}{ii} features. The strong \ion{Si}{iv} lines in the spectrum of BD~$+35\degr$3995 likely reflect a slightly higher temperature. See Fig.~\ref{fig:bsgs} for the corresponding Ia standard and Fig.~\ref{fig:b07is} for a comparison of HD~190919, which was Walborn's standard, to stars of similar spectral types.  \label{fig:temporlum} } 
\end{figure*}

For further illustration, in Fig.~\ref{fig:temporlum} we show a number of stars classified as B0.7 by \citet{walborn71}. They are all located in the Cygnus region, at similar distances from the Sun, and we can thus assume that metallicity is about the same for all of them. They all fulfil our condition for B0.7, as they display a very weak, but clearly visible \ion{He}{ii}~4686\,\AA\ line and no trace of the other \ion{He}{ii} lines. They also fulfil the classical criterion of having \ion{Si}{iii}~4553\,\AA\ slightly stronger than \ion{Si}{iv}~4089\,\AA. However, direct comparison of the Balmer lines shows that HD~194057 cannot have the same luminosity as BD~$+35\degr$3995, while the latter cannot be substantially more luminous than HD~193076. In fact, both HD~190919 (Walborn's B0.7\,Ib standard) and BD~$+35\degr$3995 are connected to the open cluster NGC~6871\footnote{HD~190919 is not listed as an astrometric member of NGC~6871 by \citet{hunt23}, but it is located at the same distance, about 6~arcmin to the south of the cluster core, and its proper motions suggest it has recently been ejected from that area.} and they have almost identical magnitudes and colours. Accepting the cluster parameters from \citet{hunt23}, both have $M_V\approx -5.3$, which would definitely put them at luminosity class II according to Walborn's own calibration \citep{walborn72}, together with HD~193076.

Why was then BD~$+35\degr$3995 classified as a much more luminous star? Very likely because its \ion{Si}{iv} lines are stronger (when compared to the neighbouring \ion{He}{i} lines) than those of HD~190919 and HD~193076. If our interpretation is correct, this is not reflecting higher luminosity, but higher temperature, still within the B0.7 type. Indeed, there may be other physical reasons contributing to varying \ion{Si}{iii}/\ion{Si}{iv} ratios, such as microturbulence, and so the actual appearance of an early B star may depend on subtle effects. In fact, when we look at the physical characteristics of stars classified B0.7\,Ib and B0.7\,II in Table~\ref{bigtable}, they do not appear well separated. Nevertheless, we believe that classifying HD~190919 as B0.7\,II is more coherent and may help some inconsistencies to disappear.

%The same can be said of the brightest standards, as HD~216411 appears definitely later than $\kappa$~Cas, but it is also more luminous 

%Another interesting case is  the luminous supergiant HD~173438, which lies in between B0.5 and B0.7.

\subsubsection{Luminosity class IV}
\label{sec:iv}

Our sample contains a fair number of standards for luminosity class IV, but they are unevenly spread across the different spectral subtypes. The main reason for this is the scarcity of these stars, due to their own nature. While for sun-like stars, luminosity class IV is used for subgiants, among B-type stars, luminosity class IV objects are recognised as main sequence \citep{mkk}. This luminosity class is generally assigned to stars around the turn-off of young open clusters. Examples among the reference clusters discussed in Appendix~\ref{app:clusters} are the B6\,IV stars in the Pleiades and the B5\,IV stars in IC~4725. Unfortunately, the clusters that would be needed to populate this class for other types are either too faint for the instrumentation used (e.g. NGC~869 for B1\,IV and B1.5\,IV or NGC~663 for B2\,--\,2.5\,IV) or not visible from the North (e.g. NGC~6067 for B7\,IV). 
 
Luminosity IV B-type stars represent a transitional stage in the life of a star and are not very numerous. Even the populous clusters just mentioned contain only a handful of them. They are therefore rare in the field. Despite this scarcity, luminosity class IV was profusely used by \citet{lesh68}. Intriguingly, her magnitude calibration (in her table~3) does not show a significant difference between stars classified as V and IV at most spectral types. After re-observing and reclassifying a substantial fraction of her sample, we have to conclude that the features that led to classification as luminosity class IV on photographic plates are rarely related to higher intrinsic luminosity. We believe that this issue has mostly been resolved in the new grid presented here, with the inclusion of a few field stars that seem to fulfil these requirements.
 
 \subsubsection{Luminosity class II}
 \label{sec:ii}
 
 The number of luminosity class II standards is very small, and none is considered primary. Many of the stars initially classified as luminosity class II have turned out to be either true supergiants with high rotation or binaries. For example, the B1\,II standard HD~1383 is in reality an SB2 system with two (twin) supergiants \citep{boyajian06}. The other B1\,II standard, HD~199216, is almost identical to the B1\,Ib standard $\zeta$~Per. Likewise, $\iota$~CMa is better classified B3\,Ib rather than B3\,II. 
 
 Nevertheless, we have found a fair number of replacements. A suitable standard for B3\,II is HD~194779. Another example with slower rotation is HD~36212, although this object is later and more luminous, close to the limit with a true supergiant. Likewise, we identify HD~31327 as B2\,II and $\varepsilon$~CMa as B1.5\,II. There are also quite a few stars in the Per~OB1 association that can be classified as luminosity class II, among which HD~14053 is B0.7\,II (and can be compared to HD~13969, which is B0.7\,III). This star is clearly less luminous than the other stars classified as B0.7\,II (see discussion above on the difficulties in this range), but not much fainter than HD~193076, a member of the open cluster IC~4996, which also contains HD~193007, B0.5\,II, and so helps to define this luminosity class.

 The situation is not so clear for later spectral types. We identify a few examples of B5\,II, such as HD~191243 or HD~175156, but find no objects of similar luminosity at later types. Examination of a large sample of field B7\,III and B8\,III giants finds that all have absolute magnitudes between $\approx-1.5$ and $-2.0$, with the single exception of Alcyone, which, as discussed in Appendix~\ref{app:clusters}, is about a magnitude brighter, and we have classified as II--III. At spectral type B9, we also find a few objects that are much brighter than the typical giant, and we also classify them as II--III, but with absolute magnitudes also around $-3$, they are very far away from the expectations for a supergiant.

 \subsubsection{Luminosity class Iab}
 \label{sec:iab}

 There is not a single early-B Iab standard in \citetalias{jm53} or \citetalias{mk73}, meaning that stars classified as Iab are those falling in between the Ia and Ib standards. This results in a degree of ambiguity, as standards of a given class may have rather different luminosities. The traditional approach to standard usage considers that any star that falls within a given luminosity class is a suitable standard, even if it is not necessarily the most representative of the class (Walborn, priv. comm.). Thus, we find that the B1\,Ib standard $\zeta$~Per is only slightly more luminous than B1\,II stars, while the B2\,Ib standard 9~Cep is so luminous that it leaves very little room for any B2\,Iab object between it and the B2\,Ia stars. As mentioned, HD~15690 in NGC~957\footnote{Although the star is in the cluster and has a parallax compatible with membership, its proper motion along Right Ascension deviates considerably from the cluster average.} is marginally more luminous (and somewhat earlier) than 9~Cep, and could be B2\,Iab-Ib. The same situation is found at B2.5\,Iab. The difference in luminosity between the B2.5\,Ib standard 3~Gem and the B2.5\,Ia 55~Cyg does not allow for an intermediate class. In fact, 3~Gem and 55~Cyg look quite different because the former is significantly earlier, not very distinct from 9~Cep. All these spectra are shown in Fig.~\ref{fig:b2p5sgs}.
 
Contrarily, the B1\,Iab standard HD~13854, in Per~OB1, is almost as bright as the B2\,Ia stars HD~14143 and 10~Per, in the same association. This is evident not only in its spectrum, where all the luminosity indicators place it very close to Ia, but also in its apparent magnitude, similar to those of the Per~OB1 Ia stars. In fact, HD~13854, which has been consistently classified as B1\,Iab by \citet{morgan55}, \citet{lesh68}, \citet{walborn71}, and \citet{lennon92}, has an H$\beta$ profile that is already variable and clearly asymmetric because of wind effects. Contrarily, $\rho$~Leo, which was originally given as B1\,Ib, but is now accepted as B1\,Iab, has a perfectly symmetric H$\beta$ profile. If the \textit{Hipparcos} distance for $\rho$~Leo is accepted, both stars have about the same intrinsic magnitude, demonstrating the importance of mass loss in the shape of Balmer lines at high luminosities.

\subsubsection{Hypergiants}
\label{sec:hyper}

The most luminous stars are classified as Ia supergiants. They are very scarce, but they can be seen to very large distances because of their huge intrinsic luminosity \citep[typical absolute magnitudes around or somewhat above $-7$\,mag; e.g.][]{humphreys84}, and thus several are known. In a study of the most luminous stars in the Magellanic Clouds, \citet{feast60} introduced the notation Ia-0 for early-type stars\footnote{This notation had already been used by Philip Keenan for cool supergiants of extreme luminosity, such as RW~Cep \citep[e.g.][]{keenan42}.} that were more luminous than the typical supergiant (which they called super-supergiants). In later years, the name hypergiant was preferred \citep[e.g.][who used it for stars brighter than $M_{V}= -8$~mag]{vangenderen83}, and the notation Ia$^{+}$ became widely used. Defining the spectral type in terms of absolute luminosity may make moderate sense in the Magellanic Clouds, but it will not work in the Milky Way; more importantly, is contrary to the philosophy of the MK system, which requires a morphological definition.

A tentative defining characteristic was given by \citet{clark12_MW}, who state that they are "distinguished from normal BSGs by the presence of (P~Cygni) emission in the Balmer series", but this must be qualified. Most luminous B supergiants display emission in H$\alpha$ and some present weak and variable emission in H$\beta$ \citep[e.g. HD~306414, B0.7\,Ia, although this is the counterpart of an X-ray transient][]{lorenzo14}. Contrariwise, HD~183143, which is one of the proposed hypergiants, does not always show a P-Cygni profile in H$\beta$, but a variable profile (see an example in Fig.~\ref{fig:b7seq}). According to \citet{vangenderen83}, at a given temperature, supergiants with P-Cygni profiles (associated with higher mass-loss rates) tend to be more luminous than those without them. \citet{walborn15} identify the presence of broad emission wings in H$\beta$ as a characteristic of most luminous blue variables and blue hypergiants. In some cases, superposition of a narrow P-Cygni profile results in a characteristic 'Prussian helmet' morphology in this line (as seen in the spectrum of HD~190603 in Fig.~\ref{fig:b2ia}, where the extended wings may have been over-normalised by our automated procedure; see Fig.~7 in \citealt{clark12_MW} for comparison). The peculiar morphologies of H$\beta$ in early-B\,Ia supergiants can be seen in Fig.~\ref{fig:b0b1Ia}. In late-B supergiants, the effects of mass loss are less conspicuous, and the H$\beta$ line generally has a more standard morphology, as can be seen in Fig.~\ref{fig:a0sgs}.

Notwithstanding the importance of these characteristics, direct comparison of the B1.5\,Ia/B1.5\,Ia$^{+}$ pair HD~14956/HD~190603 in Fig.~\ref{fig:b2ia} shows very noticeable morphological differences in terms of the intensity and width of the Balmer lines and the relative intensities of the \ion{O}{ii} and \ion{Si}{iii} lines with respect to the \ion{He}{i} spectrum, i.e. the main luminosity diagnostics. The same can be said of the B1\,Ia/B1\,Ia$^{+}$ pair HD~148688/HD~169454 shown in figs.~20 and~21 of \citet{walborn90}. This is fully consistent with the idea that a hypergiant should be identified by resorting only to the photographic classification region, which does not include H$\beta$. Sticking to comparison with the few examples available, we  consider that a hypergiant is a star whose luminosity diagnostics indicate a significantly higher luminosity than a typical Ia supergiant of the same spectral type. With this criterion, despite the very prominent P-Cygni profile in H$\beta$, HD~13256 is not a hypergiant, as its spectrum is very similar to those of HD~216411 (B0.7\,Ia) and HD~148688 (B1\,Ia), both of which have H$\beta$ decidedly in absorption, but HD~169454 is, because of the weakness of its Balmer lines and extreme values of the ratio of \ion{Si}{iii}~4552\,\AA\ (and the \ion{O}{ii} spectrum) to \ion{He}{i}~4387\,\AA. The case of HD~183143, however, remains ambiguous. Although its Balmer lines are very weak compared to Ia stars of similar spectral type, there are no pure luminosity criteria at B7 that could be used to definitely call it a hypergiant, hence the classification B7\,Ia$^{(+)}$. A set of stars that may be classified as hypergiants according to these criteria is presented in \citet{deBurgos23}.

\subsection{Classification and stellar parameters}\label{quantities}

In the Introduction, we mentioned, among the primary motivations for creating a new grid of B-type standard stars, the existence of inconsistencies between luminosity classifications by different authors, and the apparent lack of a good correlation between luminosity class and intrinsic brightness. The first point is naturally addressed by the definition of a new luminosity criterion that incorporates the well-known effect of gravity on the extension of the wings of hydrogen lines, while the second may be used to test the improved representativity of the new grid. With a new objective criterion for luminosity class, differences between classifiers should be minimised. 

\begin{figure}[!t]
\resizebox{0.45\textwidth}{!}{\includegraphics{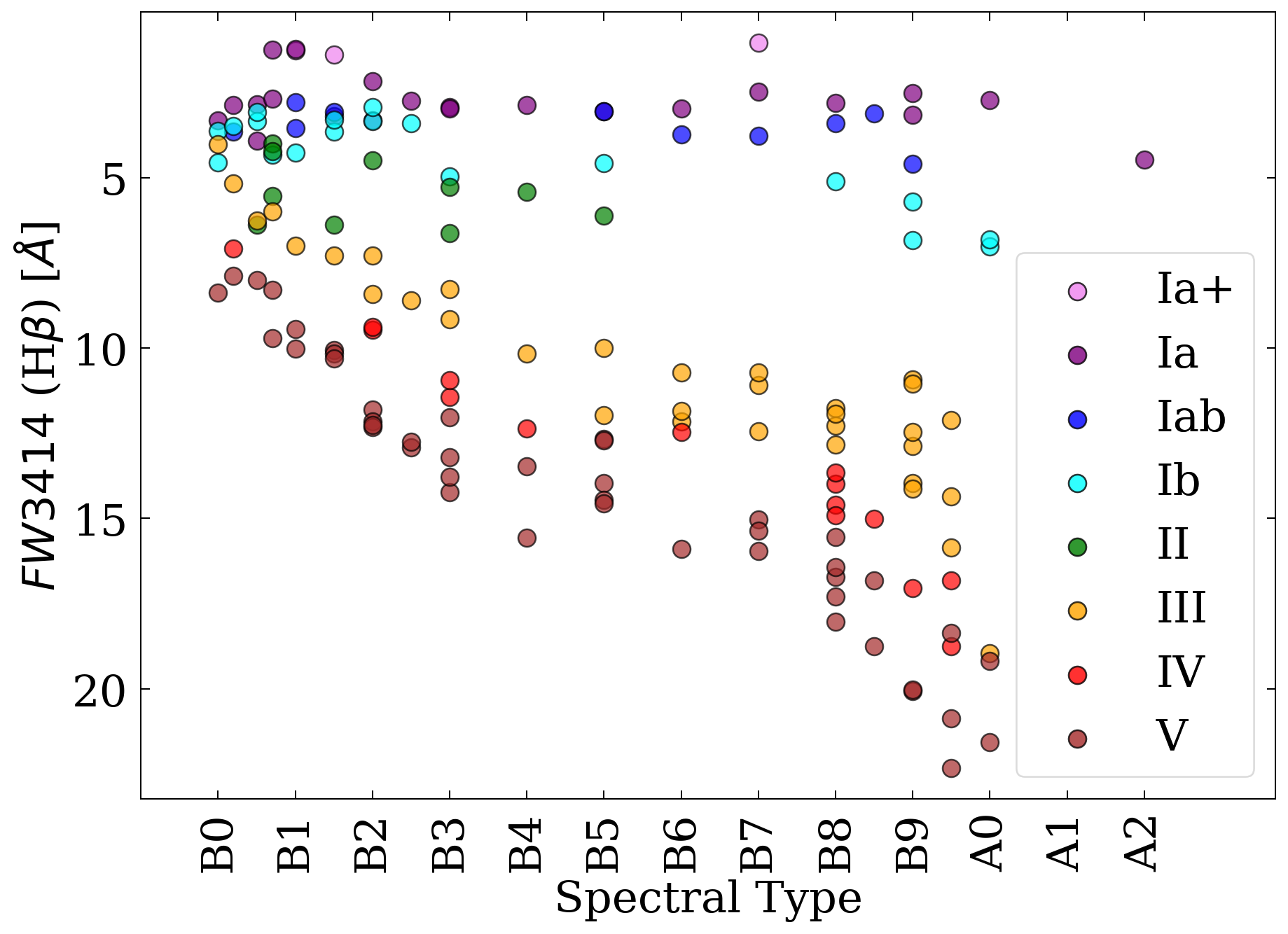}}
\resizebox{0.45\textwidth}{!}{\includegraphics{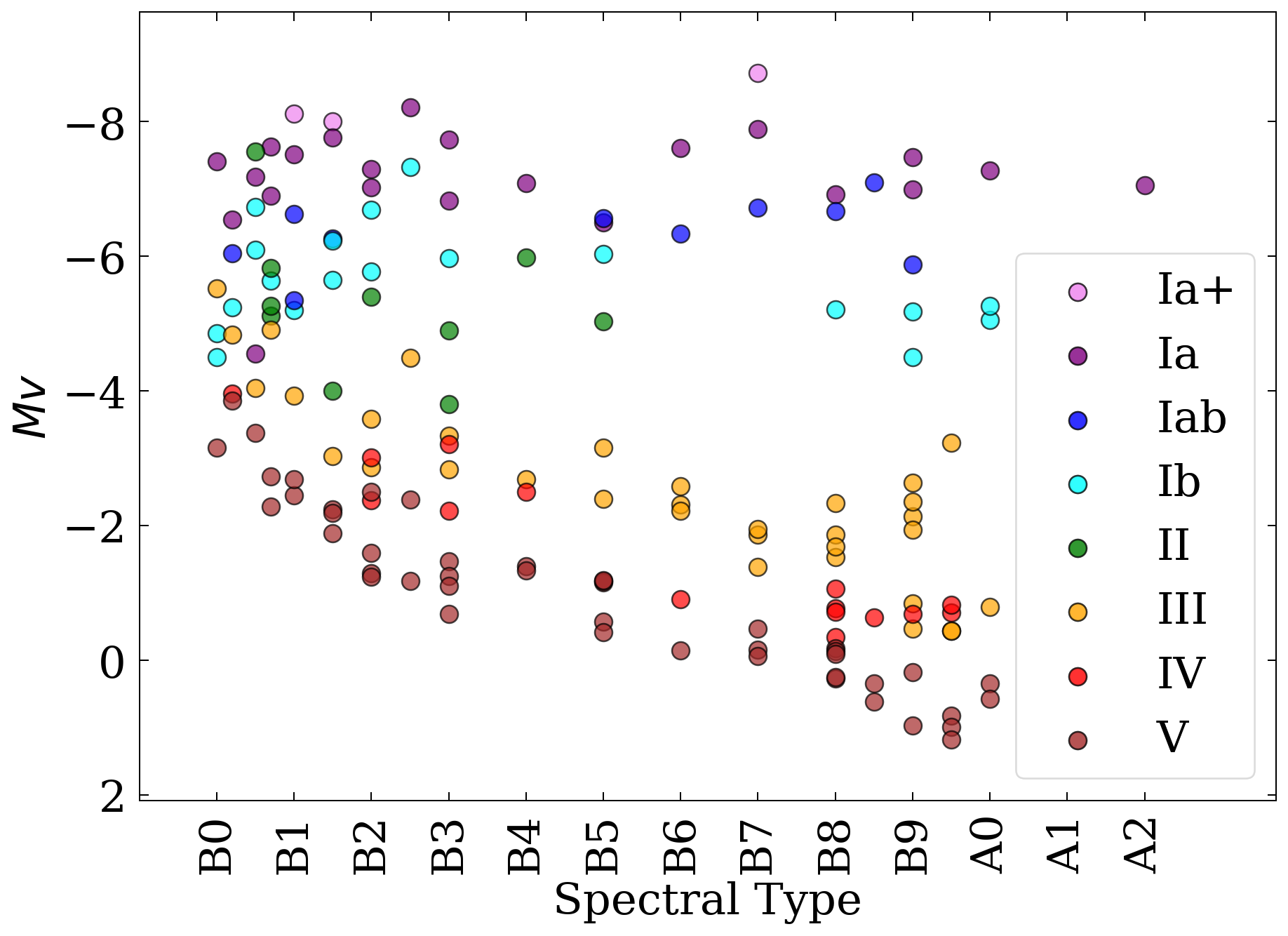}}
\centering
\caption{Distribution of the sample of 158 stars comprising the revised grid of B-type standards for spectral classification in a \fwhb\ (top) and $M_{V}$ (bottom) vs. spectral type diagram. Different colours are used to indicate luminosity class.  
\label{fig_SpT_FW3414+Mv}} 
\end{figure}

In \citet{deBurgos23}, we proposed the quantity \fwhb, defined as the difference between the width of the H$\beta$ line measured at three-quarters and one-quarter of its line depth, as an easily measurable spectroscopic diagnostic to select blue supergiants and filter out other sources with higher surface gravities. The top panel of Fig.~\ref{fig_SpT_FW3414+Mv} shows the behaviour of this quantity as a function of the spectral type for the stars making out our final list of standards, separating the stars by luminosity class. This figure is complemented by the bottom panel, where a rough approximation to the stellar intrinsic magnitude ($M_{V}$) is considered in the ordinate axis. These estimates have been done by adopting the distance modulus corresponding to the \textit{Gaia} geometric distances provided\footnote{Except for a few bright targets for which the \textit{Gaia} mission has not provided parallaxes yet. In these cases, we assumed distances from the revised \textit{Hipparcos} catalogue \citep{van2007}.} by \cite{Bailer-Jones2021}, the calibration of intrinsic $B-V$ colours, ($B-V$)$_{\rm 0}$, as a function of spectral type and luminosity class proposed by \cite{fitzgerald}, and the apparent magnitudes $B$ and $V$ quoted in the Simbad Astronomical Database \citep{Wenger00}. We have assumed, since this is intended only as a rough estimate, a constant value of $R_{V}=3.1$ when computing $A_{V}$ from E($B-V$)\,=\,($B-V$)$\,-\,$($B-V$)$_{\rm 0}$.

%Starting from spectral type B2, there is a clear separation between supergiants and the rest of the stars at \fwhb\,$\approx$\,7.5\,\AA, with the gap growing in extent for later spectral types. % Usarlo para justificar la elección
%Before B2-type, we can still separate the stars by the luminosity class, however it is not so obvious. This exercise highlights the utility of this easily measurable quantity to identify supergiant stars. TO BE CONTINUED BY SSIMON ...

%The bottom panel of Fig.~\ref{fig_SpT_FW3414+Mv} shows a similar plot, with a rough approximation to the stellar intrinsic magnitude in the ordinate axis. This approximation has been calculated simply by assuming the \textit{Gaia} distance as distance modulus and $A_{V}=3.1\cdot E(B-V)$ as the extinction, with the colour excess ...

The good correlation between \fwhb\ and the luminosity class is not surprising, as this quantity is an indirect measurement of the width of the line wings. The correlation with $M_{V}$ is also good and, especially for the less luminous stars, much better than in previous versions of the system \citep[cf.][]{jaschek98}. For types earlier than B2, there is a continuum in both plots, but for later types there is a growing gap between the supergiants and the rest of the stars. In addition, these plots also allow the visualisation of some of the effects that have been discussed in the previous subsections: the existence of early-B stars whose luminosity class, based on line-ratio criteria, is at odds with a more objective indicator of luminosity (Sect.~\ref{sec:earliest}), or the existence of luminosity class II stars on the supergiant side of the divide for spectral types earlier than B5. 
The widening gap between the supergiants and the main sequence (populated by objects of luminosity classes V, IV, and III) at later types is accompanied by the disappearance of these LC II objects. 
Aside from the obvious physical explanation, namely that the astrophysical parameters of such objects would correspond to intermediate-mass stars in the Hertzsprung gap, there is also a methodological reason: the less luminous B8\,--\,A0 Ib supergiants have luminosities and line widths comparable to those of type II stars in the early-B range. As the spread along the vertical axis increases, a given luminosity class is displaced in relation to earlier spectral types, so that the whole parameter space is represented. There is a hint that the gap may be closing again as we approach the A-type stars, but this should be checked.

\section{Summary and outlook}

The grid of B-type standards presented in this work represents an important improvement with respect to previous efforts, mainly because of a much higher degree of internal consistency, and also because of a better mapping on to physical properties. A number of minor inconsistencies may remain, which are in all likelihood intrinsic to the system and very possibly due to the need to retain the anchor points and the low number of supergiant stars available to choose our standards. In our final list (Table~\ref{bigtable}), we provide standards for all the subtypes that have been commonly used, but we note that a very fine grid is probably not necessary in modern Astrophysics. Given the availability of tools that allow a quick estimation of actual stellar parameters in a very short time \citep[e.g.][among some others]{SimonDiaz2011, holgado18, rubke23, deBurgos2023b}, spectral classification will only be a valuable tool when conditions are such that the available data (i.e. the spectra) do not provide diagnostics for accurate parameter determination.

We do not recommend the use of spectral types B8.5 and B9.5 for supergiants, and find little justification to use them at lower luminosities. Likewise, we do not believe that differentiation between B6 and B7 is meaningful within the system. The interpolated subtypes B0.2 and B0.7, \citep[introduced by][]{walborn71} can be defined for all luminosity classes, but it is unlikely that they fully achieve the purpose that led to their introduction, namely,\ resolving inconsistencies between luminosity and temperature classification in the B0\,--\,1 range (see discussion in Sect.~\ref{sec:earliest}). In fact, even the classical B2.5 subtype is poorly defined in terms of standards. Finally, the interpolated B4 type can be defined for all luminosity classes and provides a fine subdivision of B5 that may make sense in some contexts. In any event, the grid is sufficiently robust to allow classification to any level of accuracy that seems reasonable, given the quality of the available spectroscopic data.

Now that the grid is well defined, we may consider the extension of the classification process to other spectral regions. In a forthcoming paper, we will examine the possibility of obtaining a (at least, rough) classification by using the \textit{Gaia} RVS range and the improvements that may result from the addition of red spectra. {This will include an exploration of the possibility of overcoming the limitations of classification by combining it with quantitative measurements of equivalent widths and line widths}. In parallel, we are working on the determination of stellar parameters for the whole grid, with the aim of providing an accurate mapping of spectral classification into the theoretical HR diagram, as already done by \citet{holgado18} for  O stars.

% \begin{figure}
%\resizebox{\columnwidth}{!}{\includegraphics{fig_Mv_FW3414.png}}
%\centering
%\caption{. } \label{fig_FW3414+Mv}
%\end{figure}

%
%   \begin{figure*}
%   \centering
%\include[\textwidth]{B8blue}
   %%%\includegraphics{empty.eps}
   %%%\includegraphics{empty.eps}
%\resizebox{\textwidth}{!}{\input{B8blue}}
%   \caption{Colour-magnitude diagram for 2MASS data in circles of radius $3\arcmin$ (large symbols) and %$7\arcmin$ (small symbols) around the position of Be~51. Circles represent objects selected as possible early-type stars, while squares are candidate luminous stars (selected as described in the main text). Filled symbols represent stars with spectra.\label{fig:raw}}
              %
%    \end{figure*}
%

%%%%%%%%%%%%%%%%%%%%%%%%%%%%%%%%%%%%%%%%%%%%%%%%%%%%%%%%%%%%%%%%%%%%%%%%%%%%%%%%%
\begin{acknowledgements}
%%%%%%%%%%%%%%%%%%%%%%%%%%%%%%%%%%%%%%%%%%%%%%%%%%%%%%%%%%%%%%%%%%%%%%%%%%%%%%%%%
Naturally, this work is dedicated to the memory of Nolan Walborn, master classifier. Throughout his career, he held an unyielding belief in the power of spectral morphology to reveal the underlying stellar physics. We are well aware that he would have probably been upset by some of the changes that we introduce, but our procedure has tightly adhered to his teachings.\\t

A work like this is based on the collection of a huge number of spectra. We are extremely grateful to all the collaborators who have contributed to this effort throughout the years by participating in observing campaigns or otherwise. Among them, we would like to cite Drs. S.~R.~Berlanas, I.~Camacho, R.~Dorda, C.~Gonz\'alez-Fern\'andez, G.~Holgado, M.~Mongui\'o and K.~R\"ubke. Dr.~Jes\'us Ma\'{\i}z-Apell\'aniz contributed a number of spectra obtained through his observing programmes.\\

This research is based on observations made with the Mercator Telescope, operated by the Flemish Community at the Observatorio del Roque de los Muchachos (La Palma, Spain)
of the Instituto de Astrof\'isica de Canarias.
This research is partially supported by the Spanish Government Ministerio de Ciencia e Innovaci\'on and Agencia Estatal de Investigaci\'on (MCIN/AEI/10.130~39/501~100~011~033/FEDER, UE) under grants PID2021-122397NB-C21/C22. IN acknowledges the financial support of MCIN with funding from the European Union NextGenerationEU and Generalitat Valenciana in the call Programa de Planes Complementarios de I+D+i (PRTR 2022), project HIAMAS, reference ASFAE/2022/017, and from the Generalitat Valenciana under grant PROMETEO/2019/041.

This research has made use of the Simbad, Vizier and Aladin services developed at the Centre de Donn\'ees Astronomiques de Strasbourg, France. This research has made use of the WEBDA database, operated at the Department of Theoretical Physics and Astrophysics of the Masaryk University.

%%%%%%%%%%%%%%%%%%%%%%%%%%%%%%%%%%%%%%%%%%%%%%%%%%%%%%%%%%%%%%%%%%%%%%%%%%%%%%%%%
\end{acknowledgements}
%%%%%%%%%%%%%%%%%%%%%%%%%%%%%%%%%%%%%%%%%%%%%%%%%%%%%%%%%%%%%%%%%%%%%%%%%%%%%%%%%

%%%%%%%%%%%%%%%%%%%%%%%%%%%%%%%%%%%%%%%%%%%%%%%%%%%%%%%%%%%%%%%%%%%%%%%%%%%%%%%%%
\bibliography{bins,class,clusters,obstars,gaia,catalogues,ssimon}
%%%%%%%%%%%%%%%%%%%%%%%%%%%%%%%%%%%%%%%%%%%%%%%%%%%%%%%%%%%%%%%%%%%%%%%%%%%%%%%%%

\begin{appendix} %First appendix
\label{app:tables}

\section{A new grid of classification standards}
\label{app:grid}

{In the following, we discuss the criteria used to define the different spectral subtypes and historical standards. For each subtype, we present the main characteristics and propose a number of revised and new standard stars.}

\subsection{ B0 type}
\label{sec:b0}

The B0 stars are considered together with the late-O stars in \citet{sota11}, and their classification is still decided by the comparison of \ion{He}{i} and \ion{He}{ii} lines. At B0, all three \ion{He}{ii} lines, $\lambda\lambda$~4200, 4542 and 4686, are present (though the former two are weak), but \ion{Si}{iii}~4552\,\AA\ is much stronger than \ion{He}{ii}~4542\,\AA. Alternatively, the conditions \ion{He}{i}~4144\,\AA\ $>>$ \ion{He}{ii}~4200\,\AA\ or \ion{He}{i}~4388\,\AA\ $>>$ \ion{He}{ii}~4542\,\AA\ can be used. The main luminosity criterion is the strength of the \ion{He}{ii}~4686\,\AA\ line, which decreases with luminosity class. The strength of the \ion{Si}{iv}~4089, 4116\,\AA\ lines compared to neighbouring \ion{He}{i} lines can also be used \citep[see][]{sota11}. All the luminosity criteria are valid at approximately solar metallicity, but are quite strongly dependent on chemical composition, as they all involve comparison of an $\alpha$ element and He.

The type is defined by two well-established standards in the system, $\tau$~Sco (HD~149438; B0\,V) and $\varepsilon$~Ori (HD~37128; B0\,Ia). It is worth noting that for some time, $\tau$~Sco was given as B0.2\,V \citep[e.g.][]{walborn71}, but then returned to B0\,V in \citet{sota11}. At the same time, the other primary standard, $\upsilon$~Ori (HD~36512) was moved to O9.7\,V. The other two \citetalias{jm53} standards, HD~206813 and HD~207538, were reclassified by \citet{sota11} as O9.5\,IV-V and O9.7\,IV respectively. {In consequence, we follow \citet{sota11} in taking $\tau$~Sco as primary B0\,V standard, although it is a well-known magnetic star.}

Widely used standards for higher luminosity are HD~48434 (B0\,III) and 69~Cyg (HD~204172; B0\,Ib). However, if the interpolated subtype B0.2 is used, the latter should be classified as such (see next). As an alternative, we can use HD~164402, although \citet{walborn76} considers that its nitrogen spectrum is abnormally weak. Another B0\,Ib star is HD~205196, but we must note that its \textit{Gaia} parallax suggests that its intrinsic magnitude is unusually low for a supergiant. More on this in Section~\ref{sec:earliest}.

Some historical standards are disused. The \citetalias{jm53} B0\,III standard 1~Cam (HD~28446) is a very fast rotator, and should definitely be classified as O9.7. A second \citetalias{jm53} B0\,Ia standard, 15~Sgr (HD~167264) was reclassified as O9.7\,Iab by \citet{walborn71}, and keeps this classification in \citet{sota14}.
 
\subsection{ B0.2 type}
\label{sec:b0p2}

The interpolated subtype B0.2 was introduced by \citet{walborn71}, and it is pretty similar to B0. The \ion{He}{ii} lines are all weaker, with \ion{He}{ii}~4200\,\AA\ hardly noticeable by now. \ion{Si}{iv}~4116\,\AA, which was comparable to \ion{Si}{iii}~4553\,\AA\ at B0, is now decidedly weaker. The luminosity criteria are as at B0.

\begin{figure}
\resizebox{\columnwidth}{!}{\includegraphics{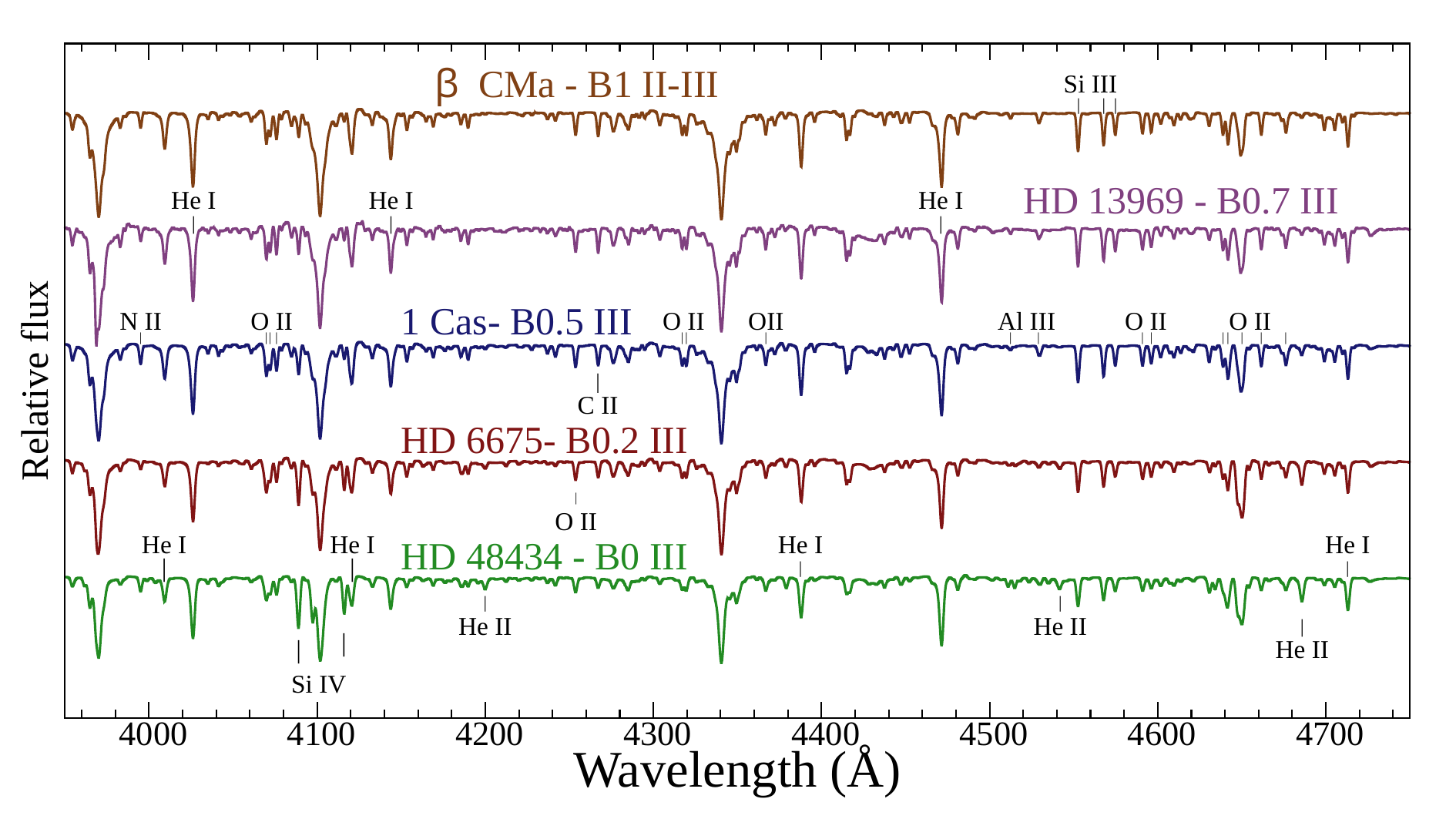}}
\centering
\caption{Spectral sequence of giants between B0 and B1, illustrating the progressive disappearance of the \ion{He}{ii} lines and the smooth inversion of the \ion{Si}{iii}/\ion{Si}{iv} ratios. Lines used in classification criteria are shown on the top and bottom spectrum. Other metallic lines \citep[after][]{kilian91} are indicated on the middle spectra. See Fig.~\ref{fig:rotcheats} for more line identifications and compare to Fig.~\ref{fig:b0b1Ia}, where a sequence at much higher luminosity over the same spectral range is shown. \label{fig:b0b1iii} } 
\end{figure}

\citet{sota11} give three standards for this type: HD~2083 (B0.2\,V), $\phi^{1}$~Ori (= HD~36822; B0.2\,IV) and HD~6675 (B0.2\,III). There have never been supergiant standards for this type, and \citet{walborn90} did not use this type for supergiants. In modern times, however, it has been customary to make use of this subtype at all luminosities. 

\citet{lennon92} give 69~Cyg (HD~204172) as an example of a B0.2\,Ia supergiant. \citet{map18}, however, classify it as B0.2\,Iab, which seems more in line with the primary standards (it was originally classified B0\,Ib; see previous section). A good example of a B0.2\,Ia star would be HD~171012 (also included in \citealt{map18}). This seems to be a very luminous object. HD~16808 would be an example of B0.2\,Ib.

%We have the case of HD~23675. It is B0.2III and has $M_{V}=-5.8$. The solution is going for B0.5II. This looks possible and suggests a non-negligible degree of degeneracy.

\subsection{ B0.5 type}
\label{sec:b0p5}

The same trends continue for B0.5. This interpolated subtype has been in use since the inception of the MKK system \citep{mkk}. Now \ion{He}{ii}~4542\,\AA\ is hardly noticeable. The defining criterion for the subtype is \ion{Si}{iv}~4089\,\AA~$\simeq$ \ion{Si}{iii}~4553\,\AA. The \ion{O}{ii} spectrum is growing in strength. These criteria are illustrated in Fig.~\ref{fig:b0b1iii}. As $\lambda$4686 is the only remaining \ion{He}{ii} line, \ion{Si}{iv}~4089\,\AA/\ion{He}{i}~4144\,\AA\ and \ion{Si}{iii}~4553\,\AA/\ion{He}{i}~4387\,\AA\ are the main luminosity criteria.

\begin{figure}
\resizebox{\columnwidth}{!}{\includegraphics{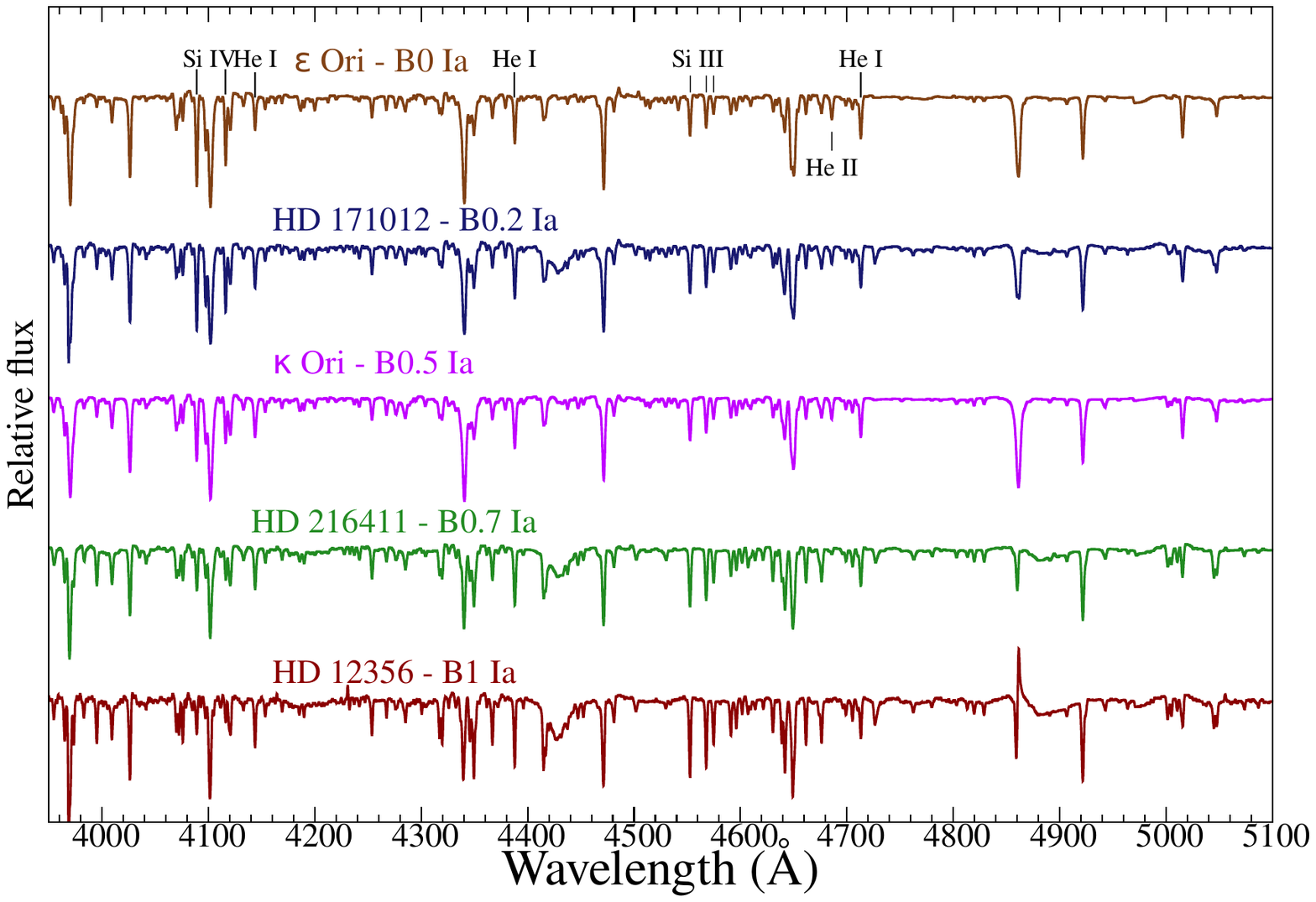}}
\centering
\caption{Spectral sequence for luminous supergiants between B0 and B1, illustrating the extreme change in the \ion{Si}{iii}/\ion{Si}{iv} ratios. The main \ion{He}{i} lines against which the Si lines are compared for luminosity classification are indicated, as well as the \ion{He}{ii}~4686\,\AA\ line, which decreases in strength as luminosity increases. Based on the shape of the Balmer lines, one would be tempted to assume that the two Orion stars are significantly less luminous than the other three, although this effect might instead be related to weaker mass loss. Compare to Fig.~\ref{fig:b0b1iii} for the same sequence at lower luminosity. \label{fig:b0b1Ia} } 
\end{figure}

In the \citet{mk73} system, the type is defined by the primary standards $\kappa$~Ori (HD~38771; B0.5\,Ia) and $\varepsilon$~Per (HD~24760; B0.5\,III). The latter is not useful as a standard for a number of reasons (see Appendix~\ref{epsper}). To replace it as B0.5\,III standard, \citet{sota11} propose 1~Cas (HD~218376). The \citetalias{jm53} standard $\kappa$~Aql (HD~184915) is a very fast rotator and a Be star, and so not appropriate. The dwarf standard is HD~36960 (B0.5\,V), a young star in Orion. A second \citetalias{jm53} B0.5\,V standard, 40~Per (HD~22951) would fit better as B0.7\,V, if this interpolated type is used.  

A second \citetalias{jm53} B0.5\,Ia standard, HD~194839, is similar to $\kappa$~Ori, though perhaps slightly more luminous. Additionally, \citet{lennon92} list two B0.5\,Ib stars, HD~192422 and 26~Cep (HD~213087). Both are displayed in Fig.~\ref{fig:b07is}. However, we must note that the use of B0.2 for supergiants makes the definition of B0.5 somewhat difficult, as both $\kappa$~Ori and HD~192422 would be bordering the earlier subtype.

\subsection{ B0.7 type}
\label{sec:b0p7}

As mentioned, the B0.7 type was introduced by \citet{walborn71}, who felt that it was necessary to resolve some discrepancies in the classification criteria observed when the resolution of the spectra was increased to 1.2\,\AA\ (similar to that used here). A substantial number of B1 stars were moved to the new subtype, including the two original \citetalias{jm53} B1\,Ia standards, $\kappa$~Cas (HD~2905) and HD~216411, which are now B0.7\,Ia. \citetalias{keenan85} does not accept this change and still lists $\kappa$~Cas as the B1\,Ia standard (cf. Section~\ref{sec:b1p5} and Appendix~\ref{kapcas}; see their spectra in Fig.~\ref{fig:bcsgs}).

\begin{figure}
\resizebox{\columnwidth}{!}{\includegraphics{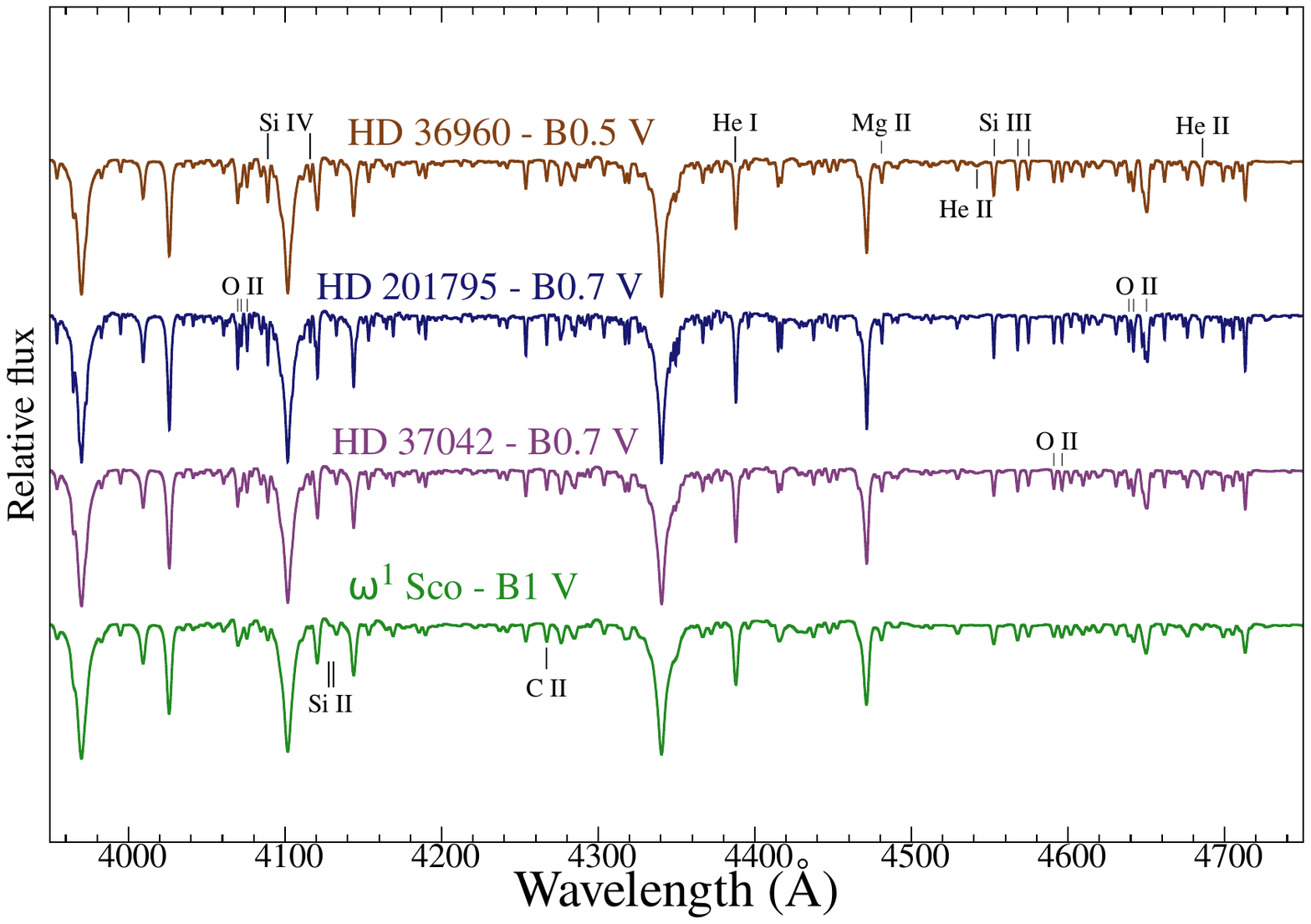}}
\centering
\caption{Main sequence stars traditionally listed as B0.7\,V compared to the standards for the neighbouring types. Note the faster rotation in $\omega^{1}$~Sco. The main features used for classification in this spectral range are marked on top of the spectrum of HD~36960. Other features of interest are indicated close to the other spectra. \label{fig:b0p7ms} } 
\end{figure}

\begin{figure}
\resizebox{\columnwidth}{!}{\includegraphics{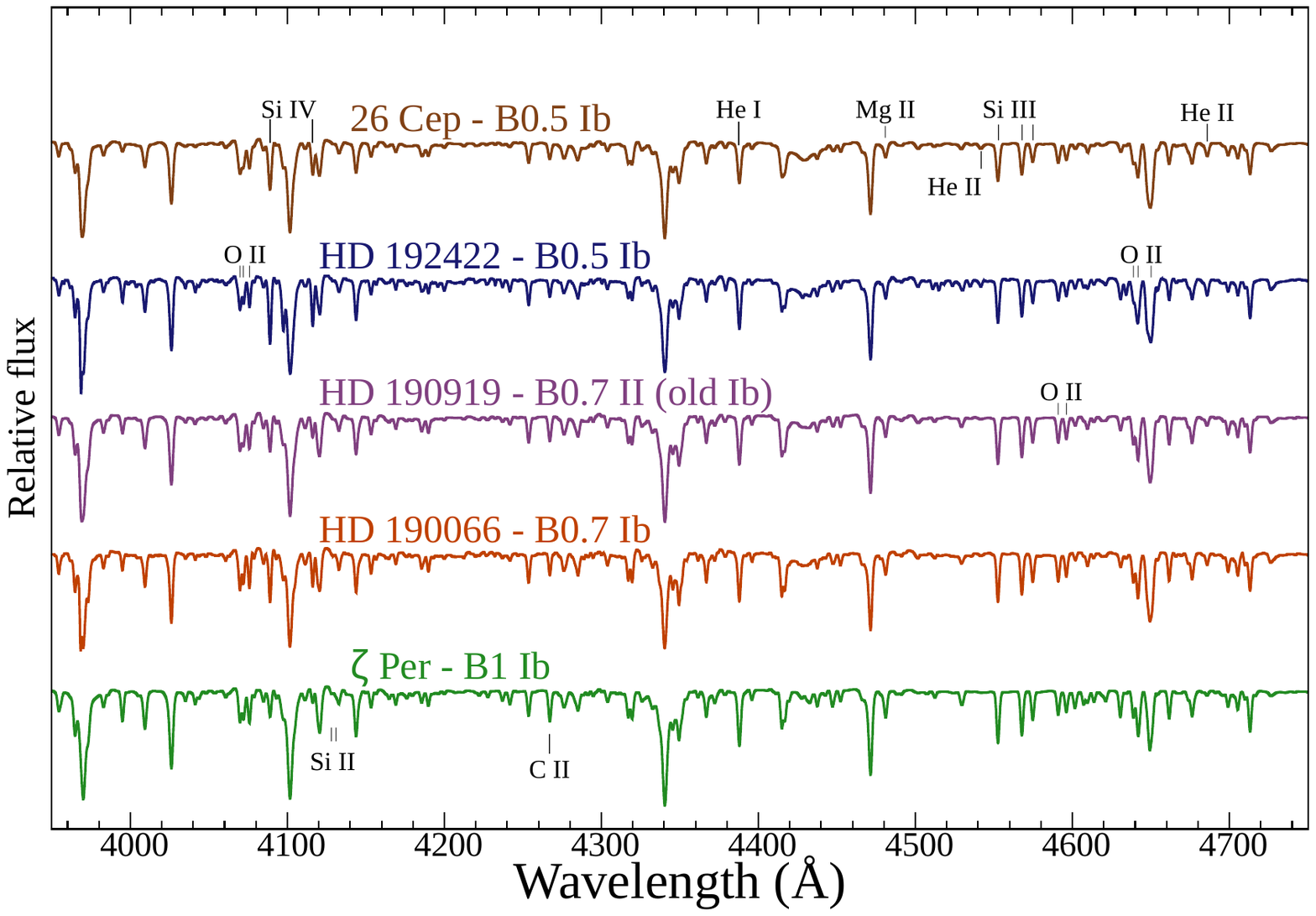}}
\centering
\caption{Low-luminosity supergiants around B0.7. The main features used for classification in this spectral range are marked on top of the spectrum of 26~Cep. Other features of interest are indicated close to the other spectra. HD~190919, the B0.7\,Ib standard suggested by Walborn, is almost identical to HD~193076, which \citet{walborn71} classifies as B0.7\,II and is shown in Fig.~\ref{fig:temporlum}. Our proposed replacement, HD~190066 lies halfway between the B0.5\,Ib and B1\,Ib standards in terms of temperature indicators and is closer to them in luminosity. Compare to Fig.~\ref{fig:temporlum}, and note that HD~192422 is clearly earlier than 26~Cep and would be close to B0.2\,Ib.\label{fig:b07is} } 
\end{figure}

In fact, the definition of B0.7 on photographic plates was not straightforward\footnote{\citet{walborn71} presents three main criteria for classification in the O9\,--\,B1 range on photographic plates, and B0.7 can only be differentiated from B1 by one of them, the ratio  \ion{Si}{iii}~4553\,\AA/\ion{Si}{iv}~4089\,\AA, which goes from approximately equal at B0.7\,V to slightly stronger at B1\,V, and from very slightly stronger at B0.7\,I to very clearly stronger at B1\,I. In addition, among Ia supergiants only, the ratio \ion{Si}{iii}~4553\,\AA/\ion{He}{i}~4387\,\AA\ goes from approximately equal to slightly stronger.}. In modern spectra, the B0.7 type may be easily defined by the presence of a weak \ion{He}{ii}~4686\,\AA\ line, while all the other \ion{He}{ii} lines have disappeared. At very high S/N, the \ion{He}{ii}~4686\,\AA\ line may be guessed in most B1 standards, but at B0.7, it should be clearly identifiable (although weak). The overall aspect of the spectrum is intermediate between B0.5 and B1.

Although the subtype is widely used, there are some difficulties in the choice of standards. As discussed in Appendix~\ref{epsper}, the star used by \citet{walborn71} to define the type, $\varepsilon$~Per, is very similar to the new B0.5\,III standard 1~Cas, except for effects related to its faster rotation. In Fig.~\ref{fig:b0p7ms}, we show the two stars used by N.~Walborn to illustrate the B0.7\,V type, HD~201795 and HD~37042, compared to the standards for B0.5\,V (HD~36960) and B1\,V ($\omega^{1}$~Sco). There is a smooth progression in the features used for classification: the weakening of \ion{He}{ii}~4686\,\AA\ and the increase of the \ion{Si}{iii}~4553\,\AA/\ion{Si}{iv}~4089\,\AA\ ratio with decreasing temperature. However, the effects are very subtle, and the B0.7\,V spectra do not differ significantly from the B0.5\,V standard. The strength of \ion{He}{ii}~4686\,\AA\ is only marginally weaker in the B0.7\,V stars, and \ion{He}{ii}~4542\,\AA\ is already hardly noticeable in the B0.5\,V standard.  

A similar situation can be seen in Fig.~\ref{fig:b07is}, where we display Ib supergiants. Again, Walborn's B0.7\,Ib standard, HD~190919, is intermediate between the B0.5\,Ib and B1\,Ib stars, but now it is only subtly different from the B1\,Ib standard $\zeta$~Per\footnote{HD~190919 was a B1\,Ib standard in \citetalias{jm53}.}. Nevertheless, there is a very evident change in both the \ion{Si}{iii}~4553\,\AA/\ion{Si}{iv}~4089\,\AA\ ratio and the intensity of \ion{He}{ii}~4686\,\AA\ between the B0.5 stars and the B1 standard. It thus seems that a B0.7 grouping can be accommodated, at least for supergiants, but the standards chosen to represent the type so far are not well suited to define it.

For stars that define the type better, we can look at $\xi^1$~CMa (HD~46328), which \citet{walborn90} used to illustrate B0.7\,IV and the B0.7\,Ia star HD~216411, which is shown in Fig.~\ref{fig:bsgs}, together with the B0.5\,Ia and B1\,Iab standards (we cannot show any B1\,Ia standards; see next and the discussion in Sect.~\ref{sec:earliest}). The progression in temperature is obvious, although HD~216411 is rather more luminous than any of the other two. Walborn's preferred standard for B0.7\,Ia is HD~152235, a southern hemisphere star, more similar to $\kappa$~Cas in luminosity. A suggestion for B0.7\,Ib would be HD~190066, which is also shown in Fig.~\ref{fig:b07is}. This is both hotter and more luminous than HD~190919, and was thus classified B0.7\,Iab in \citet{map18}. However, in view of the discussion in Sect.~\ref{sec:earliest}, we choose here to reduce the luminosity class of both stars by one subclass, moving HD~190919 to B0.7\,II.

\subsection{ B1 type}
\label{sec:b1}

\begin{figure}
\resizebox{\columnwidth}{!}{\includegraphics{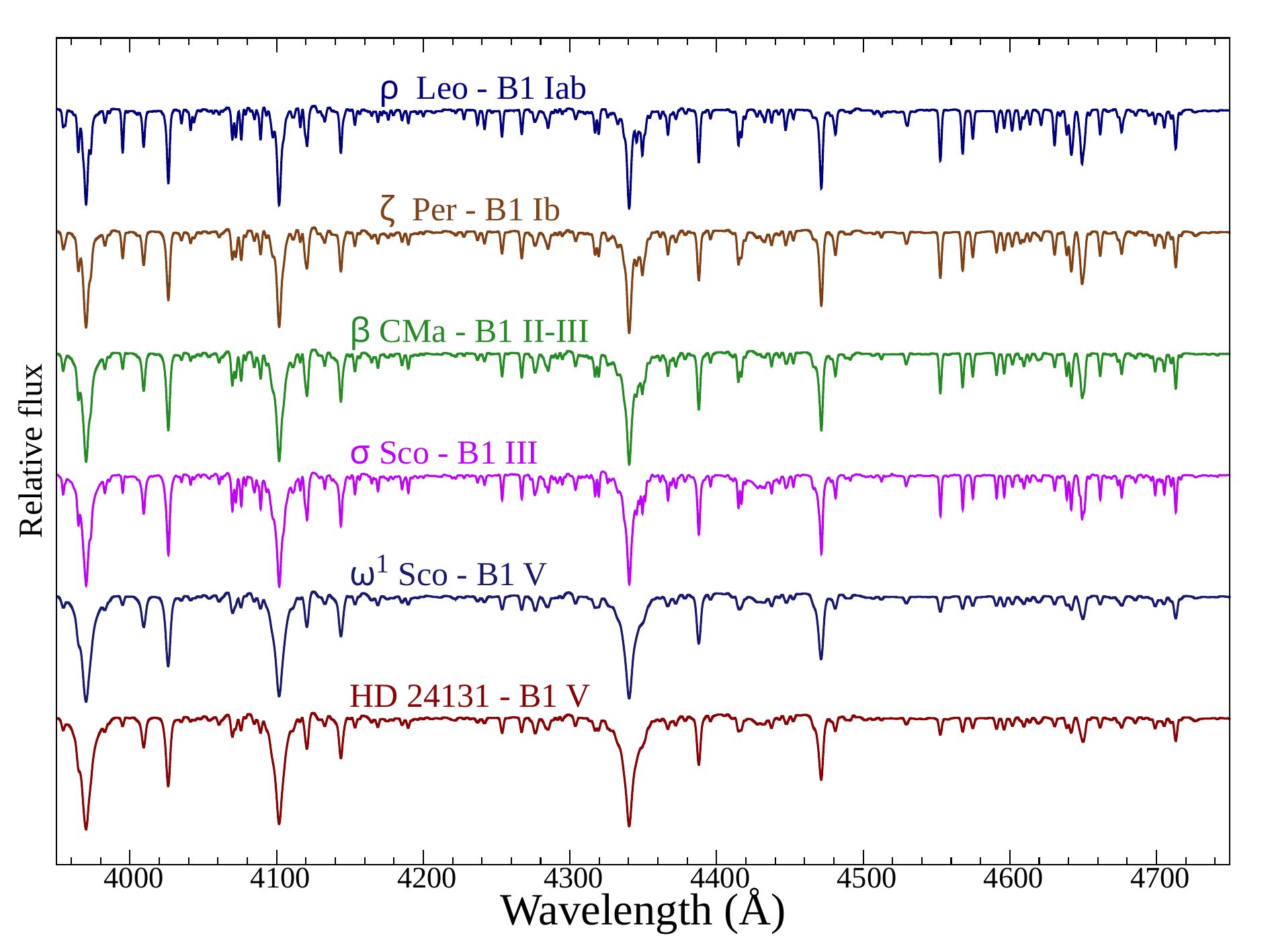}}
\centering
\caption{Luminosity sequence illustrating the properties of spectral type B1. The SB nature of $\sigma$~Sco is not perceived in this spectrum. In fact, the differences with respect to $\beta$~CMa are subtle. $\beta$~CMa is the only primary standard with an intermediate luminosity class in the B-type range. See Fig.~\ref{fig:b0b1Ia} for a more luminous supergiant.\label{fig:B1seq} } 
\end{figure}

Spectral type B1 is characterised by the disappearance of \ion{He}{ii}. \ion{Si}{iv}~4089\,\AA\ is still visible (except in fast-rotating dwarfs), but the appearance of the spectrum (beyond the Balmer and \ion{He}{i} lines) is dominated by \ion{O}{ii} lines, which reach a maximum in strength, and the \ion{Si}{iii}~triplet. The \ion{C}{ii}~4267\,\AA\ line, which was weak at earlier types, becomes as strong as a number of flanking \ion{O}{ii} lines. Some \ion{N}{ii} lines are comparable in strength to the \ion{O}{ii} lines, but many early-B stars present quite noticeable anomalies in the abundances (and hence line strength) for CNO elements \citep[see e.g.][]{Morel2006, Hunter2009}. 
As \ion{Si}{iv} has weakened and \ion{He}{ii}~4686\,\AA\ is no longer visible, the main luminosity criterion at this spectral type is the ratio \ion{Si}{iii}~4552\,\AA/\ion{He}{i}~4387\,\AA, together with the increasing strength of the \ion{O}{ii} spectrum. These characteristics can be observed in Fig.~\ref{fig:B1seq}.

The primary standards for the type were 42~Ori (HD~37018) and $\omega^{1}$~Sco (HD~144470) for B1\,V \citep{mk73}, and $o$~Per (HD~23180) for B1\,III. Unfortunately, both 42~Ori and $o$~Per are SB2 (see Appendix~\ref{oper}). In addition, the main component of 42~Ori is earlier than $\omega^{1}$~Sco, and may correspond better to the interpolated type B0.7. Also,  
$\omega^{1}$~Sco is a moderate rotator ($v_{\mathrm{rot}}\approx100\:\mathrm{km}\,\mathrm{s}^{-1}$). A somewhat less fast rotator is HD~24131, a member of the Per~OB2 star forming region that has sharp, well-defined lines (see Fig.~\ref{fig:B1seq}), although it is hotter than $\omega^{1}$~Sco, and close to B0.7\,V. A narrow-lined example would be HD~36591, although it is more luminous than the standard and close to the limit of luminosity class IV.

An alternative B1\,III standard is $\sigma$~Sco (HD~147165). Unfortunately, this is also a SB2 at high resolution, and therefore not recommended. Another primary giant standard is $\beta$~CMa (HD~44743), but this is given as B1\,II--III in \citetalias{keenan85}. As such, it is the only B-type standard with an intermediate luminosity class in the original system. The primary standard for B1 supergiants  is $\zeta$~Per (HD~24398) at B1\,Ib. The runaway supergiant $\rho$~Leo (HD~91316) and the Per~OB1 star HD~13854 \citep{walborn71} are good examples of the B1\,Iab type.

The conversion of the two original \citetalias{jm53} B1\,Ia standards to B0.7\,Ia by \citet[see Appendix~\ref{kapcas}]{walborn71} leads to the absence of any B1\,Ia standard in the northern hemisphere. \citet{lennon92} do not show any B1\,Ia supergiant in their atlas. Examples of B1\,Ia supergiants are given in \citet{walborn90}, but both are southern hemisphere objects not visible from La Palma. The only northern star that has regularly been classified as B1\,Ia is HD~13256. This is a very luminous star, with prominent P-Cygni emission in H$\beta$ (see Fig.~\ref{fig:b0b1Ia}), and so not a good choice for a standard. An example of a B1\,Ia$^{+}$ hypergiant is HD~169454 (see the Discussion for further details).

\subsection{ B1.5 type}
\label{sec:b1p5}
 
The intermediate spectral type B1.5 was already defined by \citetalias{jm53}. Its most well-known representative is the hypergiant HD~190603 (B1.5\,Ia$^{+}$; \citealt{clark12_MW}). The overall characteristics are similar to those of B1, but \ion{Si}{iv} is never seen in dwarfs and very weak in supergiants. The \ion{Si}{ii}~4129\,\AA\ doublet can now be guessed -- especially in luminous stars -- between the \ion{He}{i} lines at $\lambda$4121 and $\lambda$4144, although still very weak. In stars with normal CNO morphology, \ion{C}{ii}~4267\,\AA\ is noticeably stronger than neighbouring \ion{O}{ii} lines.

\begin{figure}
\resizebox{\columnwidth}{!}{\includegraphics{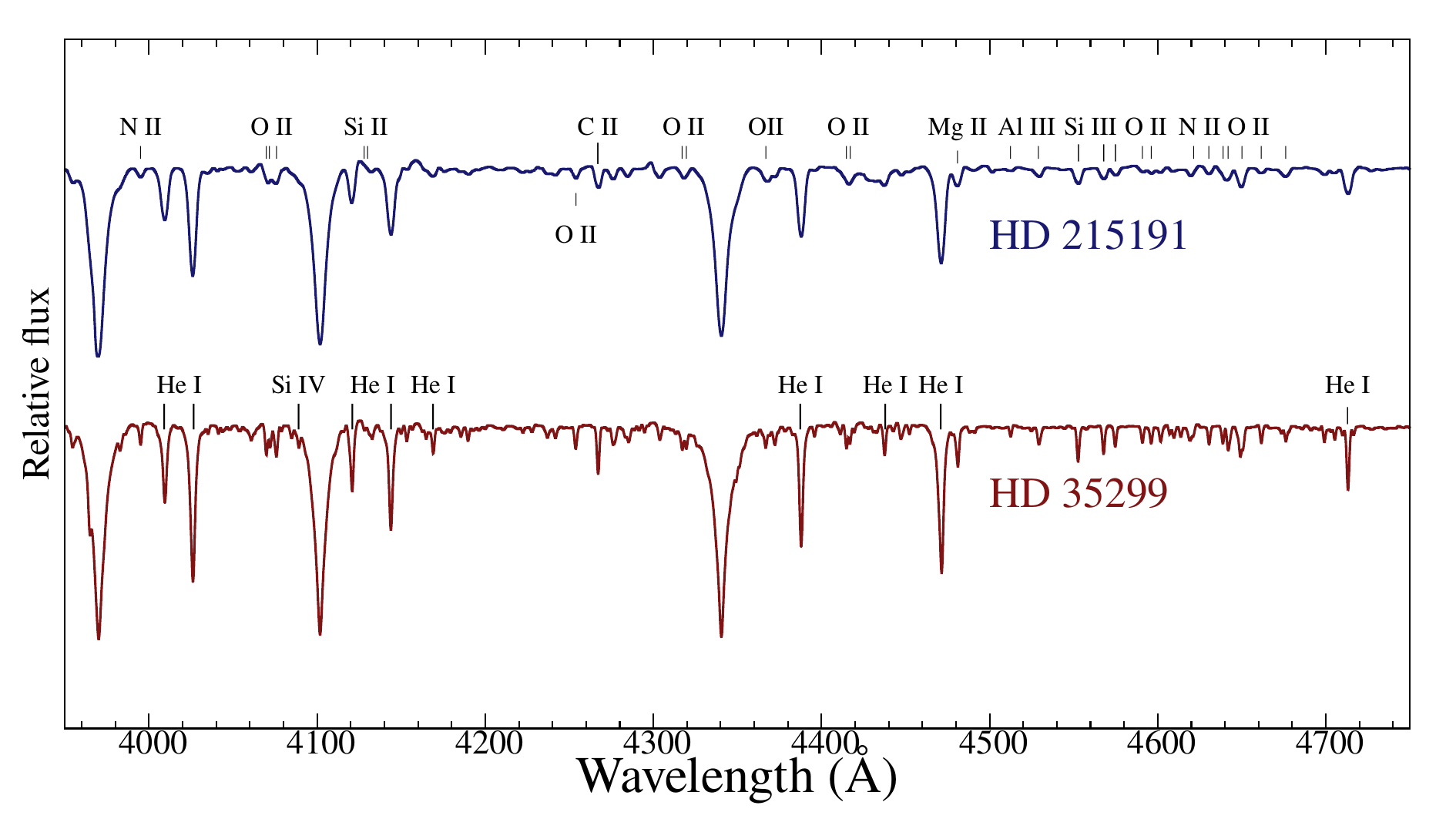}}
\centering
\caption{Comparison of the main B1.5\,V standard HD~215191 ($v\,\sin\,i\approx200\:\mathrm{km}\,\mathrm{s}^{-1}$) and a narrow-lined star of the same spectral type. Line identification after \citet{kilian91}.  \label{fig:two1p5v_stds} }
\end{figure}

Unfortunately, in modern spectra, some of the traditional standards seem ill-placed. HD~154445 (B1.5\,V \citealt{walborn71}) is almost indistinguishable from the B1\,V standard $\omega^{1}$~Sco. HD~215191 \citep{lesh68} is a better representative of the class, but it is a very fast rotator. Conversely, HD~194279 (B1.5\,Ia in \citetalias{jm53}) is too similar to the B2\,Ia standards to assign a different type, as also remarked by \citet{lennon92}. The star HD~14956, which has been classified as B2\,Ia by different authors, was suggested by \citealt{lennon92} as a B1.5\,Ia supergiant. This classification seems tenable, based on the relative strengths of lines of different Si ions. We show its spectrum in Fig.~\ref{fig:b2ia}, compared to two B2\,Ia supergiants from Per~OB1 and HD~190603. The \textit{Gaia} EDR3 distance for HD~14956 suggests that it is significantly more distant than Per~OB1, and thus may be a high-luminosity star. A somewhat less luminous star is HD~5551, which we classify as B1.5\,Iab.

\begin{figure}
\resizebox{\columnwidth}{!}{\includegraphics{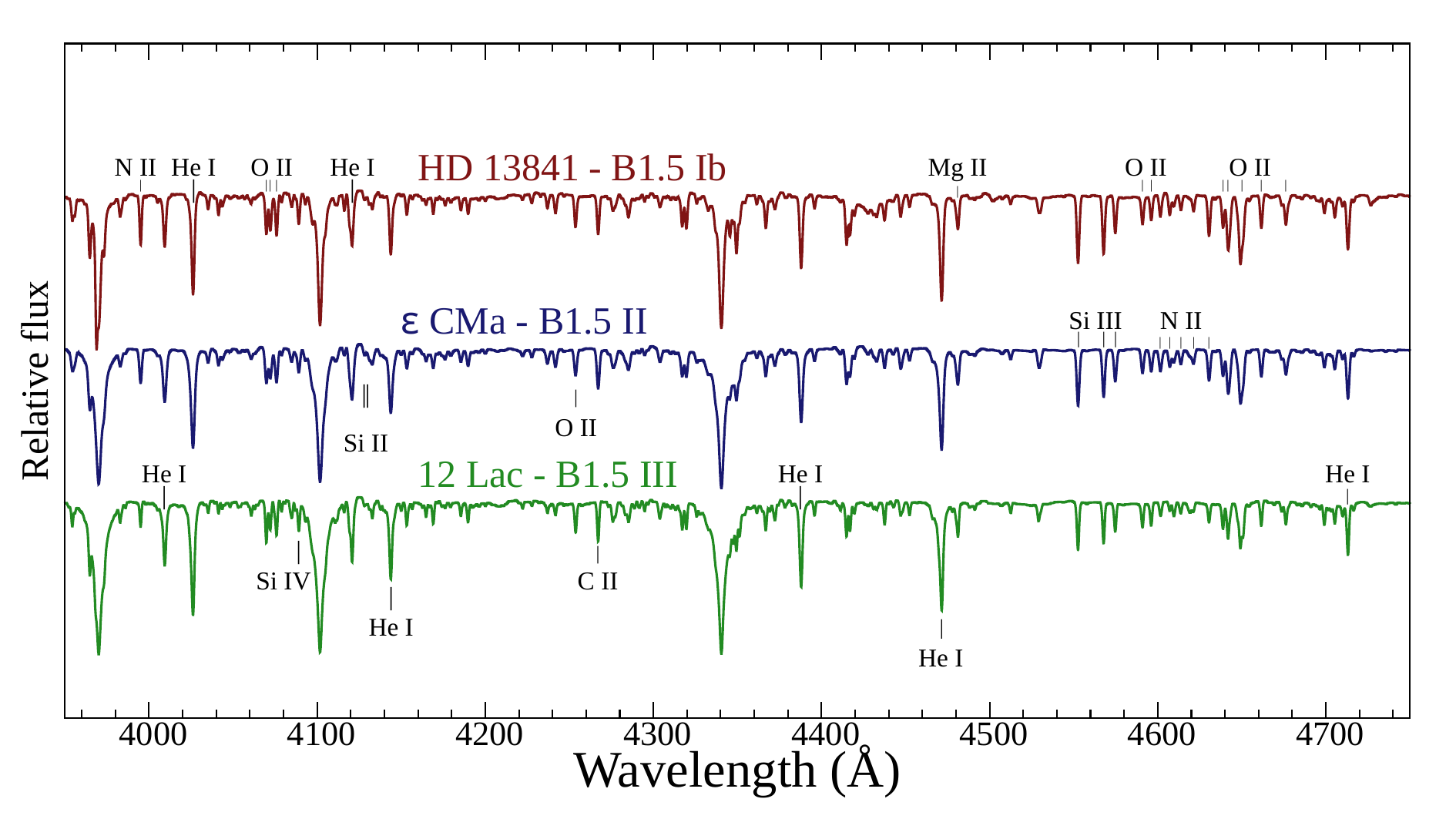}}
\centering
\caption{Luminosity sequence for moderately luminous B1.5 stars, showing the intermediate luminosity of $\epsilon$~CMa. The main criterion is the ratio of \ion{Si}{iii}~4552\,\AA\ to \ion{He}{i}~4387\,\AA. The \ion{O}{ii} and \ion{N}{ii} lines are also very sensitive to luminosity. Note: HD~13841 is close to B1. \label{fig:epsCMa} } 
\end{figure}

The Per~OB1 star HD~13841 or HD~193183 are good examples of B1.5\,Ib, although the former is very close to B1. For a better definition of the type at lower luminosity, we suggest as B1.5\,V secondary standards the Orion stars HD~35299 and HD~37744, which have sharp, well-defined lines, and thus complement HD~215191 (see Fig.~\ref{fig:two1p5v_stds}). For giants, we can list 12~Lac (HD~214993). Although originally given as B2\,III by \citetalias{jm53}, it was reclassified as B1.5\,III by \citet{walborn71}. Its spectrum can be seen in Fig.~\ref{fig:epsCMa}. In addition, we propose $\varepsilon$~CMa (HD~52089) as B1.5\,II standard. This object was originally listed as B1\,II in \citet{mkk}, but was later moved to B2\,II in \citetalias{jm53}. \citet{lennon92} argue that the star is clearly earlier and classify it as B1.5\,II, as do \citet{walborn90}. Our spectrum fully supports this classification (Fig.~\ref{fig:epsCMa}). 
 %Contrarily, HD~194279, which was a \citetalias{jm53} B1.5\,Ia standard, was shown by \citet{lennon92} to belong to B2\,Ia.

 \subsection{ B2 type}
 \label{sec:b2}
 
 By B2, the \ion{O}{ii} spectrum and the \ion{Si}{iii} triplet have weakened so much that they may not be discerned in fast-rotating dwarfs. In contrast, the \ion{Si}{ii}~4129\,\AA\ doublet is now clearly visible between the flanking \ion{He}{i} lines. In stars with normal CNO morphology, \ion{C}{ii}~4267\,\AA\ is now much stronger than any flanking lines. At this type, \ion{He}{i}~4009\,\AA\ is noticeably stronger when compared to the neighbouring \ion{He}{i}~4026\,\AA\ than at any other type. The \ion{N}{ii} spectrum also reaches its maximum at this spectral type. The \ion{N}{ii} lines grow strongly with increasing luminosity, and they can be very strong in luminous supergiants of types B1.5\,--\,B2.5. Lack of these strong lines, in most cases together with enhanced carbon lines, leads to the BC designation (see Fig.~\ref{fig:bcsgs}).
 
 There are quite a few primary standards for this type, but not all are free of complications. For B2\,V, the main standard is $\beta^2$~Sco (HD~144218), which has narrow, sharp metallic lines. Although this is not evident at our resolution, the star is an SB2 at high resolution. The dagger standard in \citet{mk73} was 22~Sco (HD~148605), which is a fast rotator ($v_{\mathrm{rot}}\approx180\:\mathrm{km}\,\mathrm{s}^{-1}$). However, this star is almost identical to the primary B2.5\,V standard $\sigma$~Sgr (see below). A more moderate rotator is $\xi$~Cas (HD~3901), but this star is also later, and it rather corresponds to B3\,IV. A better example for B2\,V would be HD~208947, but this star is classified as an SB2. Two good examples of narrow-lined B2\,V stars are given by \citet{simon10}, namely HD~36285 and HD~36629, and these may be taken as references.  Among other stars proposed as B2\,V standards by different authors, $o$~Cas (HD~4180) is a  strong Be star, HD~42041 is clearly an SB2 even at moderate resolution, while HD~191746 is of higher luminosity and should be classified B2\,IV, as done by \citet{lesh68}. 
 
\begin{figure}
\resizebox{\columnwidth}{!}{\includegraphics{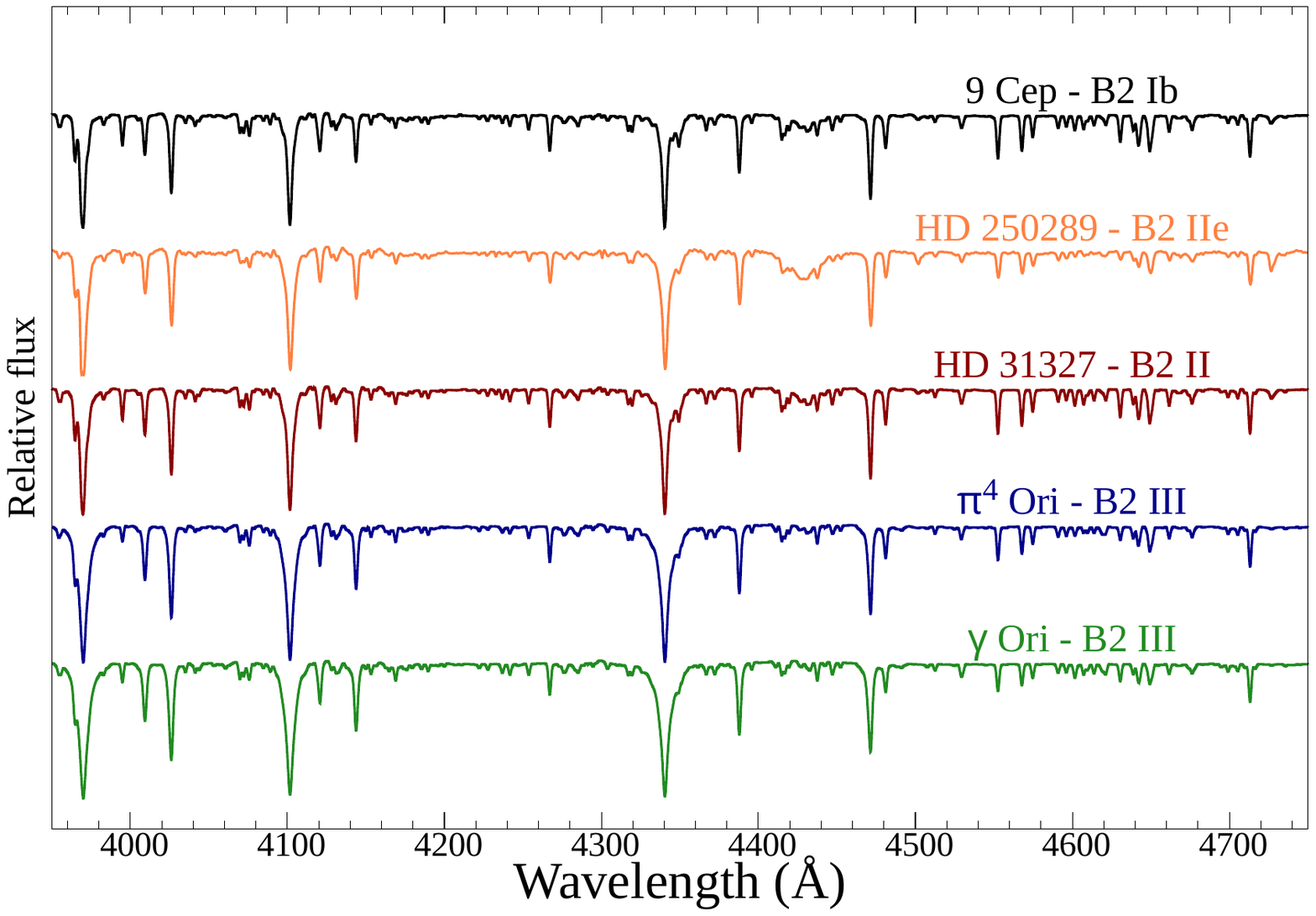}}
\centering
\caption{Luminosity sequence for moderately luminous B2 stars. Although the wings of the Balmer lines fully justify the luminosity class II for HD~31327, other luminosity criteria (e.g. the ratio of \ion{Si}{iii}~4552\,\AA\ to \ion{He}{i}~4387\,\AA \ or other metallic to \ion{He}{i} lines) show only subtle differences with respect to 9~Cep. The overall spectrum of HD~31327 is much more closely aligned with that of the supergiant than those of the giants. The spectrum of this slow rotator may also be compared to that of HD~250289 ($v\,\sin\,i\approx100\:\mathrm{km}\,\mathrm{s}^{-1}$), which may be one of the most luminous Be stars known. The nitrogen lines are unusually weak in this object.  \label{fig:b2lums} } 
\end{figure}

 There are already two standards for B2\,IV in \citetalias{mk73}, $\gamma$~Peg (HD~886) and $\zeta$~Cas (HD~3360). Their spectra are extremely similar. At B2\,III, the main standard is $\gamma$~Ori (HD~35468). A second standard proposed by \citetalias{keenan85} is $\pi^{4}$~Ori (HD~30836). This object is somewhat more luminous than $\gamma$~Ori, but has weaker \ion{N}{ii} lines (Fig.~\ref{fig:b2lums}). The B2\,Ib standard is 9~Cep (HD~206165) and the B2\,Ia primary standard is $\chi^2$~Ori (HD~41117). This is a very luminous star, but there are several other examples of the class. Apart from HD~194279, mentioned in the previous Section, useful examples of B2\,Ia are HD~14143, right in the centre of NGC~869, or 10~Per (HD~14818), also in Per~OB1.

To complement this set of standards, we can suggest HD~31327 for B2\,II. Although previously given as B2\,Ib or B2.5\,Ib \citep{lesh68},  \citet{lennon92} suggested a lower luminosity. This is fully supported by the width of its Balmer line. A star of comparable luminosity is HD~250289, in the open cluster NGC~2129. This object is one of the most luminous Be stars known, and a very fast rotator for a supergiant. Fig.~\ref{fig:b2lums} illustrates the properties of all these stars. A sensible example for B2\,Iab would be HD~15690 \citep{walborn71}, but this is only very marginally more luminous than 9~Cep, and better classified as Iab-Ib (see Sect.~\ref{sec:iab} for a discussion and Fig.~\ref{fig:b2p5sgs} for reference).

 \subsection{ B2.5 type}
 \label{sec:b2p5}
 
 The interpolated type B2.5 was also used by \citetalias{jm53}. When we reach the B2.5\,V type, most metallic lines have become very weak in dwarfs. The exceptions are the \ion{Si}{ii}~4129\,\AA\ doublet and \ion{Mg}{ii}~4481\,\AA\ which are both growing, and \ion{C}{ii}~4267\,\AA, which remains strong in stars with standard morphology. In supergiants, the \ion{N}{ii} spectrum is now noticeably stronger than \ion{O}{ii}. 
 
\begin{figure}
\resizebox{\columnwidth}{!}{\includegraphics{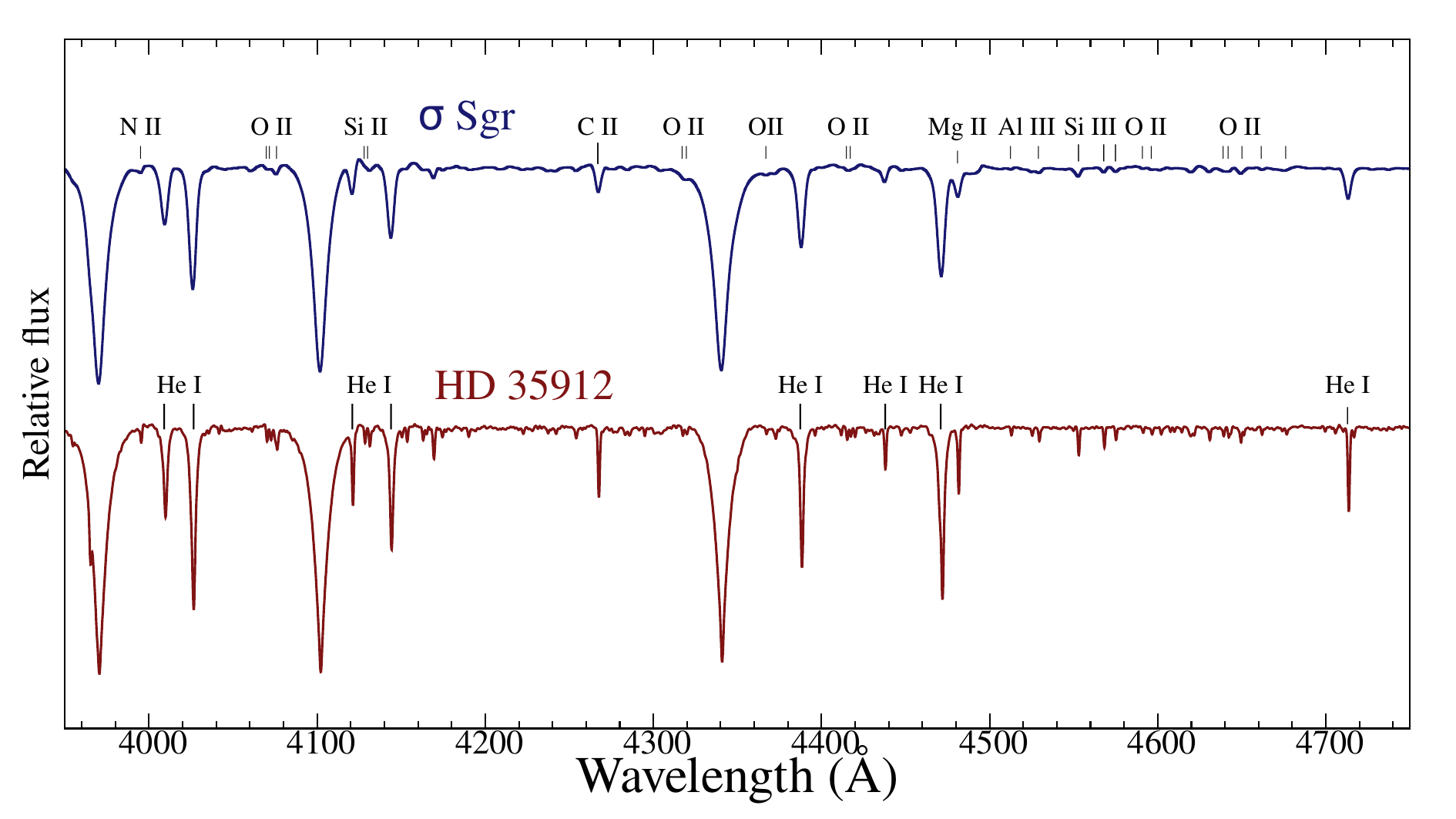}}
\centering
\caption{Comparison of the classical B2.5\,V standard $\sigma$~Sgr ($v\,\sin\,i\approx120\:\mathrm{km}\,\mathrm{s}^{-1}$) and a narrow-lined star of the same spectral type. Line identification after \citet{kilian91}.  \label{fig:two2p5v_stds} }
\end{figure}

 The primary standard for the type is $\sigma$~Sgr (HD~175191), B2.5\,V, which is a moderately fast rotator ($v_{\mathrm{rot}}\approx140\:\mathrm{km}\,\mathrm{s}^{-1}$). An even faster rotator is 22~Sco (HD~148605), which was given as B2\,V in \citetalias{mk73}, but is essentially indistinguishable from $\sigma$~Sgr. In any event, the differences between B2\,V and B2.5\,V stars, once we have taken into account rotation effects, are minimal. As narrow-lined examples of B2.5\,V, we can cite the Orion stars HD~35912 and HD~36430, from \citet{nieva13}. Fig.~\ref{fig:two2p5v_stds} shows 22~Sco and HD~35912, illustrating their very different overall appearances. \citet{lesh68} gives HD~32612 as B2.5\,IV standard, but this is indistinguishable from the B2\,V stars. \citet{walborn71} gives as B2.5\,III standard $\pi^{2}$~Cyg (HD~207330), which fulfils our criteria for this type and has a low rotational velocity ($v_{\mathrm{rot}}\approx35\:\mathrm{km}\,\mathrm{s}^{-1}$). A very similar star is 35~Aqr (HD~210191).

 For the supergiants, the basic standard is 55~Cyg (HD~198478), although this is still given as B3\,Ia in \citetalias{keenan85}, which does not use B2.5 (see Appendix~\ref{o2cma}). A less luminous supergiant is 3~Gem (HD~42087), which is generally given as B2.5\,Ib. However, it must be noted that the difference in luminosity with respect to 55~Cyg appears too small to justify two full subclasses. Moreover, although 3~Gem is later than the B2\,Ib standard 9~Cep, the difference is marginal (see Sect.~\ref{sec:iab} for further discussion and Fig.~\ref{fig:b2p5sgs} for reference). 
 
 \subsection{ B3 type}
 \label{sec:b3}
 
 By B3, all \ion{N}{ii} and \ion{O}{ii} lines have disappeared in dwarfs, except for some extremely narrow-lined stars observed with very high S/N (see discussion in Sect.~\ref{sec:issues}). From now on until we reach B9, stars of low luminosity with a moderate to high projected rotational velocity (\vsini\,$\gtrsim$\,100~\kms) are only going to display three metallic features:  \ion{Si}{ii}~4129\,\AA, \ion{C}{ii}~4267\,\AA\ and \ion{Mg}{ii}~4481\,\AA. The ratio between \ion{Mg}{ii}~4481\,\AA\ and the neighbouring \ion{He}{i}~4471\,\AA\ is going to increase smoothly with the spectral type. Unfortunately, this ratio is also quite sensitive to luminosity and, as discussed above, to rotational velocity; therefore, it cannot be used as a primary criterion. The strength of all \ion{He}{i} lines decreases with spectral type. Not all of them, however, weaken at the same rate. \ion{He}{i}~4009\,\AA\ and 4121\,\AA\ weaken faster than other lines, and this effect starts to be visible at B3. \ion{Si}{ii}~4129\,\AA\ is now clearly seen at all luminosity classes, even in fast rotators. In luminous stars, its intensity is comparable to that of \ion{Si}{iii}~4552\,\AA.
 
 There are a large number of primary B3\,V standards covering a wide range of rotational velocities. In Figure~\ref{fig:btreses}, we show HD~178849, $\eta$~Aur (HD~32630), 29~Per (HD~20365), and $\eta$~Uma (HD~120315; see the discussion in Sect.~\ref{sec:issues}). Both $\eta$~Aur and $\eta$~Uma are considered anchor standards by \citet{garrison94}. 29~Per is a member of the $\alpha$~Per cluster, which must be close to the end of the main sequence. There is also a primary standard for B3\,IV, $\iota$~Her (HD~160762; displayed in Fig.~\ref{fig:rotcheats}). For B3\,III, the historical standard is HD~21483, a moderate rotator ($v_{\mathrm{rot}}\approx130\:\mathrm{km}\,\mathrm{s}^{-1}$).
 
 In \citetalias{jm53}, $\iota$~CMa (HD~51309) is given as B3\,II. According to our criteria, it is definitely B3\,Ib, a classification that had already been given by \citet{lennon92}. An alternative for B3\,II would be HD~36212, although this object is quite luminous and lies close to the bright edge for the class. Moreover, it is decidedly earlier than $\iota$~CMa, and \citet{walborn71} classified it as B2.5\,II. Direct comparison with 3~Gem, however, favours B3\,II, although it is a borderline case. For B3\,Ia, $o^2$~CMa (HD~53138) is the traditional standard, and also an anchor point. Another choice is HD~14134, in NGC~869.
 
 \subsection{ B5 type}
 \label{sec:b5}
 
 By B5, the \ion{He}{i} lines are becoming weaker. \ion{He}{i}~4009\,\AA\ decreases in strength much more quickly than its neighbour \ion{He}{i}~4026\,\AA, and this is one of the main differences with respect to B3. \ion{Si}{ii}~4129\,\AA\ has become much stronger with respect to \ion{He}{i}~4144\,\AA\ (\ion{He}{i}~4121\,\AA\ has also weakened considerably), while \ion{C}{ii}~4267\,\AA\ has started to decrease in strength. In supergiants and stars of lower luminosity with narrow lines, features corresponding to \ion{Fe}{ii} start to be seen. While in the supergiants the \ion{Si}{iii} and \ion{N}{ii} lines are still quite stronger than \ion{Fe}{ii}, these lines are no longer seen in stars of lower luminosity.

 \begin{table*}
\caption{Stars that have been used as standards for low-luminosity B5, along with some new recommendations. The stars in bold are our recommended standards. }             % title of Table
\label{tab:b5lows}      % is used to refer this table in the text
\centering                          % used for centering table
\begin{tabular}{l c c c l}        % centered columns (4 columns)
\hline\hline                 % inserts double horizontal lines
\noalign{\smallskip}
HD& Other & Old & New & Notes\\    % table heading 
 &Name& Type & Type &\\
 \noalign{\smallskip}
\hline                        % inserts single
\noalign{\smallskip}
4142  & 68~Cas & B5\,V        & \textbf{B5\,V} & Moderately fast rotator \\      
4742  & $\nu$~And         & B5\,V        & B5\,V+ & SB2; do not use\\            
34759  & $\rho$~Aur & B5\,V & B5\,V      & \textbf{B4\,V}, if used; broad line wings      \\        
161572&  $-$      & $-$        &  \textbf{B5\,V }& Member of IC~4665              \\
36936  & $-$ & B5\,V      & B5\,V & Very broad wings\\  
198183 & $\lambda$~Cyg            & B5\,V        &      B5\,III-IV & Sometimes Be\\      
147394 & $\tau$~Her               & B5\,IV & \textbf{B5\,V}       & Slow rotator  \\ 
170475 & $-$  & $-$ & B5\,IV & In IC~4725; may be CP \\ 
22928  & $\delta$~Per             & B5\,III      & B5\,III      & Be and SB2; do not use \\
34503  & $\tau$~Ori               & B5\,III        & B6\,III & SB2 \\             
41692  & & B5\,III & B5\,III & Slow rotator; \textbf{B4\,III} if used \\
184930 & $\iota$~Aql    & B5\,III                 & B6\,IV   \\ 
170682 & $-$   & $-$ & \textbf{B5\,III}      & Brightest member of IC~4725\\    
211924 & 30~Peg & $-$ & B5\,III &\\
\noalign{\smallskip}
\hline                                   %inserts single line
\end{tabular}
\end{table*}
 
The luminous supergiant $\eta$~CMa (HD~58350) is the anchor for the spectral type (B5\,Ia). Another \citetalias{jm53} standard is 5~Per (HD~13267), but this object is clearly a binary with a hotter companion. Two other \citetalias{jm53} standards that we are not continuing are HD~167838 (B5\,Ia) and $\chi$~Aur (HD~36371; B5\,Iab). These two objects seem decidedly earlier than $\eta$~CMa, and could be used to define the B4 spectral type, if the need to interpolate it is felt (see Appendix~\ref{sec:b4}); moreover, with our criteria, their luminosities should be switched. 
 
\citetalias{keenan85} uses 67~Oph (HD~164353) as B5\,Ib standard. \citet{lennon92} argued that this star is not sufficiently luminous to classify as a supergiant, and suggest B5\,II.  According to our criteria, it is indeed somewhat less luminous than other Ib standards of similar spectral type, although perhaps not a whole luminosity class. The distance to this object is very poorly known. Its \textit{Gaia} EDR3 parallax determination $\varpi=1.37\pm0.29$~mas has a RUWE=2.5, indicating a very poor solution. If we were to accept this value, its distance would be around 800~pc \citep{bj21}, implying a luminosity consistent with a supergiant, but the error bars are very large. \citet{lennon92} suggest HD~7902 as standard for B5\,Ib. However, our spectrum shows that this object is only very slightly less luminous than $\eta$~CMa, and in fact its membership in NGC~457 puts its luminosity around $M_V=-6.7$, which is compatible with a B5\,Iab classification. A more suitable B5\,Ib comparator may be HD~9311, in NGC~581. If we compare 67~Oph to these stars, it is noticeably earlier, and so perhaps could be better classified as B4\,Ib-II, if this intermediate type is used (see Appendix~\ref{sec:b4}). A second star that \citet{lennon92} classify as B5\,II, HD~191243, definitely merits this classification by comparison to the others.
 
Among stars of lower luminosity, there has historically been a strong degeneracy between luminosity class and rotational velocity, which has led to many changes in the stars proposed as standards through time. The dagger standard is $\rho$~Aur (HD~34759), although this objects is discernibly earlier than most other standards (see Appendix~\ref{app:others} for a discussion of the B4 spectral type). The \citetalias{keenan85} standard is HD~36936, which is a fast rotator ($v\sin\,i\approx210\:\mathrm{km}\,\mathrm{s}^{-1}$) and has broader Balmer line wings than other standards.  Given its location in the Orion star forming region, it is likely a ZAMS star. The slow-rotation standard $\nu$~And (HD~4727) is an SB2, and its lines are clearly asymmetric at our resolution, which renders it inadequate. In fact, the dagger standard for B5\,IV, $\tau$~Her (HD~147394), is identical to the primary in $\nu$~And, and should be reclassified as a slow-rotation B5\,V star, although somewhat more luminous than the average for this type. As an anchor for all the B5\,V standards, we choose HD~161572. This object is a member of Melotte~20 and, as such, it should be classed as a typical main sequence star. Reassuringly, it is identical to the \citetalias{jm53} standard 68~Cas (HD~4142), which has the same rotational velocity ($v\sin\,i\approx180\:\mathrm{km}\,\mathrm{s}^{-1}$).
 
The \citetalias{keenan85} standard for B5\,III is $\tau$~Ori (HD~34503), but this object looks decidedly later than the other standards, and may be better classified as B6\,III (\citetalias{keenan85} does not use B6). Other \citetalias{jm53} standards are $\iota$~Aql (HD~184930) and $\delta$~Per (HD~22928). The former is decidedly later and appears less luminous than the others. It is very similar to the B6\,IV standard 19~Tau (see below). The latter has been reported as a Be star and a SB2 \citep{morrell92}, rendering it inadequate as a standard.

\begin{figure}
\resizebox{\columnwidth}{!}{\includegraphics{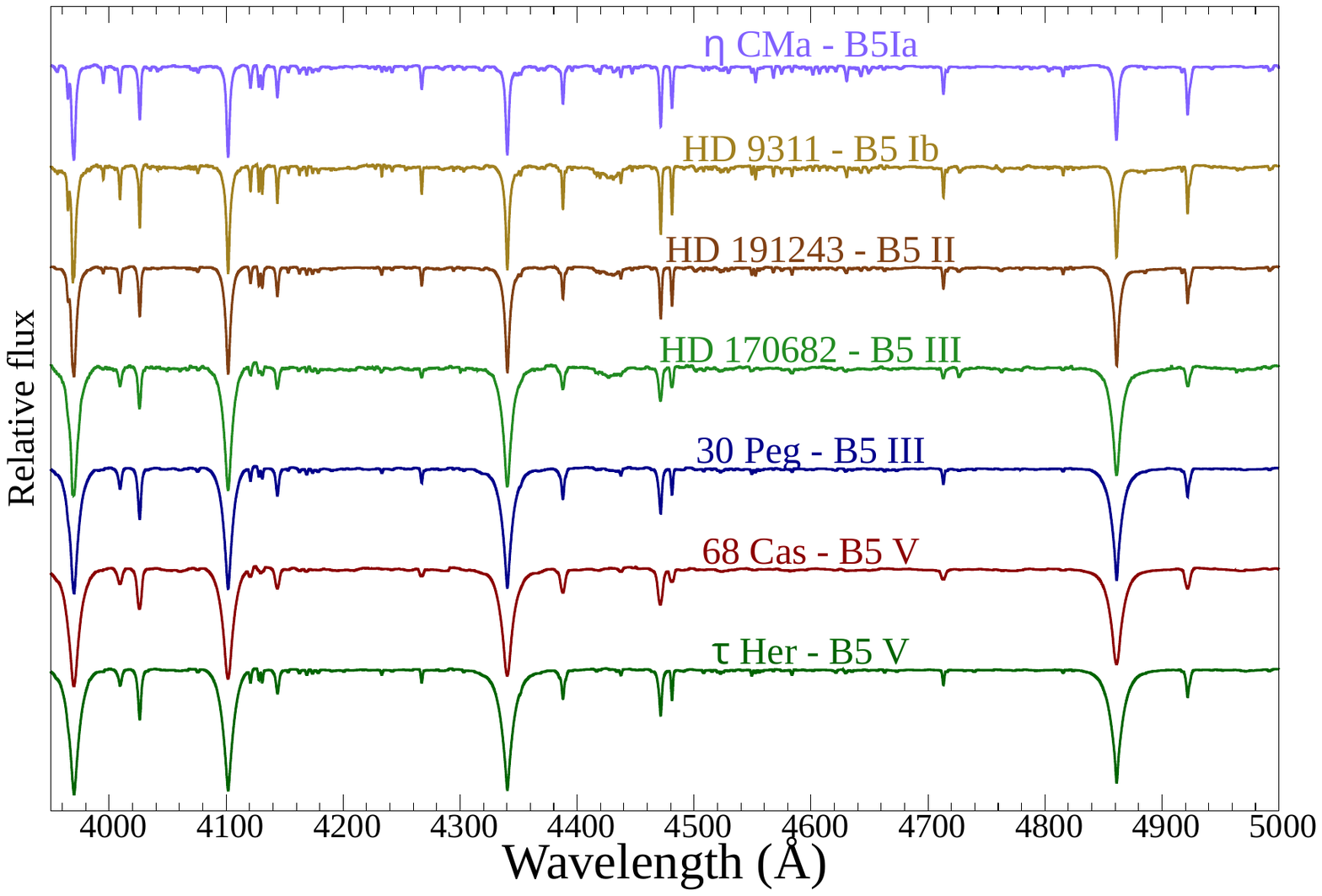}}
\centering
\caption{Luminosity sequence illustrating the properties of spectral type B5. The B5\,Ia supergiant $\eta$~CMa is the anchor standard. The pairs of dwarf and giant standards illustrate the effects of rotation. Compare 68~Cas to the fast-rotation B3\,V standard $\eta$~UMa in Fig.~\ref{fig:btreses}; the main (and almost only) difference is the weakening of all the \ion{H}{i} lines. \label{fig:B5seq} } 
\end{figure}

For a better anchorage of luminosity classes at B5, we compared the standards to the brightest stars in the nearby open cluster IC~4725, which are reported to be of this spectral type. The two brightest cluster members HD~170682 and HD~170719 are comparable in luminosity to $\delta$~Per, although the latter is later, and better classified as B6\,III. The next two brighter objects, HD~170475 and HD~170836 are both B5\,IV, but both may have some chemical peculiarity, with HD~170836 certainly meriting the B5\,IVp classification. We therefore do not consider them useful standards. A good slowly rotating B5\,III example is HD~41692, but this object may be classified as B4\,III, if this intermediate type is used. Another example of B5\,III is 30~Peg (HD~211924), which is somewhat less luminous than HD~170682. We summarise the changes proposed in Table~\ref{tab:b5lows}, and provide an updated luminosity sequence in Fig.~\ref{fig:B5seq}. 
 
\subsection{ B6 and B7 types}
\label{sec:b67}
 
The original \citet{mkk} system went straight from B5 to B8. Intermediate types B6 and B7 were introduced later in \citetalias{jm53}, but their use has rarely been systematic. For instance, B6 is not used in \citetalias{keenan85}, but a B6 dagger standard (19~Tau; B6\,IV) is given in \citetalias{mk73}. There is a strong observational limitation leading to this lack of definition: the whole temperature range is characterised by spectra very poor in features, especially if we consider moderate or high rotation. The features responding to the decrease in temperature, namely, the ratios of \ion{Si}{ii}~4129\,\AA\ to  \ion{He}{i}~4144\,\AA\ and \ion{Mg}{ii}~4481\,\AA\ to \ion{He}{i}~4471\,\AA,\ and the overall weakening of the \ion{He}{i} lines, are also linked to luminosity. Perhaps, as a consequence, a substantial number of standards have been moved between luminosity classes along the years. A luminosity classification through the width of the Balmer lines is a mandatory first step towards an accurate spectral typing. {Stars later than B5 can be identified by the quite evident weakening of \ion{He}{i}~4009\,\AA\ when compared to \ion{He}{i}~4026\,\AA\, as well as the strong \ion{Mg}{ii}~4481\,\AA\ in relation to \ion{He}{i}~4471\,\AA.} As the temperature decreases, the \ion{Fe}{ii} spectrum grows in strength, while the \ion{N}{ii} spectrum has virtually disappeared in the B6 supergiants.

As mentioned, the only primary standard for the B6 type is 19~Tau (HD~23338; B6\,IV). This is one of the three B6\,IV stars marking the main sequence turn-off in the Pleiades (see Appendix~\ref{sec:mel20}). Other useful standards for the type are 30~Sex (HD~90994; B6\,V) and HD~15497 (B6\,Ia), a luminous supergiant in Per~OB1. Another supergiant in Per~OB1, HD~17145, is a suitable standard for B6\,Iab. On the main sequence, B6 stars are characterised by the intensity of \ion{Si}{ii}~4129\,\AA\ being similar to that of \ion{He}{i}~4144\,\AA. Because of its dependence with luminosity, the Si doublet is stronger than the He line in supergiants. As mentioned in the previous Section, the former B5 standard $\iota$~Aql should be classified as B6\,IV, while the second brightest star in IC~4725, HD~170719, would be useful for B6\,III. Comparison to these stars suggests that the B5\,III standard $\tau$~Ori is better classified as B6\,III. 

The B7 type is better characterised {in terms of standards, although no specific criterion can be given to separate it from B6, beyond a gradual increase in the general trends discussed at the start of this session}. The primary standard \citepalias{keenan85} for B7\,V is HD~21071, a member of the $\alpha$~Per cluster (see Appendix~\ref{sec:mel22}). As such, it is not very far away from the ZAMS. Another useful standard is 16~Tau (HD~23288), a member of the Pleiades. This star is close to the end of the main sequence, and a fast rotator. As can be seen in Fig.~\ref{fig:combinedHR}, these two B7\,V stars occupy very different regions of the HR diagram, but have very similar intrinsic brightness. Nevertheless, HD~21071 is extremely similar to 30~Sex (see Fig.~\ref{fig:bms}), and could perhaps be better classified as B6\,V (note again that \citetalias{keenan85} did not use this type). A somewhat more representative narrow-lined B7\,V star is 49~Eri (HD~29335).

Another Pleaides star, 20~Tau (HD~23408), is a useful standard for B7\,III, because of its narrow lines. This object has sometimes been classified as B8\,III, and is indeed borderline between the two types. The primary (\citetalias{mk73} dagger) standard for B7\,III, $\beta$~Tau (HD~35497) is less luminous, and comes close to the limit with class IV. A fairly typical sharp-lined B7\,III star is HD~1279. The only supergiant standard is HD~183143 (HT~Sge), which was given by \citetalias{jm53} as the example for B7\,Ia. As seen in Fig.~\ref{fig:bsgs}, this object is rather more luminous than the typical Ia supergiants, and indeed its \textit{Gaia} DR3 parallax suggests that it is extraordinarily luminous ($M_{V}\la-8.5$). It is thus difficult to know if its differences with respect to HD~15497 are in fact due to lower temperature alone or mainly driven by luminosity. This object shows behaviour typical of extreme stars and has been considered a hypergiant \citep{clark12_MW}. A better comparator would be HD~199478, which we reclassify as B7\,Ia. The sequence of B5\,--\,8\,Ia supergiants shown in Fig.~\ref{fig:bsgs} reveals an interesting feature that can only be seen in spectra of high S/N. The ratio of \ion{Si}{iii}~4552\,\AA\ to the neighbouring \ion{Fe}{ii} lines seems to correlate well with the spectral type -- the \ion{Si}{iii} line has all but disappeared at B8\,Ia. This criterion can only be used for supergiants (the \ion{Si}{iii} is not seen at lower luminosity) and, of course, relies on the stars having the same chemical composition (as we are comparing an $\alpha$ element to an iron-group element); it cannot be extrapolated to other metallicities. The same can be seen for somewhat less luminous stars in Fig.~\ref{midBsgsatIab}.

\begin{figure}
\resizebox{\columnwidth}{!}{\includegraphics{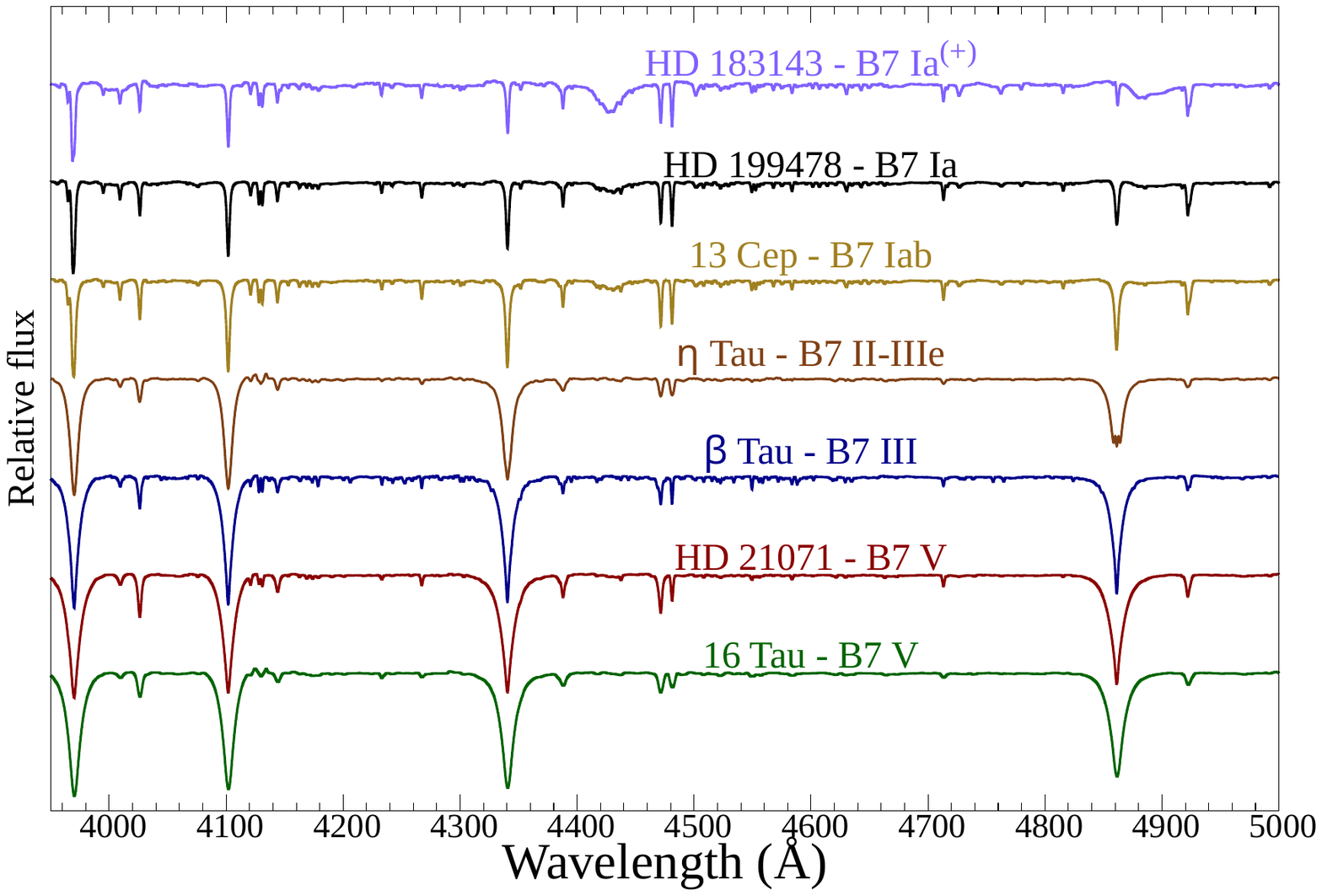}}
\centering
\caption{Luminosity sequence for B7. HD~21071 is the primary standard for B7\,V, but it is clearly earlier than 16~Tau, and may be classified as B6\,V if this spectral subtype is used. The Be nature of $\eta$~Tau, which led to its rejection as a standard, is evident in the emission infilling of H$\beta$. The star is also very noticeably more luminous than $\beta$~Tau. HD~183143, the original B7\,Ia standard, may be considered a hypergiant (see Fig.~\ref{sec:hyper}).   \label{fig:b7seq} } 
\end{figure}

The star 13~Cep (HD~208501) was the \citetalias{keenan85} standard for B8\,Ib. Its \ion{Fe}{ii} spectrum, however, indicates an earlier type, and we suggest B7\,Iab. However, as illustrated in Fig.~\ref{midBsgsatIab}, the differences between HD~17145 and 13~Cep are rather subtle. In all, the separation between B6 and B7 may not be fully justified within the system. A luminosity sequence displaying the main B7 standards can be seen in Fig.~\ref{fig:b7seq}.

\subsection{ B8 type}
\label{sec:b8}

The B8 type is defined by a ratio of \ion{Mg}{ii}~4481\,\AA\ to \ion{He}{i}~4471\,\AA\ close to unity. Since \ion{Mg}{ii}~4481\,\AA\ shows a marked dependence on effective gravity and, to a lesser extent, the location of the transition region between photosphere and wind \citep{clark12_MW}, this condition can be met at earlier spectral types for luminous supergiants. As discussed in Sect.~\ref{sec:issues}, rotational velocity affects very strongly this ratio, and thus a wide range of perceived ratios can lead to a B8 classification. A good example of this is $\alpha$~Leo (HD~87901), which previously was classified as B7\,V \citepalias{jm53}, but we here put at B8\,IV, following \citet{gray03}, who give it as B8\,IVn, to indicate its rotationally broadened lines. Narrow-lined examples of B8\,IV could be $\pi$~Cet (HD~17081, which is close to III) or HD~46075.

\begin{figure}
\resizebox{\columnwidth}{!}{\includegraphics{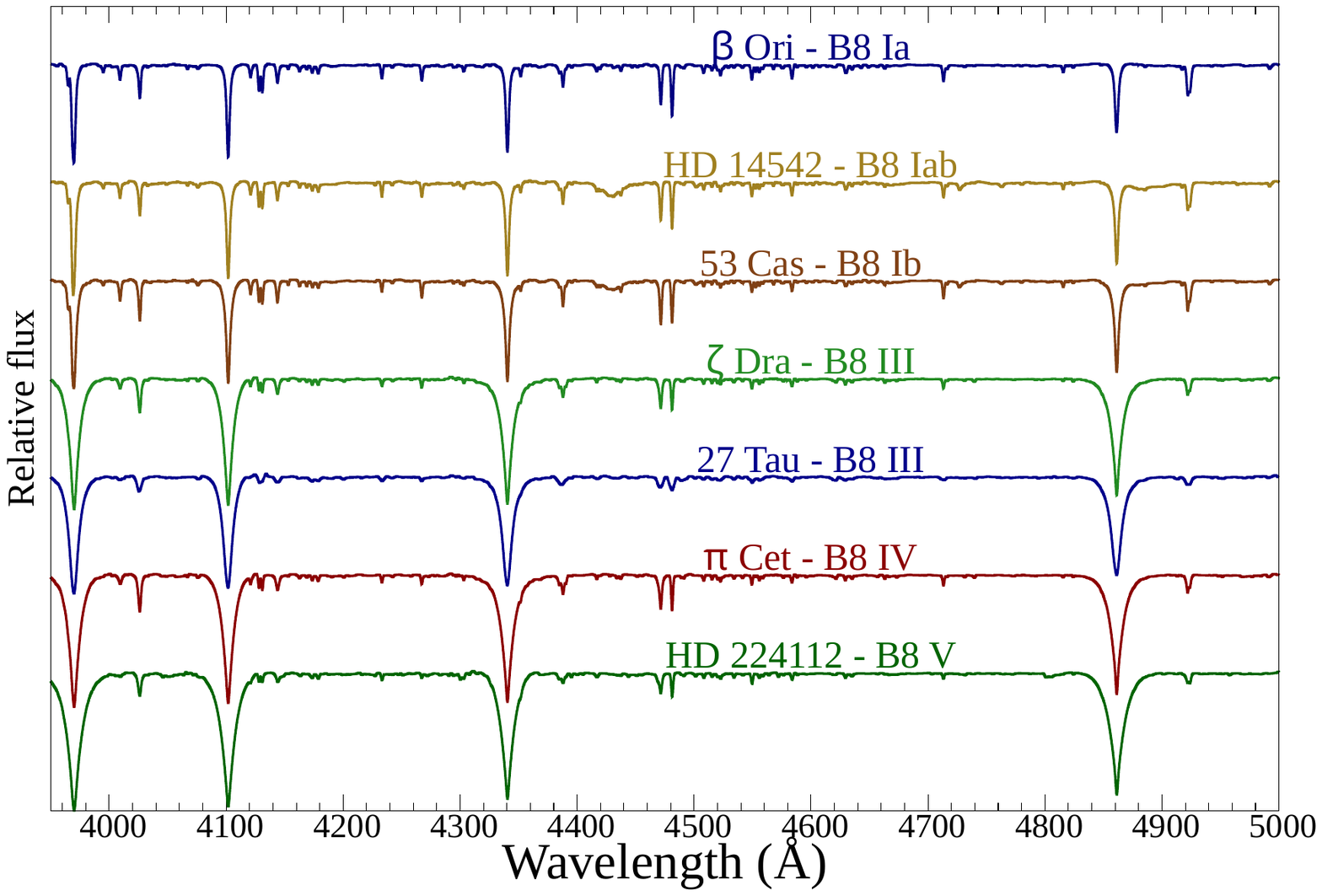}}
\centering
\caption{A luminosity sequence illustrating the properties of spectral type B8. The B8\,Ia supergiant $\beta$~Ori (Rigel) is the anchor standard. All the stars shown are slow rotators, except for 27~Tau, included so that the pair of giant standards illustrate the effects of rotation. The SB2 nature of 27~Tau is not perceived at this resolution. See Fig.~\ref{fig:mel22_late} for examples of fast rotating B8\,V stars.  \label{fig:B8seq} } 
\end{figure}

In B8 stars that rotate moderately fast, \ion{He}{i}~4121\,\AA\ is lost in the wing of H\,$\delta$ and \ion{He}{i}~4009\,\AA\ is barely seen, while \ion{Si}{ii}~4129\,\AA\, which is not resolved as a doublet, is comparable to \ion{He}{i}~4144\,\AA. In supergiants, \ion{Si}{ii}~4129\,\AA\,is clearly seen as two lines, much deeper than the neighbouring \ion{He}{i} features. The main supergiant standard for the type is  Rigel ($\beta$~Ori = HD~34085), an anchor point of the system for \citet{garrison94} at B8\,Ia. Another \citetalias{jm53} B8\,Ia standard, HD~199478, seems considerably more luminous than Rigel. When this is taken into account, it is also clearly earlier, and thus we have reclassified it as B7Ia. A third \citetalias{jm53} standard, HD~14542, is clearly less luminous (see Fig.~\ref{fig:B8seq}) and we reclassify it as B8\,Iab. A very similar star is HD~14322, also in Per~OB1. Another star of similar luminosity is HD~184943, but this is considerably later, and could be classified as B8.5\,Iab, if the need to use this interpolated subtype is felt (see Appendix~\ref{app:others}). The B8\,Ib standard is 53~Cas (HD~12301). \citet{lennon92} proposed that 53~Cas should be reclassified as B7\,II. After a direct comparison to Rigel, we find no reason to support this change. As can be seen in Fig.~\ref{fig:B8seq}, it is far too luminous not to be a proper supergiant. \citetalias{jm53} give a second B8\,Ib standard, 13~Cep (HD~208501), but direct comparison to Rigel rather suggests B7\,Iab (see Fig.~\ref{midBsgsatIab} for a spectral sequence that illustrates well this classification).

Two Pleaides stars are the dagger \citetalias{mk73} standards for the type. 18~Tau (HD~23324) is the B8\,V standard, while 27~Tau (HD~23850) is the B8\,III standard. Given its position in the cluster HR diagram (see Fig.~\ref{fig:combinedHR}), 27~Tau is brighter than typical for the type; this may partially be due to binarity. There is a second, fainter B-type star in a 291~d orbit \citep{zwahlen04}, and its lines are seen in high-resolution, high S/N spectra. It therefore cannot be used as a standard at high resolution. A number of other objects can be used to characterise B8 giants. The \citetalias{jm53} B8\,V standard $\iota$~And (HD~222173) is only slightly less luminous than 27~Tau, and would be in the low-luminosity limit for B8\,III, while the proposed B8\,II standard $\gamma$~CMa (HD~53244) is in fact a slow-rotation B8\,III\footnote{We note that, at high resolution, $\gamma$~CMa can be identified as a mild HgMn peculiar star \citep[cf.][]{woolf99}. Its luminosity is also close to the lower limit for B8 giants.}. The same can be said for another star given as B8\,II, 21~Aql (HD~179761). A third star generally given as B8\,II, $\zeta$~CMi (HD~63975), is, in our view, a B7\,III star with very low rotational velocity. Finally, the \citet{jm53} B8\,V standard $\tau$~And (HD~10205) is also a giant with moderate rotational velocity that can be used instead of 27~Tau. A good narrowed-line example is $\zeta$~Dra (HD~155763).

A secondary \citepalias{jm53} B8\,V standard is 21~Tau, also in the Pleiades, which is marginally later and less luminous than 18~Tau. A third B8\,V star in the Pleiades, HD~23753, is a very fast rotator ($v\sin\,i\approx320\:\mathrm{km}\,\mathrm{s}^{-1}$), and may serve as comparison to the other two. Another \citetalias{jm53} B8\,V standard, $\zeta$~Peg (HD~214923) is decidedly later than the two Pleiades standards, with no \ion{He}{i}~4009\,\AA\ and \ion{Mg}{ii}~4481\,\AA\ $>$ \ion{He}{i}~4471\,\AA, which would invite for the definition of a spectral type B8.5 (see Appendix~\ref{app:others}). Narrow-lined B8\,V stars are very difficult to find. An example would be HD~171301, although this object is a bit later than most stars of this type. A more typical narrow-lined B8\,V may be HD~224112, an outlying member of Blanco~1, which had already been proposed as a standard by \citet{garrison_g94}. A luminosity sequence illustrating all the issues discussed in this section is shown in Fig.~\ref{fig:B8seq}.

\subsection{ B9 type}
\label{sec:b9}

Spectral type B9 is characterised by \ion{Mg}{ii}~4481\,\AA\ much stronger than \ion{He}{i}~4471\,\AA. The \ion{He}{i} lines $\lambda$ $\lambda$ 4026, 4144 and 4471 are still seen, but very weakly. Many \ion{Fe}{ii} lines can now be seen even in dwarfs. The supergiants present a very rich metallic spectrum, while the dwarfs display extremely broad Balmer lines, {which completely dominate the appearance of the spectrum}.

The only dagger standard for the type is $\alpha$~Del (HD~196867), B9\,IV, while there are no B9 standards in \citetalias{keenan85}. Unfortunately, $\alpha$~Del is a triple system. The B9 primary is orbited by a distant close binary containing two lower-mass stars \citep{gardner21}. At high resolution, very narrow lines from the secondary binary can be seen in the red part of the spectrum, superimposed on the broad features from the primary. Nevertheless, the primary completely dominates the blue spectrum, where no signs of the secondary are seen. Therefore, $\alpha$~Del can only be considered a standard if we restrict its use to the blue. Even then, its absolute magnitude ($M_V=-0.7$) is quite bright for the spectral type. It must be close to the limit of luminosity class III, which seems supported by comparison of its lines to those of main sequence objects.

We do not have a spectrum of 22~Tau (HD~23441), the only member of the Pleiades that has been sometimes classified as  B9\,V (see Appendix~\ref{sec:mel20}) and therefore we choose as a standard for B9\,V HD~21931, a member of Melotte~20. Given the age of this cluster, this star should be close to the ZAMS. \citet{garrison_g94} proposed $\omega$~For (HD~16046) as a slowly rotating B9\,V standard, but the width of its lines implies that it is almost as luminous as $\alpha$~Del, and so it should be of luminosity class IV (see Fig.~\ref{fig:B9iii}). It is also later than $\alpha$~Del, and can be classified as B9.5\,IV if the interpolated type is used (see the discussion below). A suitable B9\,V standard with sharp lines is 134~Tau (HD~38899). This star is somewhat more luminous than HD~21931, but significantly less than $\alpha$~Del, and should be close to the bright limit for class V. 

The B9\,III standard in \citetalias{jm53} was $\gamma$~Lyr (HD~176437). The extreme weakness of its \ion{He}{i} makes it noticeably later than $\alpha$~Del, and again we should classify it as B9.5, if we use this interpolated subtype (see below). Moreover, this is a very luminous star, with an absolute magnitude approaching $M_V\approx-3$, and so it should be B9.5\,II-III. \citet{garrison_g94} give HD~178065 as the B9\,II standard, but this star is chemically peculiar (HgMn) and a binary \citep[e.g.][]{guthrie86}, and its Balmer lines do not suggest high luminosity. Contrarily, \citetalias{jm53} give 12~Gem (HD~43836) as the B9\,II standard, but this star is as luminous as any Ib supergiant. Its \textit{Gaia} distance of 1.8~kpc \citep{bj21} confirms that it is a true supergiant, and we classify it as A0\,Ib. Its spectrum, however, is characterised by weak lines, and we do not recommend its use as a standard. Another star frequently given as B9\,III, HD~53929, is definitely earlier, its strong metallic spectrum simply reflecting a very low rotational velocity. We classify it as B8\,III, even if it is somewhat later than the standards. Spectra of these objects are displayed in Fig.~\ref{fig:B9iii}.

\begin{figure}
\resizebox{\columnwidth}{!}{\includegraphics{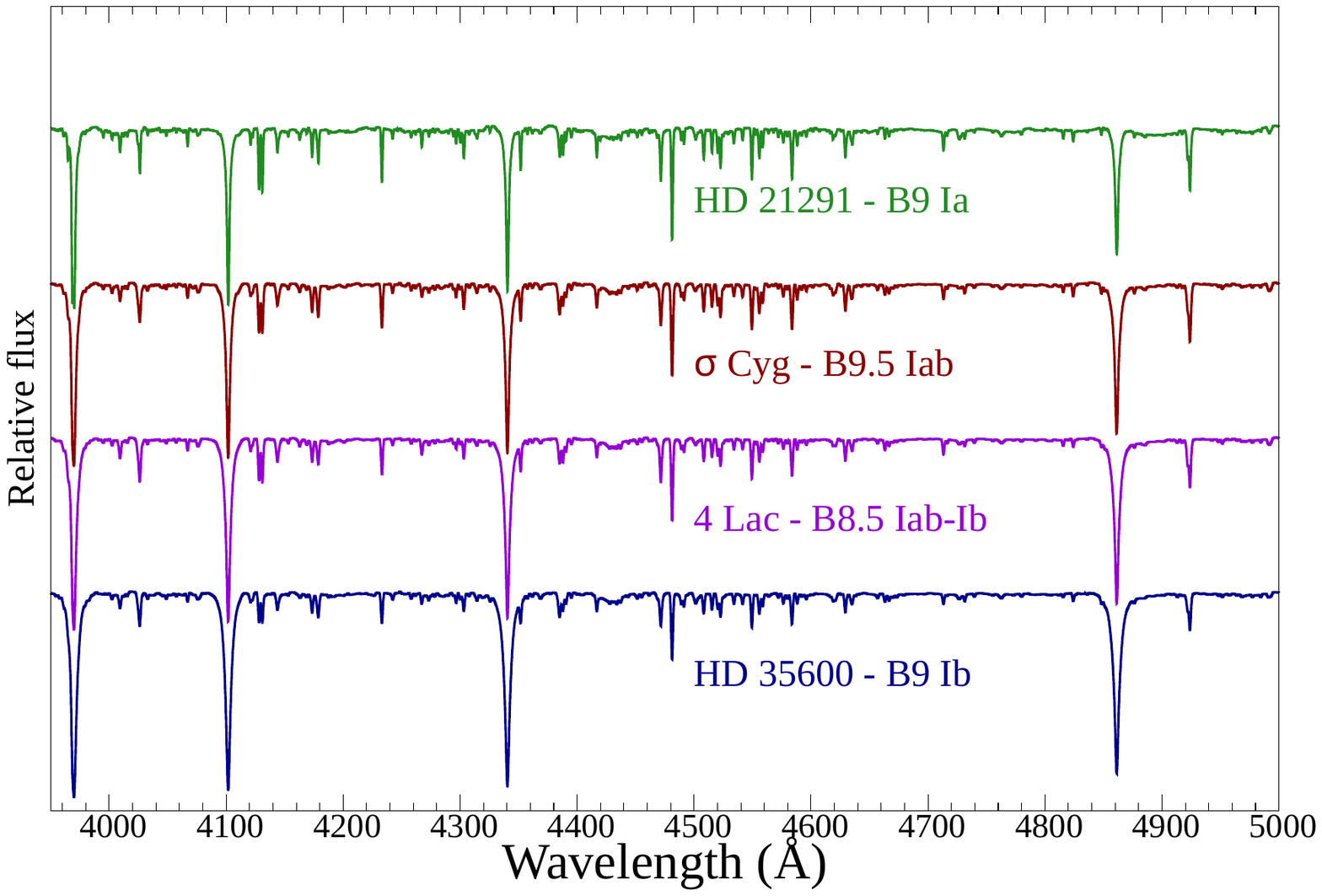}}
\centering
\caption{Supergiant standards for spectral type B9. Although the quality of the spectra permits the assignment of fractional types to 4~Lac and $\sigma$~Cyg (here shown to call attention to subtle differences), it is in all likelihood preferable to give spectral type B9 to all these objects. \label{fig:B9i} } 
\end{figure}

\begin{figure}
\resizebox{\columnwidth}{!}{\includegraphics{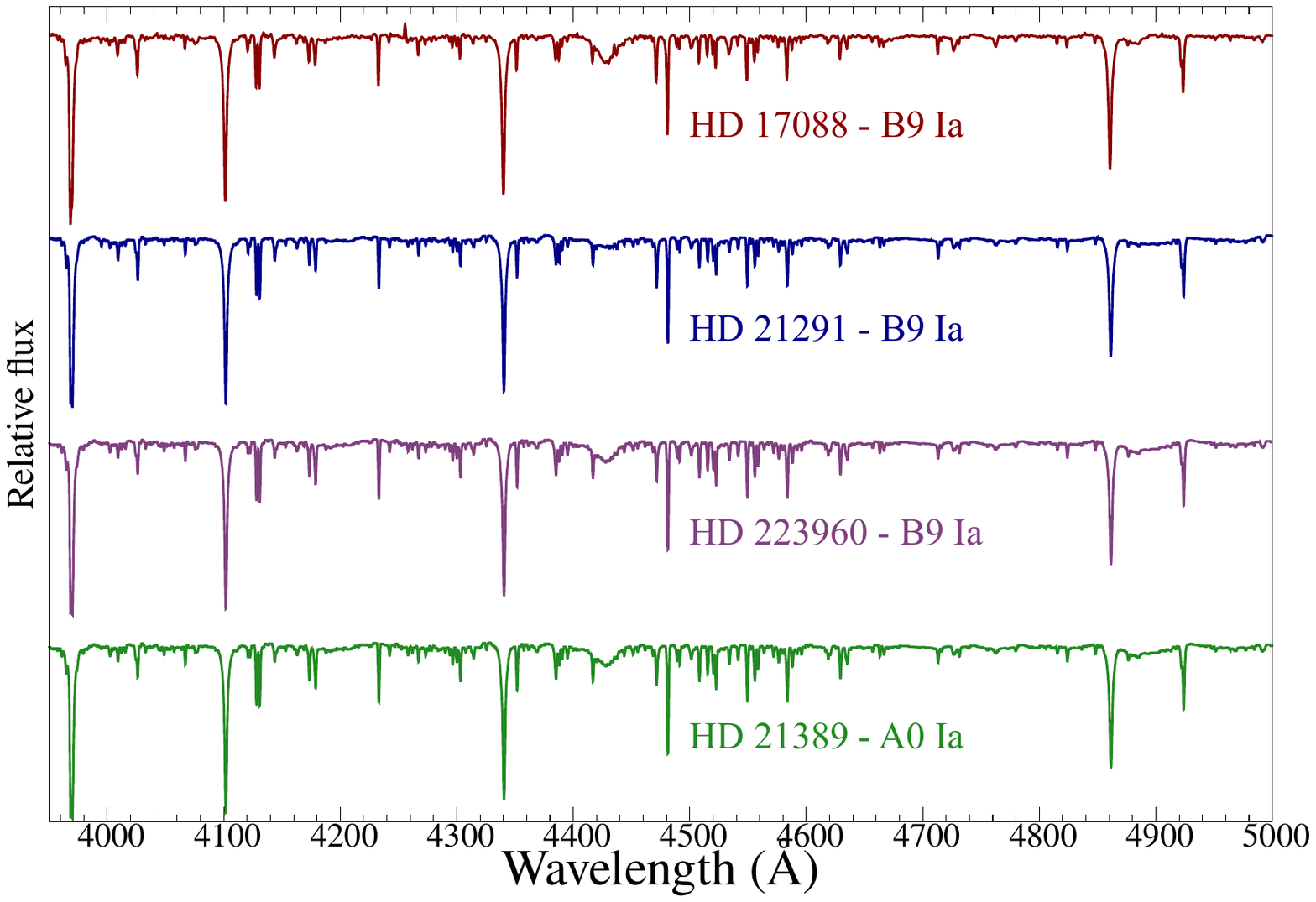}}
\centering
\caption{Stars listed as standards of B9\,Ia and A0\,Ia spectral type. The differences between the two types are subtle. Only a moderate weakening of the \ion{He}{i} lines with respect to the metallic spectrum. HD~21389 is the \citetalias{keenan85} main standard and a \citet{garrison94} anchor. By comparison, HD~223960 is earlier and notably more luminous. Although it is brighter than other Ia supergiants, it does not merit the hypergiant classification. \label{fig:a0sgs} } 
\end{figure}

Other historical standards \citepalias[from][]{jm53} are HD~35600 (B9\,Ib), $\sigma$~Cyg (HD~202850; B9\,Iab), HD~17088 and HD~21291 (both B9\,Ia). Their spectra are displayed in Fig.~\ref{fig:B9i}, where they can be seen to form a consistent luminosity sequence, based on the width of their Balmer lines (HD~17088 is shown in Fig.~\ref{fig:a0sgs}; it is almost identical to HD~21291, thus giving a well-defined anchor point for the type). The metallic spectrum in HD~202850, however, looks stronger than in the two Ia standards, while the \ion{He}{i} lines are marginally weaker, thus suggesting a later spectral type. This opens again the discussion on the need of a 9.5 spectral type (see Appendix~\ref{app:others}). Fig.~\ref{fig:a0sgs}, however, shows that there is little difference between the B9 and A0 supergiants, rendering an intermediate type likely superfluous. A second \citetalias{jm53} B9\,Iab standard, 4~Lac (HD~212593), is decidedly less luminous than $\sigma$~Cyg and can be considered the borderline between Ib and Iab. It is also marginally earlier than the others, especially as shown in the \ion{Mg}{ii}~4481\,\AA\ $>$ \ion{He}{i}~4471\,\AA\ ratio, and it could be given a type 8.5, if this interpolated value is used (see Appendix~\ref{app:others}).

\clearpage
\section{Other spectral subtypes}
\label{app:others}

As mentioned, not all the spectral subtypes that can be defined have been considered necessary throughout history. As per \citet{keenan85}, it is 'admittedly a somewhat awkward notation, for not all of the decimal subdivisions of the main HD types are equally meaningful'. As the quality of spectra has increased, more intermediate types have been added, and \citetalias{keenan85} considered O9.5, B0.5 and B9.5 full subtypes, although still not using B6. Using spectra of higher dispersion, \citet{walborn71} considered the introduction of O9.7, B0.2 and B0.7 necessary. In the main text, we have already discussed the case of B1.5 and B2.5, which are well defined, and B0.2, which lacked supergiant standards, but can be defined in terms of the presence or absence of \ion{He}{ii} lines. In addition, we have seen that the differences between B6 and B7, although noticeable, are probably too subtle to justify two separate subtypes. Here we discuss three other types that have sometimes been used, but we are not including in the primary list of standards. We stress that our not using these types does not imply that they are not well defined (in particular, B4) or that they are not necessary under any circumstances.

\subsection{ B4 type}
\label{sec:b4}

As discussed in the main text, between B3 and B8, there is a gradual decrease in the strength of \ion{He}{i} lines and a gradual shift in the ratios of \ion{Mg}{ii}~4481\,\AA\ and \ion{Si}{ii}~4128\,\AA\ with respect to these \ion{He}{i} lines. Fast-rotating main sequence stars have almost no other features, and the original MK system only considered subtypes B3, B5 and B8, with a few stars classified B6 or B7. Nevertheless, by the late 60's the group led by W.~W.~Morgan had developed criteria for all intermediate types and \citet{lesh68} used profusely the B4 type for low-luminosity stars.

%$\chi$~Aur (HD~36371; B4\,Ia)

\begin{figure}
\resizebox{\columnwidth}{!}{\includegraphics{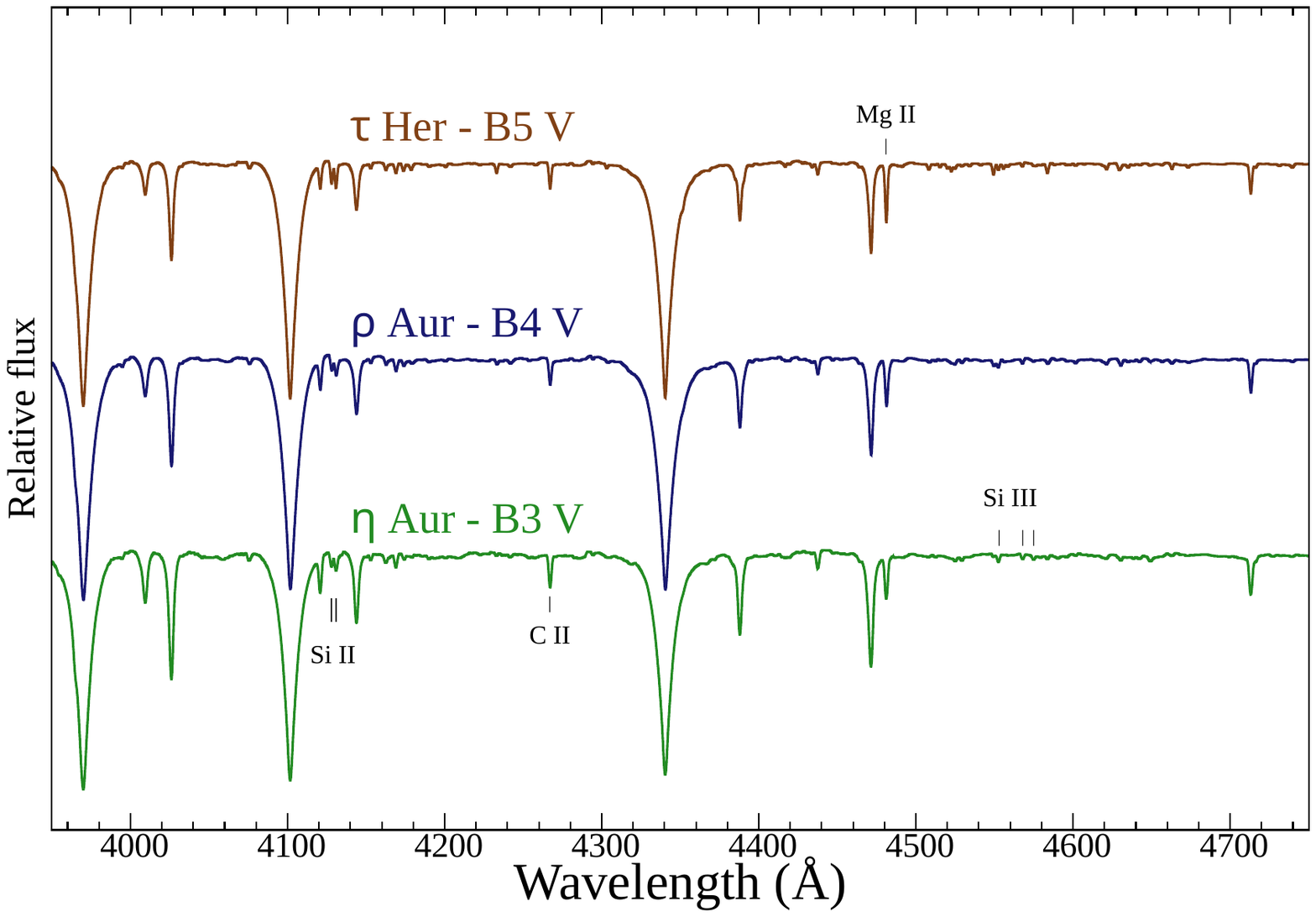}}
\centering
\caption{Some primary standards illustrating the intermediate type of $\rho$~Aur. At the very high S/N of these spectra, \ion{Si}{iii}~4553\,\AA\ is still visible in both $\eta$~Aur and $\rho$~Aur, allowing a clear comparison against neighbouring Fe lines. Reclassification of $\rho$~Aur as B4\,V is still compatible with its role as anchor standard of the MK system, as this subtype is not used in either \citetalias{mk73} or \citetalias{keenan85}. \label{fig:b4dwarf} } 
\end{figure}
%HD~216200  & 14~Lac           & B4   & III        Be

The subtype, however, does not seem well defined.
Lesh's proposed B4\,V standard, 90~Leo (HD~100600) is almost identical to the B3\,V standard $\eta$~Aur, and certainly not later. Likewise, the proposed B4\,IV standard, HD~176502, only looks very marginally later and somewhat more luminous than the low-rotation B3\,V standard HD~178849 (a classification B3\,IV-V would fit it), while her B4\,III standard, 14~Lac (= HD~216200), is a mild Be star and a moderately fast rotator ($v\,\sin\,i\approx 175\:\mathrm{km}\,\mathrm{s}^{-1}$) in an interacting binary. Among the stars classified as B4 by \citet{lesh68}, $\mu$~Eri (HD~30211) and 1~Vul (HD~180554) are probably the best examples of spectra intermediate between B3 and B5 (both would be B4\,IV, with the former coming close to the luminous edge of the box). 

Defining the B4 type is perfectly feasible with the high quality spectra that we are using. The intermediate features can be better seen at higher luminosity. As illustrated in Fig.~\ref{fig:bsgs}, $\chi$~Aur falls nicely between the B3\,Ia and the B5\,Ia anchor standards in all criteria, and it could be used to define B4\,Ia -- weirdly, it was originally classified as B3\,Ib by Morgan and then moved to B5\,Iab. Other luminous stars that could be classified as B4 are HD~167838 (B4\,Ia\,--\,Iab) and 67~Oph (B4\,Ib-II). However, it must be stressed that, if B4 is not used, all these stars would naturally fall into B5. This is further illustrated in Fig.~\ref{fig:b4dwarf}: a sensible choice for a primary B4\,V standard would be $\rho$~Aur, which happens to be the B5\,V dagger standard. This, together with all the cases discussed above, implies that B4 would not be an intermediate type between B3 and B5, but a subdivision of B5, which would require a re-arrangement of standards. If B4 is used, a slowly rotating standard B4\,V could be HD~26739.

%Is HD 25914 B4\,Ia?

\subsection{ B8.5 type}
\label{sec:b8p5}

\begin{figure}
\resizebox{\columnwidth}{!}{\includegraphics{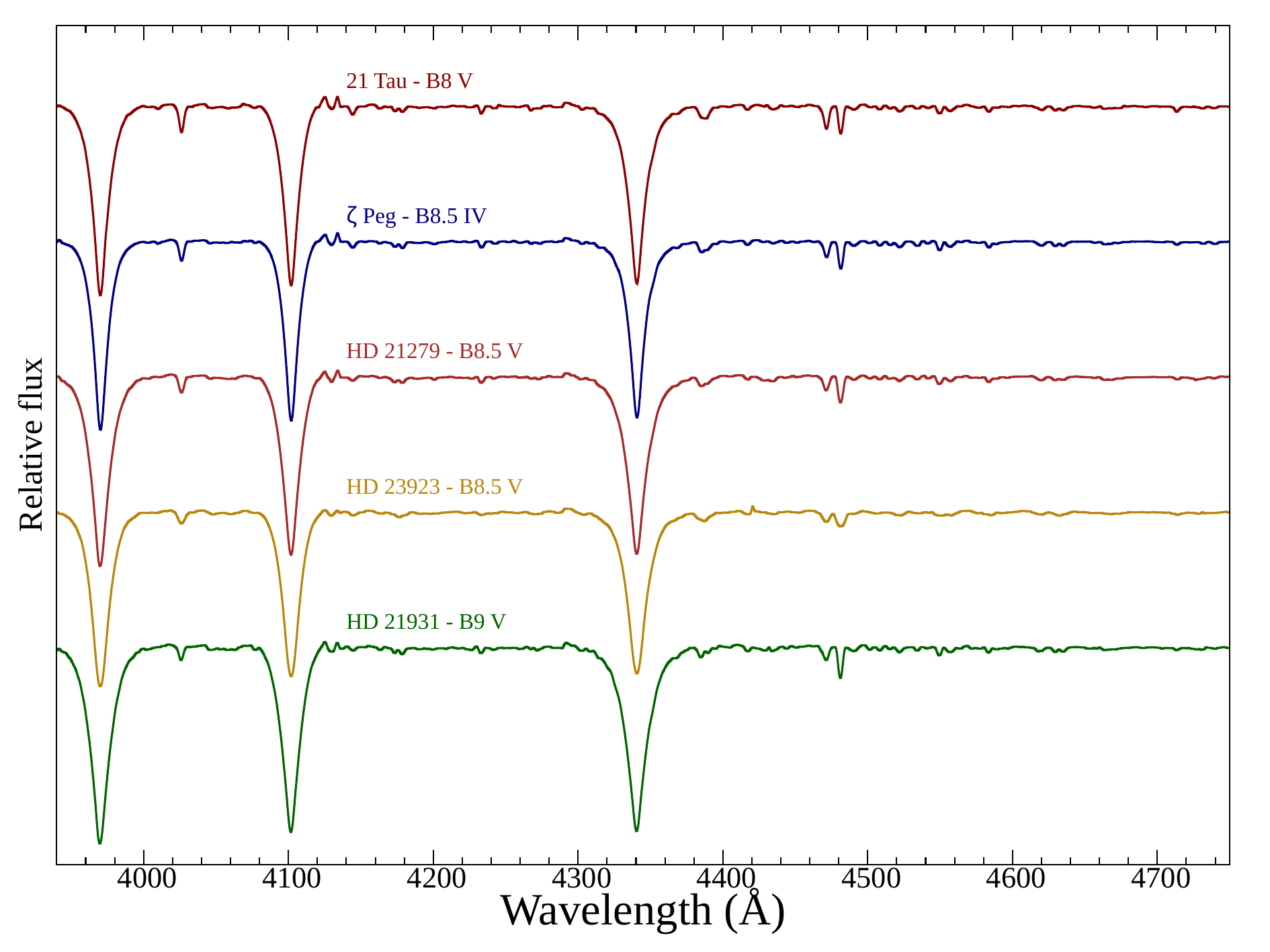}}
\centering
\caption{Some low-luminosity stars that may be classified as B8.5, compared to B8\,V and B9\,V standards. The earlier type of 21~Tau is confirmed by the presence of very weak \ion{He}{i}~4009, 4713\,\AA\ lines, which cannot be seen in the spectra of the B8.5\,V or B9\,V stars. The weakening of \ion{He}{i}~4026\,\AA\ is also obvious. $\zeta$~Peg would be close to the limit between the B8 and B8.5 subtypes. \label{fig:b8p5s} } 
\end{figure}

There are no published standards of spectral type B8.5, but this classification was widely used by W.W.~Morgan and collaborators in studies of open clusters, most notably NGC~2516 \citep{abt69}, where there are examples of B8.5 stars of all luminosity classes between III and V, and Melotte~20 \citep[the $\alpha$~Per cluster;][]{morgan71}. Since NGC~2516 is not accessible from La Palma, we may use as reference HD~21279, which is a member of Melotte~20 classified by W.W.~Morgan himself\footnote{This object is not included in Fig.~\ref{fig:combinedHR}, because it lacks photometry from \citet{harris56}.}. This allows for a direct and easy comparison to the B9\,V standard HD~21931, a member of the same cluster. Both stars are displayed in Fig.~\ref{fig:b8p5s}. The Pleaides member HD~23923, also displayed in Fig.~\ref{fig:b8p5s}, is an even faster rotator. It can be compared against other members of its cluster in Fig.~\ref{fig:mel22_late}. The former B8\,V standard $\zeta$~Peg has a higher luminosity than these stars, and we may classify it as B8.5\,IV, although it would be close to the border with B8\,IV (Fig.~\ref{fig:b8p5s}). The segregation is possible, but does not seem to provide a clearly separated grouping.

\begin{figure}
\resizebox{\columnwidth}{!}{\includegraphics{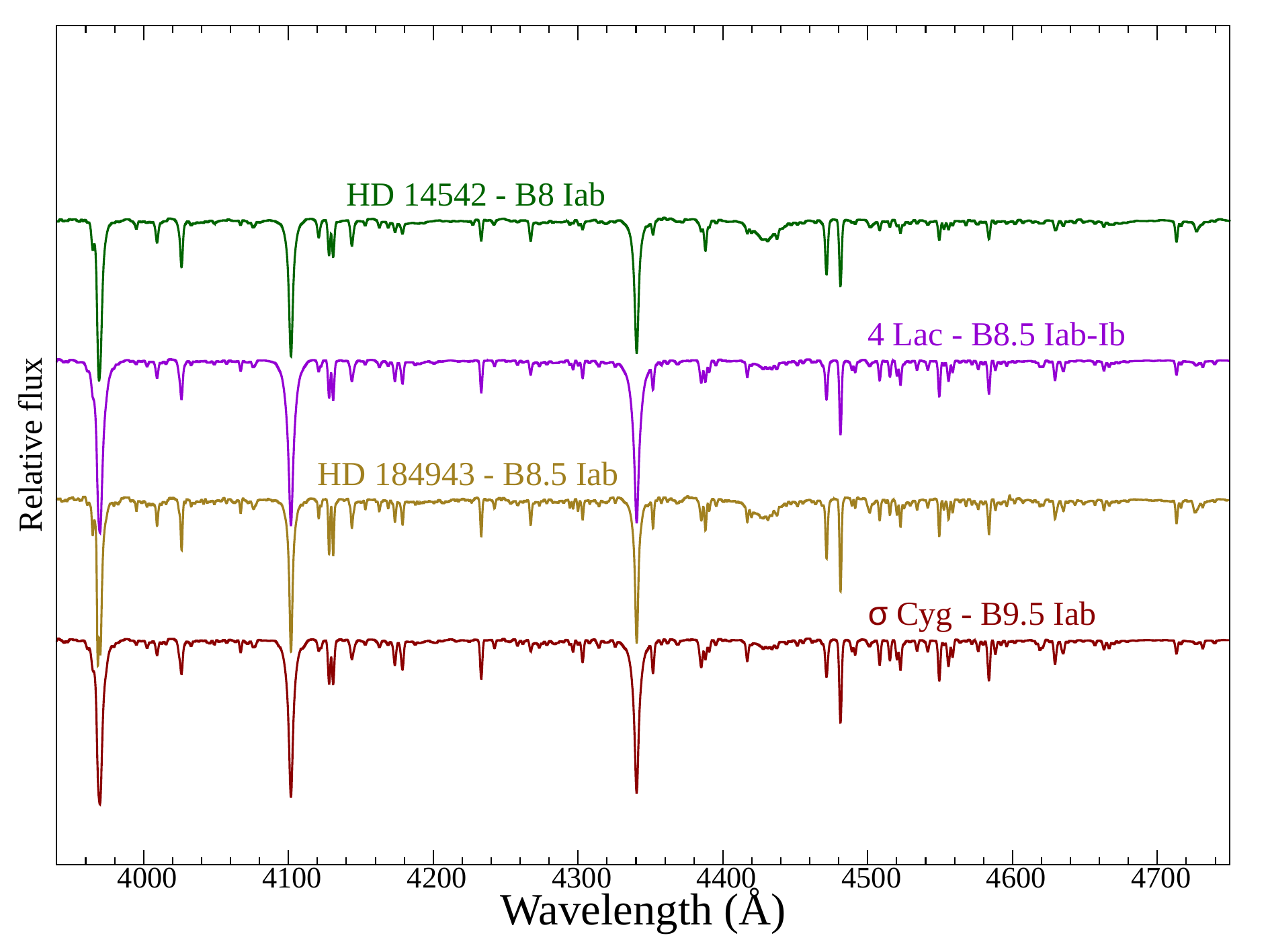}}
\centering
\caption{Some stars that may be classified as B8.5 supergiants, compared to the B8\,Iab and the B9\,Iab (B9.5\,Iab, if used) standards. \ion{He}{i} lines are much stronger than in the corresponding dwarfs (Fig.~\ref{fig:b8p5s}). The growth of the metallic spectrum in this range is spectacular, and may allow the definition of fractional subtypes, but this would then be very strongly dependent on metallicity.  \label{fig:latebIab} } 
\end{figure}

At higher luminosities, there are also stars that can be classified as B8.5, if the type is used. Objects that have been mentioned before are HD~184943 and 4~Lac. They are displayed together with other Iab standards in Fig.~\ref{fig:latebIab}. If we rely only on the traditional criterion \ion{Mg}{ii}~4481/\ion{He}{i}~4471, the stars displayed form a temperature sequence. This seems to be borne out by the evolution of the metallic spectrum, which grows enormously in this range. There is, however, no natural grouping, and other effects, such as metallicity and atmospheric structure, may impact on these features. It would be equally natural to group together HD~14542 and HD~184943 as B8\,Iab and 4~Lac and $\sigma$~Cyg as B9\,Iab (see next section about the B9.5 type). In summary, we do not recommend the use of B8.5.

\subsection{ B9.5 type}
\label{sec:b9p5}

There are no B9.5 dagger standards in \citetalias{mk73}, but W.W.~Morgan and collaborators made profuse use of this type in the analysis of different young open clusters. \citetalias{keenan85} listed HD~19805\footnote{This is erroneously given as HD~19803 in his table~I}, which is believed to be an outlying member of Mel~20, as B9.5\,V standard. This is a narrow-lined star with very broad Balmer line wings, but there is no hint of a \ion{He}{i}~4471\,\AA\ line in its spectrum, and so it must be an A-type star \citep[indeed, it was classified as A1\,V by][]{petrie69}, and its position in the cluster CMD is compatible with this classification. For the B9.5\,V standard, we adopt the \citetalias{jm53} standard $\omega^{2}$~Aqr (HD~222661). It compares very well in terms of luminosity to the B9\,V standard HD~21931 or the Pleiades B9.5\,V member HD~23873, both of which must be still in the early main sequence. \citet{morgan78} discussed the possibility of dividing dwarfs of spectral type B9\,--\,A0 in two luminosity subclasses, with Va being representative of field stars (such as the anchor A0\,V standard $\alpha$~Lyr) and Vb characterising ZAMS stars. In fact, HD~19805 is given as the prototype of B9.5\,Vb. We do not follow this criterion here and group together all dwarfs under luminosity class V, remaining conscious of the possibility of very broad Balmer lines in very young stars. \citet{garrison_g94} add $\omega$~For (HD~16046) as a slowly rotating B9.5\,V standard, but the width of its lines implies that it is almost as luminous as $\alpha$~Del, and so should be B9.5\,IV (see Fig.~\ref{fig:B9iii}). Stars of low luminosity that have been classified as B9.5 are displayed in Fig.~\ref{fig:B9v} At higher luminosity, as mentioned in Sect.~\ref{sec:b9}, $\gamma$~Lyr can be used as a B9.5 giant standard, although this is a rather luminous star, and we prefer to classify it as II--III (see Fig.~\ref{fig:B9iii}).

There are no B9.5 supergiant standards listed anywhere. Fig.~\ref{fig:a0sgs} compares all the stars given as B9\,Ia and A0\,Ia standards. \citetalias{keenan85} lists HD~21389 as the main A0\,Ia standard, and it is also one of the anchor standards in \citet{garrison94}. We therefore chose to use it as the main reference in this spectral region. The plot shows that the differences between the two subtypes are rather subtle, with only a moderate increase in the ratios of \ion{Mg}{i}~4881\,\AA/\ion{He}{i}~4471\,\AA\ and \ion{Si}{ii}~4128\,\AA/\ion{He}{i}~4026\,\AA. Then, HD~223960, which was the second A0\,Ia standard in \citetalias{jm53}, is more closely aligned with the B9\,Ia standards than with HD~21389. Its metallic spectrum is slightly stronger than those of the two B9\,Ia standards, while its \ion{He}{i} lines are marginally weaker. It must be noted, however, that its Balmer lines indicate a higher luminosity (it was classified A0\,Ia$^{+}$ by \citealt{morgan55}). Its \textit{Gaia} distance of about 2.9~kpc implies an absolute magnitude around $-7.6$~mag, which is high for Ia, in agreement with the width of its Balmer lines, which, though significantly narrower, do not justify a hypergiant classification.  Given the limited number of stars at our disposal, it would seem reasonable to reclassify $\sigma$~Cyg as B9.5\,Iab, but this seems stretching the system too much. It is probably best to leave it as B9\,Iab and not use B9.5 for supergiants (if at all).

\clearpage

\clearpage

\section{Problematic primary standards}
\label{app:gone}

A fair number of primary standards that have been widely used throughout the years have seen their spectral types changed. Some objects have had to be abandoned, due to a combination of causes. Here we list the most commonly used standards that have been removed from our list, and a few objects whose classification has been controversial, but are kept as standards.

\subsection{$\upsilon$~Ori = HD~36512}
\label{upsori}
$\upsilon$~Ori is a primary \citetalias{mk73} standard for B0\,V, together with $\tau$~Sco. $\upsilon$~Ori is clearly earlier (hotter) than $\tau$~Sco. When \cite{sota11} extended the interpolated type O9.7 to all luminosity classes by adopting the criterion \ion{He}{ii}~4542\,\AA$\;\simeq\;$\ion{Si}{iii}~4553\,\AA, $\upsilon$~Ori was moved to be the O9.7\,V standard.

\subsection{$\epsilon$~Per = HD~24760}
\label{epsper}

$\epsilon$~Per is the primary B0.5\,III standard in both \citetalias{mk73} and \citetalias{keenan85}. In contrast, \citet{walborn71} assigned $\epsilon$~Per to the newly created interpolated type B0.7\,III. Consideration of its spectrum does not support this change. The differences with the new B0.5\,III standard 1~Cas are minimal and most likely due to faster rotation. Moreover, $\epsilon$~Per shows strong line-profile variations due to non-radial pulsations, combined with single-lined binary motion \citep[see][and references therein]{tarasov95}. The companion is a solar-type star in an eccentric 14~d orbit, but there are reasons to believe that there is a second B-type star in a very wide orbit \citep{libich06}. It is therefore not adequate as a primary standard.

\subsection{$\kappa$~Cas = HD~2905}
\label{kapcas}

One of the original \citetalias{jm53} B1\,Ia standards, $\kappa$~Cas is given as the primary B1\,Ia \citetalias{keenan85} standard. However, \citet{walborn71} gives it as the defining standard for the new interpolated type B0.7\,Ia. In addition, $\kappa$~Cas has very weak N and very strong C lines for the type; thus, it can be considered a prototypical OBC star \citep{walborn76}. For this reason, its use as a standard is not encouraged. Then, the other \citetalias{jm53} B1\,Ia standard, HD~216411, was also moved to B0.7\,Ia, noting that it is very luminous and has a strong \ion{N}{ii} spectrum. Both spectra are compared in Fig.~\ref{fig:bcsgs}. The most suitable B0.7\,Ia standard would be HD~152235, but this is a southern hemisphere object and, thus, it is not included in this atlas.

\subsection{$o$~Per = HD~23180}
\label{oper}

$o$~Per is the primary \citetalias{mk73} standard for B1\,III. At our resolution, it is clearly a SB2. The secondary is of similar spectral type and lower luminosity. Although clearly weaker, it distorts all the line profiles. Therefore, $o$~Per is not suitable as a primary standard. 

\subsection{55 Cyg = HD~198478}
\label{55cyg}

55~Cyg was one of the original \citetalias{jm53} B3\,Ia standards, and it is given as the main B3\,Ia standard in \citetalias{keenan85}. When the interpolated type B2.5 is used, 55~Cyg is found to fit within this new type \citep[e.g.][]{walborn71}. The grid of \citetalias{keenan85}, however, does not include the B2.5 type, and keeps 55~Cyg as B3\,Ia. It is likely that its higher temperature made 55~Cyg look more luminous than $o^2$~CMa, explaining the demotion of the latter to Iab in this system (see next). In fact, the differences between the two are very minor.

\subsection{$o^2$ CMa = HD~53138}
\label{o2cma}

This star is given as the B3\,Ia standard in \citet{mk73}, but appears as the B3\,Iab standard in \citetalias{keenan85}. It was given again as B3\,Ia by \citet{lennon92}, where it is compared to other stars of the same type, leaving little doubt that it belongs to this class (see Appendix~\ref{55cyg} for a possible explanation). In fact, it is one of the anchor standards of \citet{garrison94}. Therefore, we have kept it as primary B3\,Ia standard.

\subsection{$\rho$ Aur = HD~34759}
\label{rhoaur}

$\rho$ Aur is the primary B5\,V standard in \citet{mk73}, where the intermediate spectral type B4 is not used. However, if B4 is used, $\rho$~Aur should belong to this type (see Appendix~\ref{sec:b4} and Fig.~\ref{fig:b4dwarf}). For this reason, we have provided other standards, more typical of the type (see Table~\ref{tab:b5lows}). If B4 is not used, $\rho$~Aur should be included as a B5\,V standards to account for the diversity within the type.

\subsection{$\tau$ Her = HD~147394}
\label{tauher}

$\tau$ Her is the primary B5\,IV standard in \citet{mk73}. However, according to our criteria, it is instead a slowly rotating B5\,V star. Its absolute magnitude is moderately higher than the average for B5\,V stars (cf. Table~\ref{tab:b5lows}), but the width of its line wings keeps it definitely as a dwarf.

\subsection{67~Oph = HD~164353}

67~Oph is the primary B5\,Ib standard. However, \citet{lennon92} argued that it was probably better classified as B5\,II. Detailed examination of its spectrum favours a classification B4\,Ib--II (see Sect.~\ref{sec:b5}). However, if we take into account that we do not strongly recommend the use of B4 (cf. Appendix~\ref{sec:b4}), 67~Oph may still be classified as B5\,Ib and used as a standard, if care is taken to note that it is close to the lower luminosity edge of the box. 

\subsection{$\tau$ Ori = HD~34503}
\label{tauori}

$\tau$ Ori is the primary B5\,III standard in \citet{mk73}. However, its line wings are broader than those of other stars of the same type, although it is almost as bright as our selected standard, HD~170682. Careful analysis suggests that it is, in reality, later, and we propose a B6\,III classification. In addition, at high resolution, it is an SB2 system (cf. Table~\ref{tab:b5lows}). 

\subsection{13 Cep = HD~208501}
\label{13Cep}

\begin{figure}
\resizebox{\columnwidth}{!}{\includegraphics{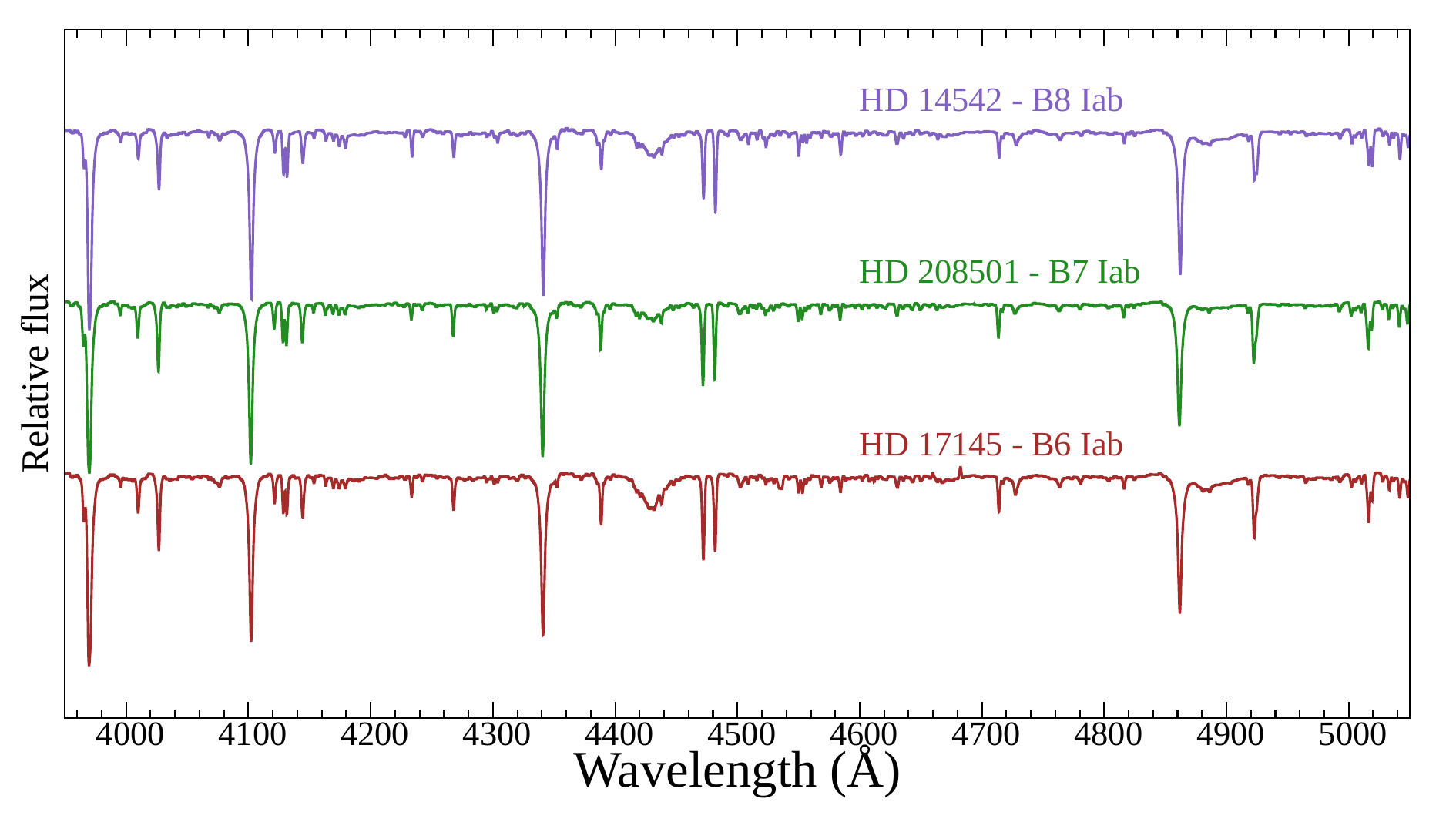}}
\centering
\caption{Temperature sequence for moderately luminous supergiants of mid B-types. At constant luminosity, the main criterion is the ratio of \ion{Si}{ii}~4129\,\AA\ to \ion{He}{i}~4144\,\AA, whose progression is clear here. The difference between the B6\,Iab and the B7\,Iab stars is, however, very small, and the secondary criterion \ion{Mg}{ii}~4481\,\AA\ to \ion{He}{i}~4471\,\AA\ would not allow a separation between them. Note that 13~Cep (HD~208501) was a B8\,Ib standard, but the width of its Balmer lines suggests higher luminosity, and it fits nicely in the sequence with this new classification. \label{midBsgsatIab} } 
\end{figure}

13~Cep is the B8\,Ib standard in \citetalias{keenan85}. However, according to our criteria, it is clearly more luminous. The weaker metallic lines are due to higher temperature. Comparison with the other \citet{jm53} standard, 53~Cas, and the B8\,Ia standard Rigel suggests a classification as B7\,Iab. The temperature sequence shown in Fig~\ref{midBsgsatIab} confirms this classification.

\clearpage

\section{Reference open clusters}
\label{app:clusters}

\begin{figure}
\resizebox{\columnwidth}{!}{\includegraphics{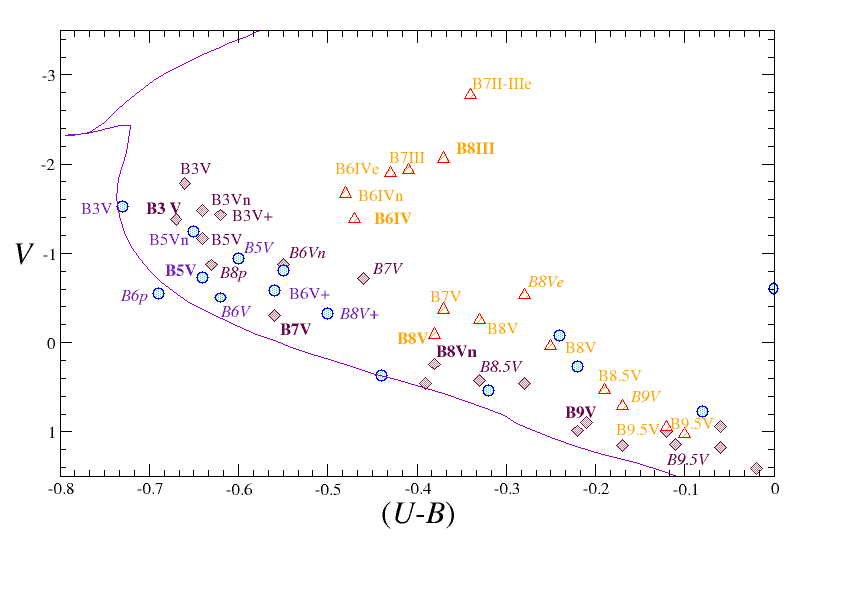}}
\centering
\caption{Combined HR diagram for stars in Melotte~20 ($\alpha$~Per cluster; maroon squares), Melotte~22 (Pleiades; orange triangles) and IC~4665 (blue circles). Stars have been dereddened individually following the standard techniques of \citet{jm53} and displaced by the distance modules calculated from TGAS. Spectral types in bold indicate standards in the main grid. Spectral types in italics are from the literature. The solid line is the observational ZAMS. \label{fig:combinedHR} } 
\end{figure}

When defining a luminosity scale for normal stars, accurate positioning of the main sequence, where stars with the highest effective gravity reside, is a fundamental first step. Since rotational velocity and other physical effects change our perception of line widths, a well-anchored reference point becomes mandatory. To this aim, we have used a number of open clusters and associations that provide us with stars conveniently close to the Zero Age Main Sequence (ZAMS). For the late-B stars, we could already count on a number of primary standards that had been taken from the Pleiades (Melotte~22). For earlier types, we have resorted to stars in the $\alpha$~Per cluster, which includes the B3\,V and B7\,V primary standards, and IC~4665. All three clusters are included in the detailed study of nearby open clusters carried out by \citet{leeuwen17} with data from the first \textit{Gaia} data release (TGAS). For a better comprehension of the evolutionary stage of each star, in Fig.~\ref{fig:combinedHR} we plot a combined HR diagram for all the B-type stars in the three clusters. In order to build it, we used $UBV$ photometry from classical sources: \citet{jm58} for Melotte~22,  
\citet{harris56} for Melotte~20, and \citet{hogg55} for IC~4665. Stars were individually dereddened by means of standard techniques based on the $Q$ index \citep{jm53,johnson58}, while absolute magnitudes were calculated by applying the mean distance modulus derived by \citet{leeuwen17}, namely $\mu=5.64$ for Melotte~22, $\mu=6.14$ for Melotte~20, and $\mu=7.75$ for IC~4665. We use the $(U-B)_0$ colour, as it has a much broader run than $(B-V)_0$ throughout the B-type range. We also plot, as a guide, the observational ZAMS from \citet{lb6}.

In an ideal world, we would also reference the luminosity class scale to a collection of open clusters where stars of the same spectral type and different luminosity classes are found. This has to some degree be done in Melotte~22, which provides primary standards for B6\,IV and B8\,III. Unfortunately, all the B-type \textit{Gaia} confirmed members of Melotte~20 and IC~4665 are dwarfs, with one exception. There are, of course, clusters that may provide very adequate spectral and luminosity sequences, most notably $h$ and $\chi$~Per, but their stars are, except for supergiants, too faint to achieve a S/N comparable to that of the main sample with the instrumentation used. A borderline case is IC~4725, which has a distance modulus  $\mu=9.2$ and is affected by relatively high extinction \citep{cantat20}. This is a rich cluster, and we have used its two brightest stars as standards for B5\,III and B6\,III. It also contains some valuable examples of B5\,IV and main sequence stars, but again they are too faint to be observed with our instrumentation at the required S/N.

\subsection{Melotte~22}
\label{sec:mel22}

\begin{figure}
\resizebox{\columnwidth}{!}{\includegraphics{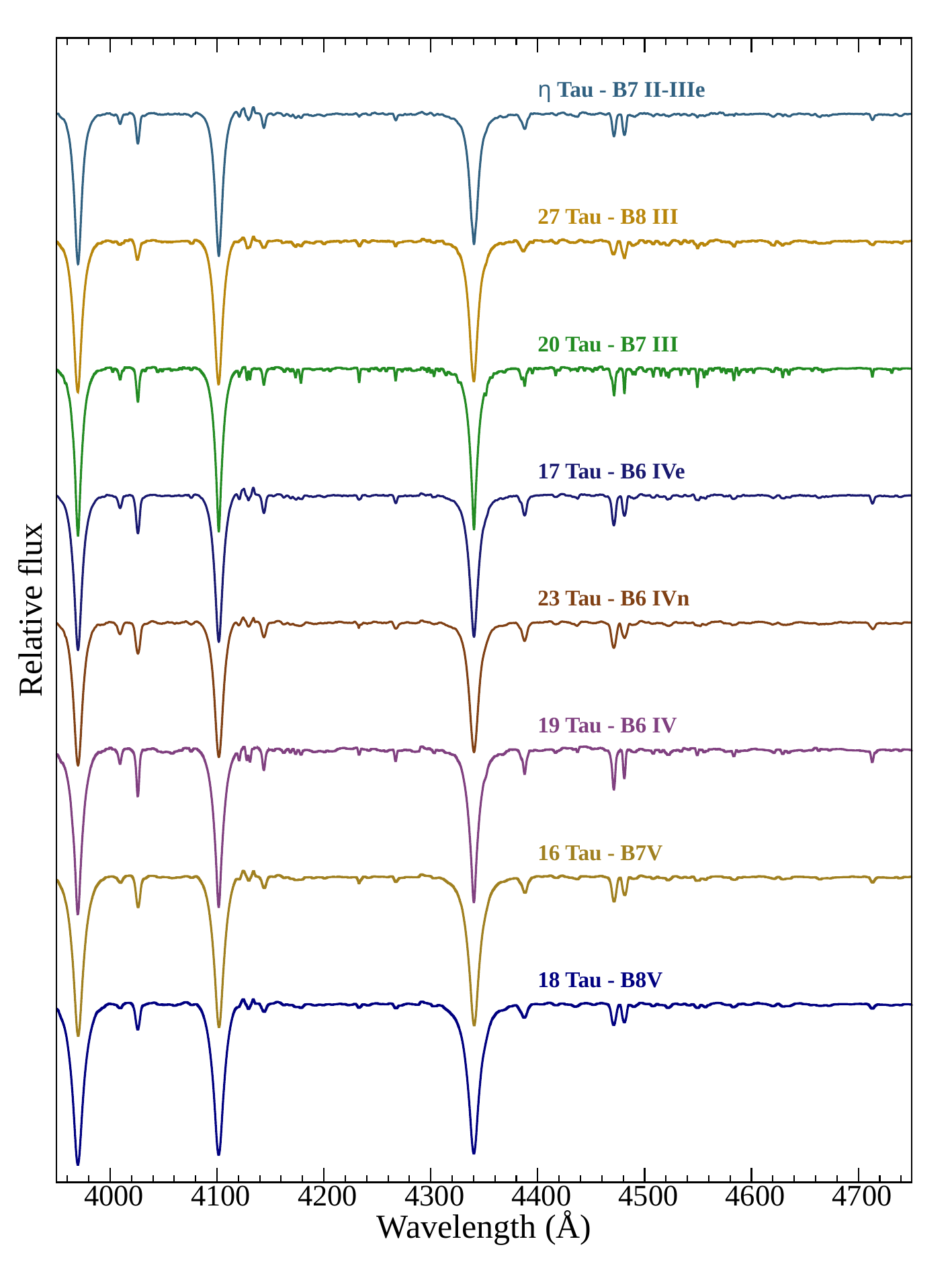}}
\centering
\caption{ Brightest members of the Pleiades (plus the B8\,V primary standard 18~Tau, also a member), ordered (from top to bottom) by brightness. 27~Tau (Atlas), 19~Tau and 18~Tau are the primary standards for their respective types, while 20~Tau is a useful standard for B7\,III, because of its narrow lines. $\eta$~Tau (Alcyone) was the original primary standard for B7\,III, but was discontinued because of its shell episodes. In addition, it appears too luminous to be representative of the type. A fainter shell star, 28~Tau (Pleione = HD~23682) was not observed. \label{mel22} } 
\end{figure}

The brightest members of the Pleiades (shown in Fig.~\ref{mel22}) include several useful standards, which can be effectively placed on an observational HR diagram. 16~Tau (B7\,V) is the hottest dwarf member. It was formerly classified as B7\,IV, but our criteria make it luminosity class V. In fact, it is still on the main sequence, even if it is much more evolved than the primary standard HD~21071, as can be seen in Fig.~\ref{fig:combinedHR}. All the B6 stars are evolved away from the main sequence. Of the three B6\,IV stars, 19~Tau is the obvious choice for a standard, as 23~Tau (HD~23480) is a very fast rotator and 17~Tau (HD~23302) is a Be star (although no emission lines are present in our spectrum).

20~Tau is a good secondary standard for B7\,III, as it is a slow rotator ($v_{\mathrm{rot}}\approx30\:\mathrm{km}\,\mathrm{s}^{-1}$) and its absolute magnitude ($M_V\approx -1.9$) is close to the average for the type. Meanwhile, $\eta$~Tau (Alcyone) was the primary B7\,III standard in \citetalias{mk73}, but it was discarded because of its frequent shell episodes. Moreover, its absolute magnitude ($M_V\approx -2.9$) is very high for the type, and our luminosity criteria place it close to class II. 27~Tau is the primary standard for B8\,III. It is a moderately fast rotator and somewhat brighter than typical for the spectral type. As discussed in Sect.~\ref{sec:b8}, it is a binary.

\begin{figure}
\resizebox{\columnwidth}{!}{\includegraphics{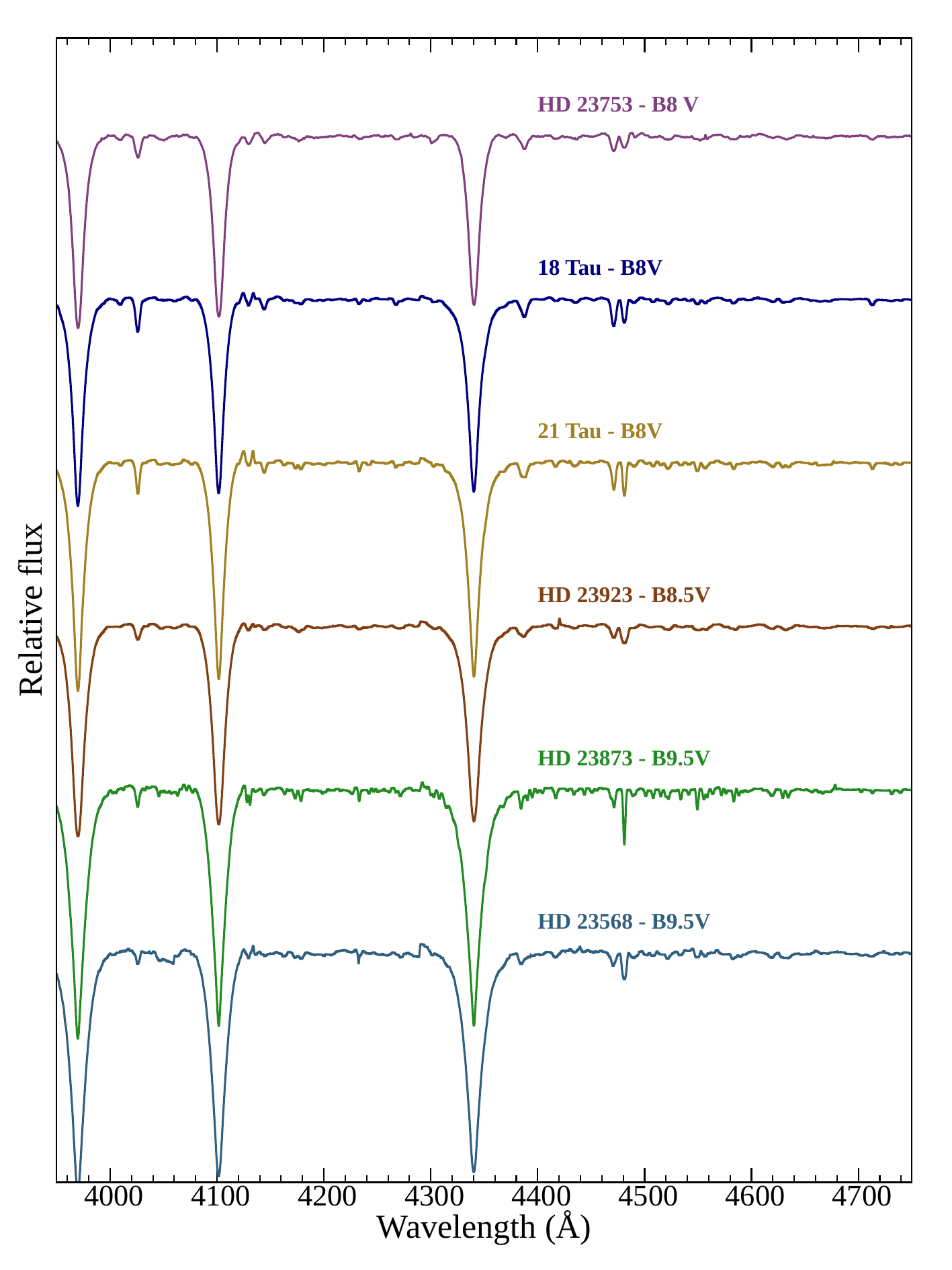}}
\centering
\caption{Other fainter members of the Pleiades ordered (from top to bottom) by brightness. The B8 spectral type is characterised by \ion{He}{i}~4471\,AA\ and \ion{Mg}{ii}~4481\,\AA\ displaying similar strengths (18~Tau is the primary standard), but their ratio changes strongly over this temperature range, allowing the definition of spectral type B8.5 (see Appendix~\ref{sec:b8p5}). HD~23873 is used as a slowly rotating B9.5\,V standard in Fig.~\ref{fig:B9v}. When compared to HD~23873, HD~23568 displays strongly the effects of rotation, although it is still a moderate rotator. \label{fig:mel22_late} } 
\end{figure}

The primary B8\,V standard is 18~Tau. 21~Tau is sometimes given as a secondary standard; it is slightly cooler and more luminous. Figure~\ref{fig:mel22_late} shows some late-B members of Melotte~22. As can be seen in Fig.~\ref{fig:combinedHR}, at the age of this cluster \citep[$\approx 125\:\mathrm{Ma}$; e.g.][]{cummings18,monteiro20, galindo22}, only the B9.5\,V stars are close to the ZAMS.

%and can be classified as B8.5\,V if this intermediate type is adopted (see Appendix~\ref{b85}). 

\subsection{Melotte~20}
\label{sec:mel20}

\begin{figure}
\resizebox{\columnwidth}{!}{\includegraphics{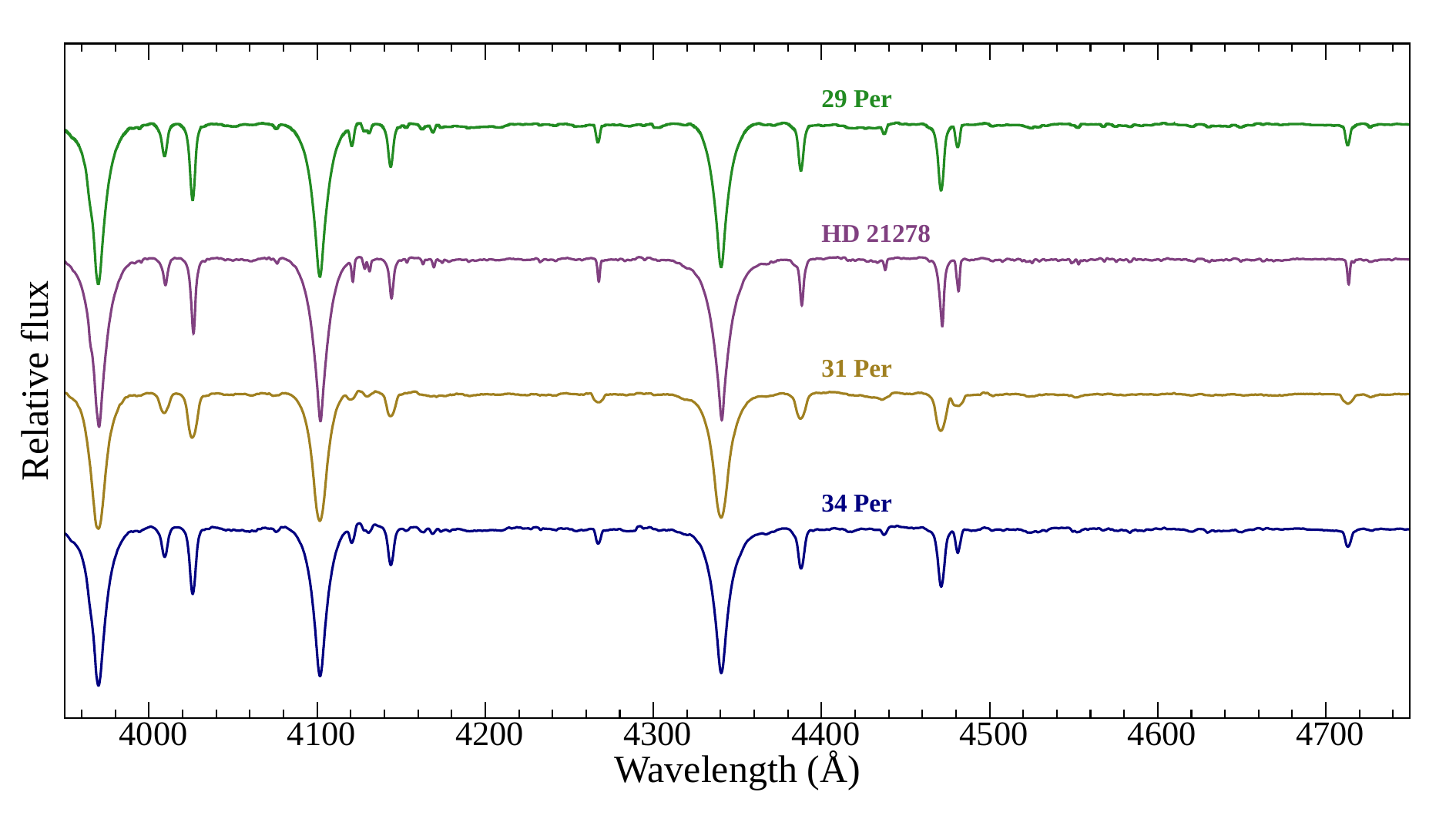}}
\centering
\caption{Four B3\,V stars from Melotte~20. 29~Per is one of the primary standards for this type. HD~21278 is a spectroscopic binary, but only the primary is seen in this spectrum.\label{fig:mel20} } 
\end{figure}

Also known as the $\alpha$~Per cluster, because of its brightest member, the F5\,Ib supergiant HD~20902, Melotte~20 is an extended cluster affected by little extinction. Its earliest members are all B3\,V, with the possible exception of $\varphi$~Per, which is a likely member, a binary containing a $\sim$B1\,V Be star and a hot subdwarf, believed to have undergone substantial mass transfer \citep{mourard15}. The four B3\,V members are shown in Fig.~\ref{fig:mel20}. They illustrate the range of stellar parameters that can be assigned to a given spectral type. 29~Per, the star closer to the ZAMS, is one of the primary standards for the type. 34~Per (HD~21428) is almost identical (very marginally later), despite being almost half a magnitude brighter. Both stars have similar rotational velocity. 31~Per (HD~20418) is a much faster rotator (and thus B3\,Vn), and has about the same brightness as 29~Per. HD~21278 is a slow rotator ($\approx50\:\mathrm{km}\,\mathrm{s}^{-1}$) and somewhat later than 29~Per, although not enough to justify a B4\,V classification. This object has repeatedly been marked as a spectroscopic binary. No secondary is seen in our spectra, but \citet{morrell92} detected it at some phases, concluding that it has a mass about half that of the primary and orbits it every 22~d.

Recently, \citet{meingast21} identified an extended corona to Melotte~20. Among their list of new members, 53~Per (HD~27936) would be the brightest member on the main sequence. \citet{lesh68} classified it as B4\,IV. We find its spectral type to lie between B3\,IV and B4\,IV, and could in fact be given as any. It is only about 0.5~mag brighter than 29~Per, the preferred B3\,V standard.

The only B5\,V member, HD~20809, is very similar to the B5\,V standard 68~Cas. Another member of the cluster is HD~21071, primary standard for B7\,V. 
The age of Melotte~20 has traditionally been estimated in the 50\,--\,60~Ma range, which agrees with ages estimated for $\varphi$~Per by modelling \citep[and references therein]{schootemeijer18}. This age is in very good agreement with a turn-off at B3\,V, corresponding to $\approx6\,M_{\sun}$ at approximately solar metallicity \citep[e.g.][]{harmanec88}. Nevertheless, several modern age determinations based on the properties of the low-mass cluster members result in ages 80\,--\,$110\:$Ma \citep[and references therein]{cummings18,pamos22,galindo22}, which is incompatible with the presence of B3\,V stars, unless they are blue stragglers formed via binary interaction. This possibility seems very unlikely for a number of reason. Firstly, $\varphi$~Per provides an example of the sort of binary interaction products that we would expect to see. The system has a current (i.e. after mass loss) total measured mass close to $11\,M_{\sun}$ \citep{mourard15}, and the best model indicates an initial mass $>7\:M_{\sun}$ for the original primary \citep[the progenitor of the subdwarf;][]{schootemeijer18}, compatible with an initial spectral type somewhat earlier than B3\,V. Secondly, as remarked, HD~21278 is a binary with two B-type stars, namely, a pre-interaction system. Finally, the position of cluster members in the HR diagram (Fig.~\ref{fig:combinedHR}) lies much closer to the ZAMS than to the members of Melotte~22, which is not much older than 100\:Ma. In view of this, an age younger than 60\:Ma seems much more consistent with the upper main sequence, and we can assume that HD~21071 is close to ZAMS. Another standard that is a cluster member, HD~21931 (B9\,V) should certainly be considered a ZAMS star. In Appendix~\ref{sec:b8p5}, we recommend the use of another member, HD~21279 as B8.5V standard, if the need to use this spectral type arises.

\subsection{IC~4665}
\label{sec:ic4665}

IC~4665 is a small, dispersed cluster, and the youngest in the sample, with an age not significantly higher than 30\:Ma \citep[e.g.][]{cargile10,randich18,galindo22}. Its earliest member is HD~161573, a slowly rotating ($\approx40\:\mathrm{km}\,\mathrm{s}^{-1}$) B3\,V star that still lies moderately close to the ZAMS. Its spectrum is pretty much identical to that of HD~178849, which has a similar rotational velocity.  The cluster contains two B5\,V stars: HD~161672 is a fast rotator $\approx190\:\mathrm{km}\,\mathrm{s}^{-1}$) that we take as primary standard for B5\,V close to the ZAMS (as its spectrum is almost identical to that of 68~Cas), while HD~161677 is an even faster rotator.

 Most bright members of the cluster are spectroscopic binaries \citep{abt72} and they present double lines at high S/N. Meanwhile, the higher distance modulus and extinction means that late-B members are considerably fainter than those in the Pleiades or Melotte~20, rendering this cluster less useful with our instrumentation.

\clearpage

\section{Additional figures}
\label{app:additional}

This appendix contains figures   useful for illustrating classification issues detailed in the main text.

\begin{figure}
\resizebox{\columnwidth}{!}{\includegraphics{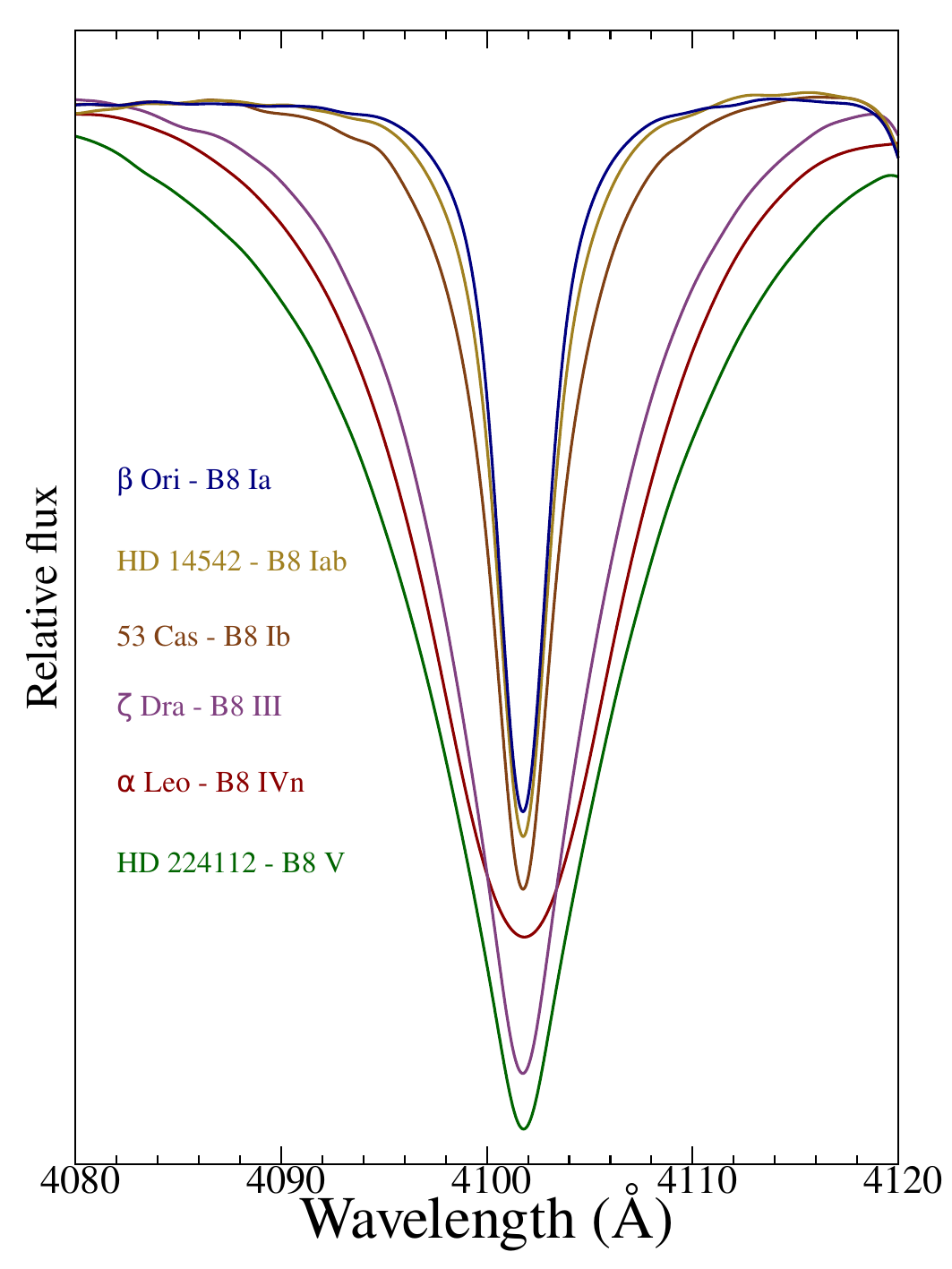}}
\centering
\caption{The H$\delta$ line for a number of B8 stars illustrates our criterion for assigning luminosity class. Compare to the B3 stars in Fig.~\ref{fig:wingsb3}. By this late type, there is a very large gap between stars close to the main sequence and the supergiants (see discussion in Sect.~\ref{sec:ii}). The stars shown here are the same as in Fig.~\ref{fig:B8seq}, except for $\alpha$~Leo, which has replaced $\pi$~Cet in order to demonstrate the  effect of fast rotation on the central part of the line profiles.  \label{fig:wingsb8} } 
\end{figure}

\begin{figure}
\resizebox{\columnwidth}{!}{\includegraphics{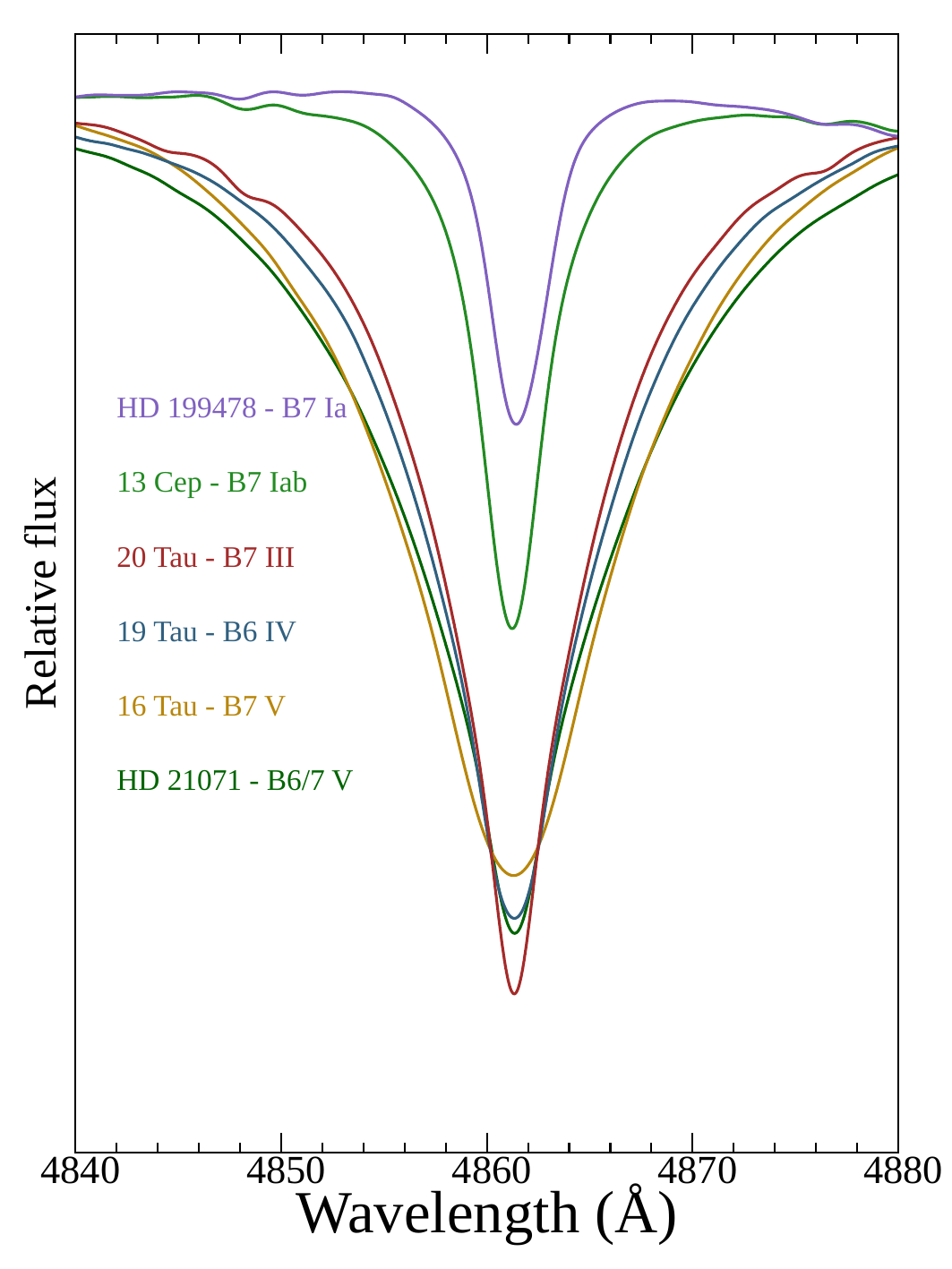}}
\centering
\caption{ H$\beta$ line for a number of B6/7 stars, demonstrating that our criterion for assigning luminosity class can be used in any Balmer line. Compare to the B3 stars in Fig.~\ref{fig:wingsb3}. The fast rotator 16~Tau has been added so that  it can be compared to HD~21071 (different line profiles and depths), and to provide a luminosity sequence within the Pleiades.   \label{fig:wingsb7} } 
\end{figure}

\begin{figure}
\resizebox{\columnwidth}{!}{\includegraphics{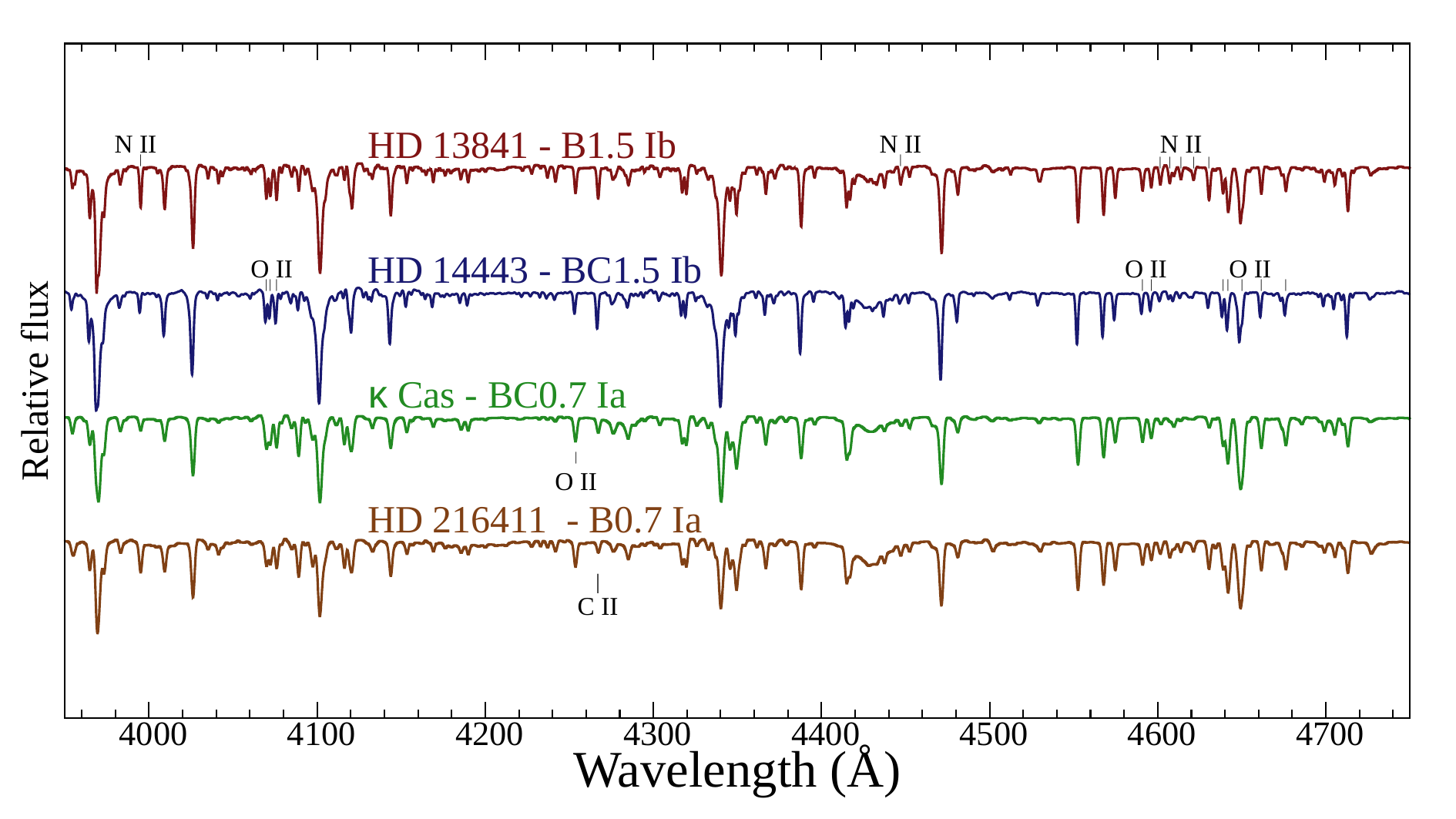}}
\centering
\caption{Spectra of two C-rich supergiants compared to standards of the same spectral types. The most obvious effect is the weakening of all \ion{N}{ii} lines, the strongest of which are marked on top of the spectrum of HD~13841. In addition, \ion{C}{ii}~4267\,\AA\ is enhanced in the spectrum of HD~14443. In $\kappa$~Cas, there are also enhanced contributions of \ion{C}{iii} to the \ion{O}{ii} complex around 4072\,\AA, and the 4650\,\AA\ blend. \label{fig:bcsgs} } 
\end{figure}

\begin{figure}
\resizebox{\columnwidth}{!}{\includegraphics{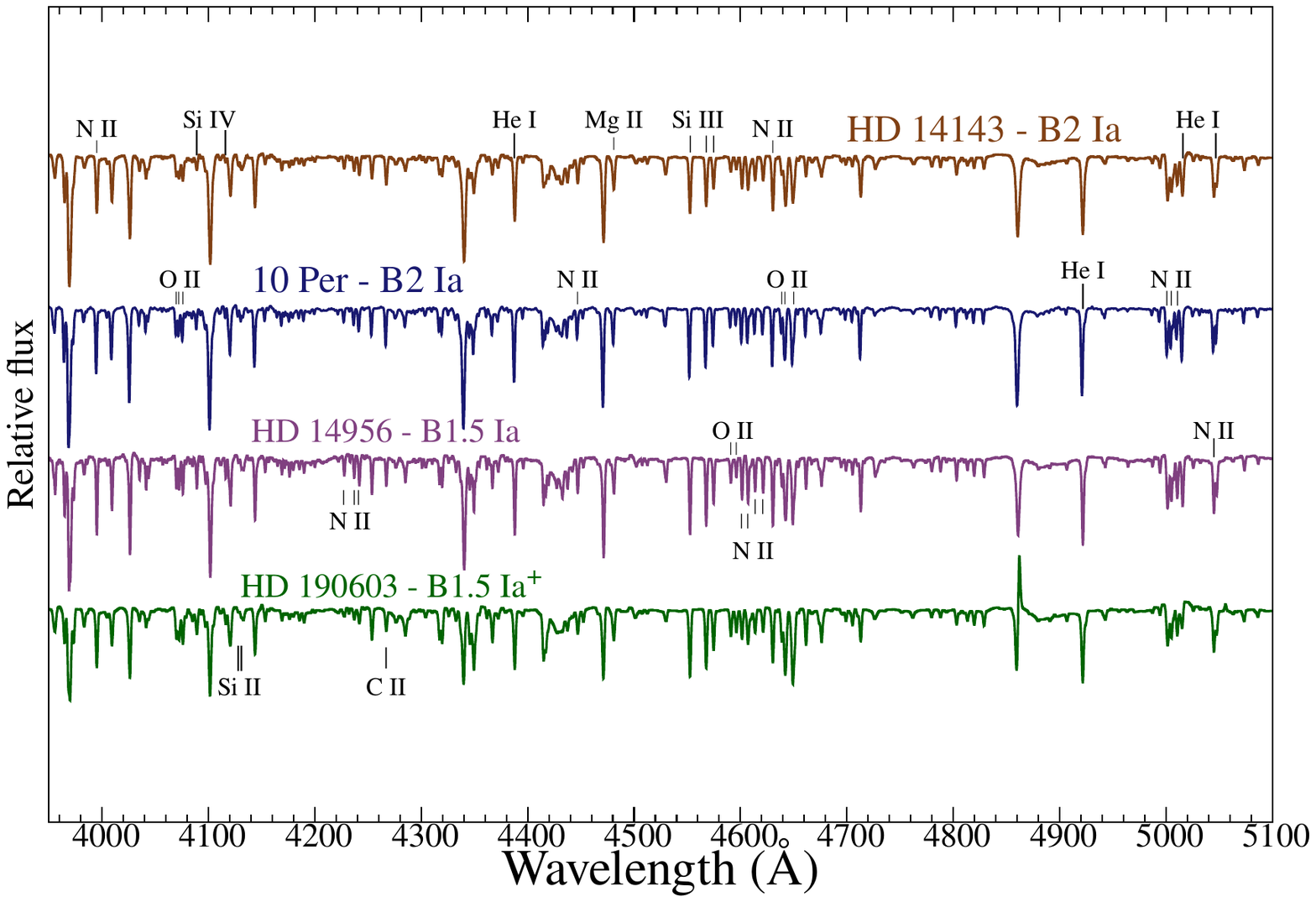}}
\centering
\caption{Spectrum of HD~14956, B1.5\,Ia compared to two B2\,Ia supergiants from Per~OB1 and the B1.5 hypergiant HD~190603. The higher temperature of the two lower spectra is reflected in the stronger \ion{Si}{iv} lines. The rest of the spectrum is almost indistinguishable. The main differences are due to the mild Carbon deficiency and Nitrogen enhancement of the two B1.5 stars (reflected in the weakened \ion{C}{ii}~4267\,\AA\ and strengthened \ion{N}{ii}~3995\,\AA\ lines). Compare to the more typical B1.5 morphology of HD~13841 in Fig.~\ref{fig:epsCMa}. Note, however, that, except for the strong \ion{N}{ii}~3995\,\AA\ line, N is much more enhanced in HD~14956 than in the hypergiant. \label{fig:b2ia} } 
\end{figure}

\begin{figure}
\resizebox{\columnwidth}{!}{\includegraphics{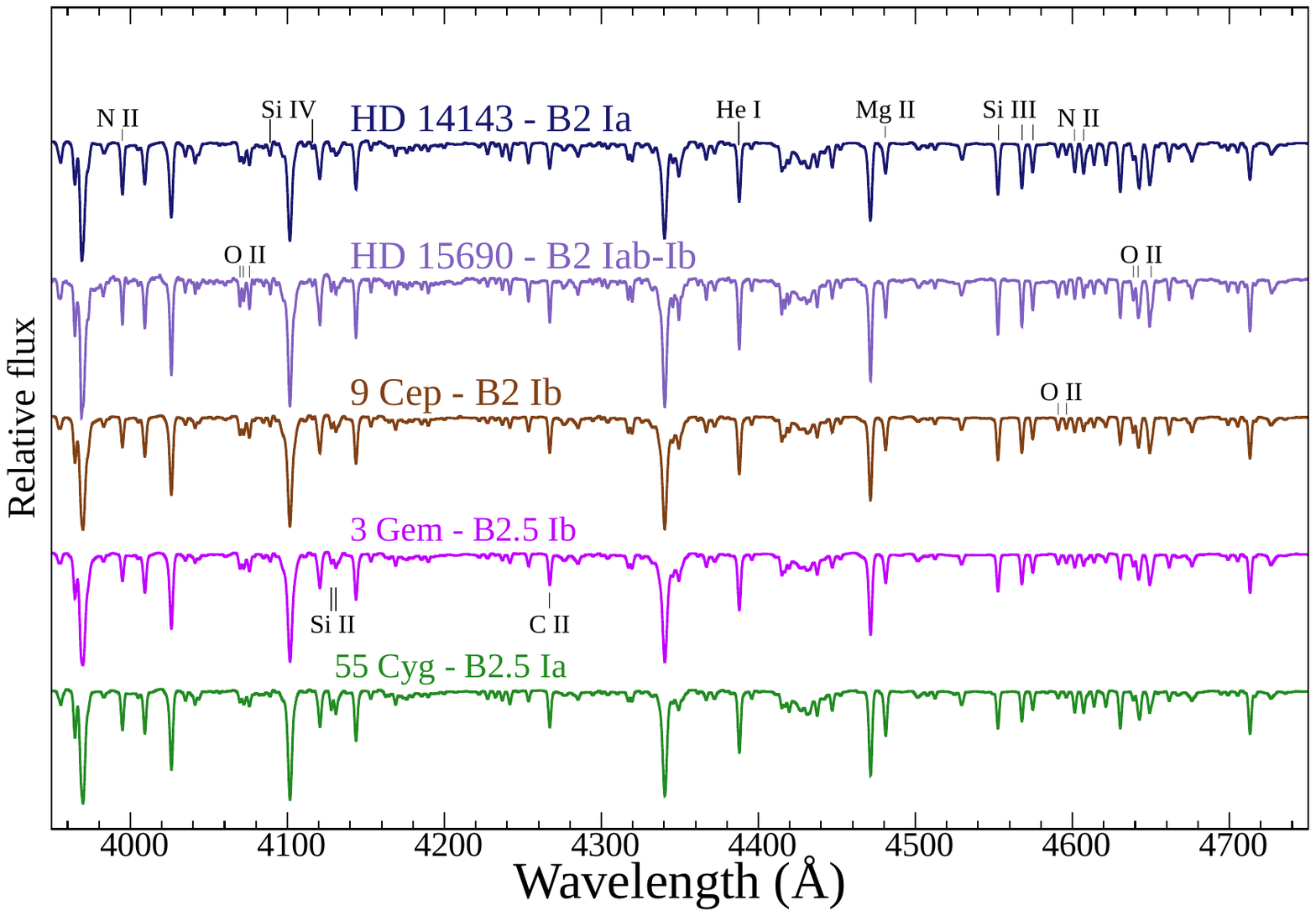}}
\centering
\caption{Some supergiants of spectral types B2 and B2.5, illustrating the difficulties in defining the Iab class in this range. The three stars in the middle are very similar among themselves. Apart from the enhanced \ion{N}{ii} in the Ia stars, the differences with respect to the Ib standards do not allow for an intermediate class.\label{fig:b2p5sgs} } 
\end{figure}

\begin{figure}
\resizebox{\columnwidth}{!}{\includegraphics{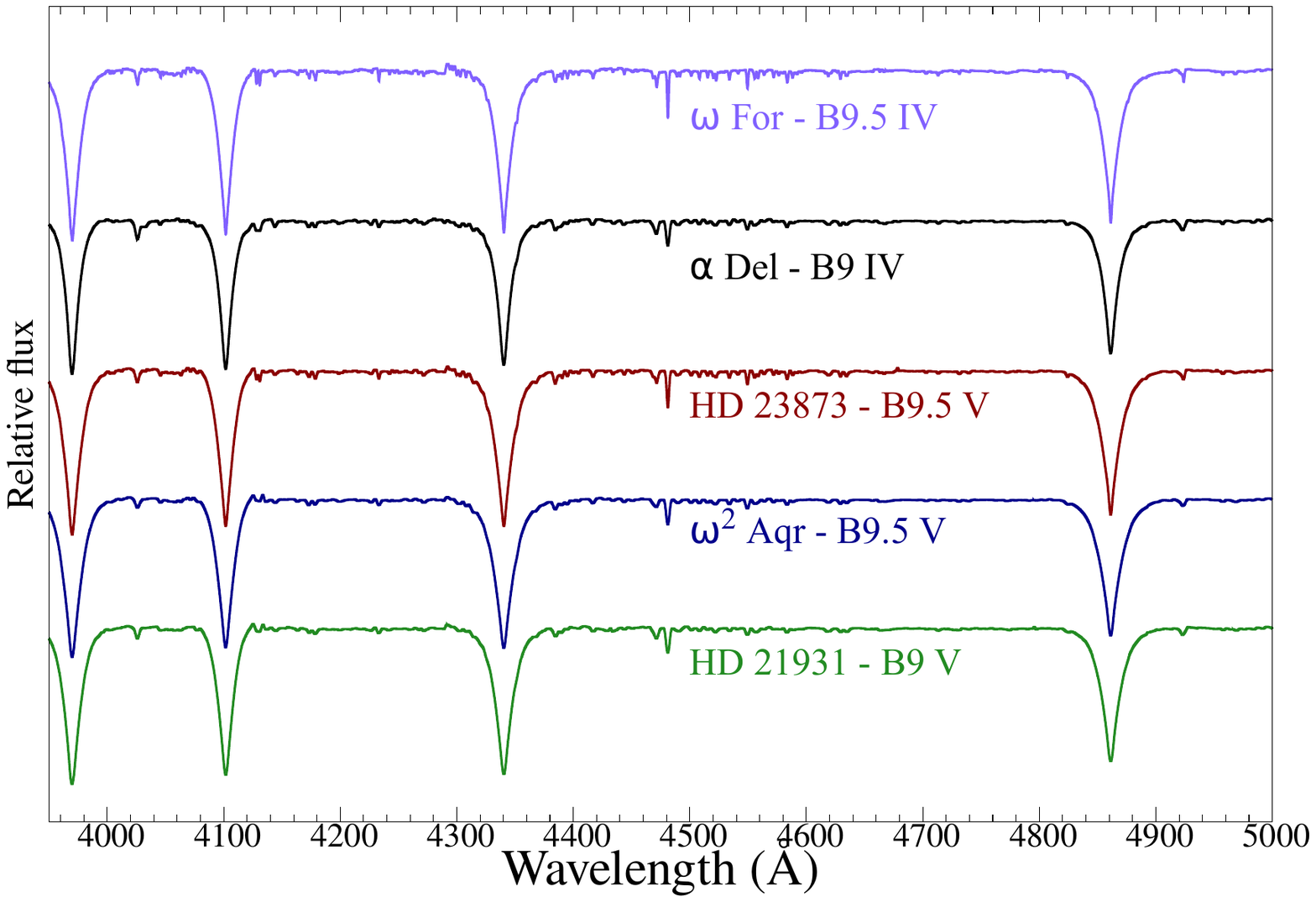}}
\centering
\caption{Stars  previously classified as B9\,V or B9.5\,V compared to the primary standard $\alpha$~Del. The spectral types indicated are those adopted. $\omega^{2}$ Aqr is a faster rotator than HD~23873. For a faster rotating B9.5\,V star, see HD~23568 in Fig.~\ref{fig:mel22_late}. The differences between B9 and B9.5 stars of similar rotational velocity are subtle.  \label{fig:B9v} }
\end{figure}

\begin{figure}
\resizebox{\columnwidth}{!}{\includegraphics{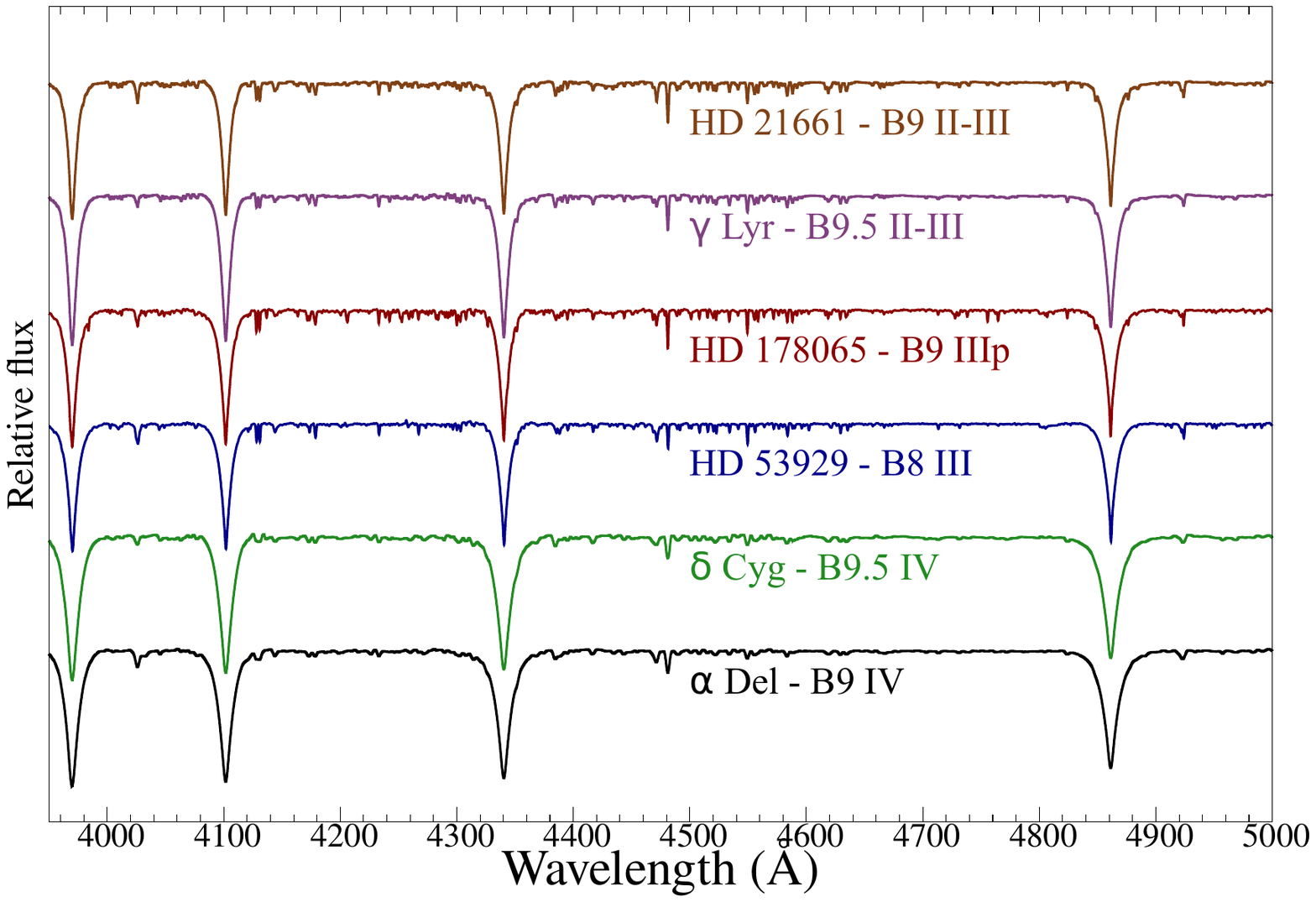}}
\centering
\caption{Stars that have been classified as B9\,III or B9.5\,III compared to the primary standard $\alpha$~Del. The spectral types indicated are those adopted. HD~178065 is a HgMn peculiar star, with a rich metallic spectrum. HD~53929 could be argued to be B8.5\,III, but again differences between subtypes are very subtle.\label{fig:B9iii} } 
\end{figure}

\clearpage

\section{New list of B-type standards}

Table~\ref{bigtable} shows the new list of standards. Columns list their HD reference, their spectral type, their \vsini\ measured after \citet{ssimon14}, their $V$ magnitude from Simbad, an estimate of their absolute magnitude (see Sect.~\ref{quantities} for details of calculation), and the \fwhb\ measurement \citep[after][]{deBurgos23}. Stars that have been identified as spectroscopic binaries, SB1 or SB2, are marked with (1) or (2) respectively in the Target Name column. Stars whose H$\beta$ line does not show a perfect absorption profile have also been marked in the \fwhb\ column, to show that the fitting performed when measuring this value is not perfect. The profiles that show a P-Cygni shape with emission in the red wing of the line are marked with (3),  profiles where the red wing of the line is filled are marked with (4), while those where the core of the line is filled with emission are marked with (5), following the criteria described in\citet{deBurgos23}. 

We note that all the B4 standards should be classified B5 if the interpolated type is not used. HD~184943 is classified as B8\,Iab, but may be used as reference for B8.5 supergiants, if the need for this interpolated subtype is felt. The A0 standards and the anchor point Deneb (HD~197345), A2\,Ia, have been added for reference.

\clearpage
\onecolumn

\begin{longtable}{@{}lllcccc@{}}
\caption{New list of B-type standards, with some early-A standards added for reference. \label{bigtable}}\\
\noalign{\smallskip}
\hline\hline
\noalign{\smallskip}
Target name & Common name & SpC & $v\,\sin\,i$ & $V$ & $M_{\rm V}$ & \fwhb \\
& & & (\kms) & (mag) & (mag) & (\AA) \\
\noalign{\smallskip}
\hline
\noalign{\smallskip}
\endfirsthead
\caption{continued.}\\
\hline\hline
\noalign{\smallskip}
Target name & Common name & SpC & $v\,\sin\,i$ & $V$ & $M_{\rm V}$ & \fwhb \\
& & & (\kms) & (mag) & (mag) & (\AA) \\
\noalign{\smallskip}
\hline
\noalign{\smallskip}
\endhead
\noalign{\smallskip}
\hline
\endfoot
HD\,37128 & $\epsilon$~Ori & B0\,Ia & 52 & +1.7 & -7.4 & 3.3 \\
HD\,205196 & ~ & B0\,Ib & 51 & +7.4 & -4.9 & 3.6 \\
HD\,164402$^{(1)}$ & ~ & B0\,Ib & 54 & +5.8 & -4.5 & 4.6 \\
HD\,48434 & ~ & B0\,III & 52 & +5.9 & -5.5 & 4.0 \\
HD\,149438 & $\tau$~Sco & B0\,V & 9 & +2.8 & -3.2 & 8.4 \\
HD\,171012 & ~ & B0.2\,Ia & 66 & +7.0 & -6.5 & 2.9 $^{(5)}$ \\
HD\,204172 & 69~Cyg & B0.2\,Iab & 58 & +5.9 & -6.0 & 3.6 \\
HD\,16808 & ~ & B0.2\,Ib & 78 & +8.6 & -5.2 & 3.5 \\
HD\,6675$^{(1)}$ & ~ & B0.2\,III & 29 & +6.9 & -4.8 & 5.2 \\
HD\,36822 & ~ & B0.2\,IV & 19 & +4.4 & -4.0 & 7.1 \\
HD\,2083$^{(2)}$ & ~ & B0.2\,V & 11 & +6.9 & -3.9 & 7.9 \\
HD\,38771 & $\kappa$~Ori & B0.5\,Ia & 47 & +2.1 & -4.6 & 3.9 \\
HD\,194839 & ~ & B0.5\,Ia & 65 & +7.5 & -7.2 & 2.9 \\
HD\,192422 & ~ & B0.5\,Ib & 50 & +7.1 & -6.7 & 3.3 \\
HD\,213087 & 26~Cep & B0.5\,Ib & 64 & +5.5 & -6.1 & 3.1 \\
HD\,193007$^{(1)}$ & ~ & B0.5\,II & 61 & +7.9 & -7.6 & 6.4 \\
HD\,218376 & 1~Cas & B0.5\,III & 29 & +4.8 & -4.0 & 6.3 \\
HD\,36960 & ~ & B0.5\,V & 25 & +4.7 & -3.4 & 8.0 \\
HD\,216411 & ~ & B0.7\,Ia & 60 & +7.2 & -7.6 & 1.2 $^{(3)}$ \\
HD\,2905 & $\kappa$~Cas & BC0.7\,Ia & 59 & +4.2 & -6.9 & 2.7 $^{(4)}$ \\
HD\,190066 & ~ & B0.7\,Ib & 60 & +6.6 & -5.6 & 4.3 \\
HD\,190919 & ~ & B0.7\,II & 43 & +7.3 & -5.8 & 4.0 \\
HD\,193076$^{(1)}$ & ~ & B0.7\,II & 60 & +7.7 & -5.1 & 4.2 \\
HD\,14053 & ~ & B0.7\,II & 23 & +8.4 & -5.3 & 5.5 \\
HD\,13969$^{(1)}$ & ~ & B0.7\,III & 36 & +8.8 & -4.9 & 6.0 \\
HD\,37042 & ~ & B0.7\,V & 33 & +6.4 & -2.3 & 9.7 $^{(4)}$ \\
HD\,201795 & ~ & B0.7\,V & 8 & +7.5 & -2.7 & 8.3 \\
HD\,169454 & ~ & B1\,Ia$^{+}$ & 47 & +6.7 & -8.1 & 1.3 $^{(3)}$ \\
HD\,13256 & ~ & B1\,Ia & 45 & +8.7 & -7.5 & 1.2 $^{(3)}$ \\
HD\,13854 & ~ & B1\,Iab & 55 & +6.5 & -6.6 & 2.8 $^{(4)}$ \\
HD\,91316 & $\rho$~Leo & B1\,Iab Ns & 47 & +3.9 & -5.3 & 3.5 \\
HD\,24398 & $\zeta$~Per & B1\,Ib & 39 & +2.8 & -5.2 & 4.3 \\
HD\,44743 & $\beta$~CMa & B1\,II-III & 26 & +2.0 & -3.9 & 7.0 \\
HD\,144470 & $\omega^1$~Sco & B1\,V & 104 & +4.0 & -2.4 & 9.4 \\
HD\,24131 & ~ & B1\,V & 80 & +5.8 & -2.7 & 10.0 \\
HD\,190603 & ~ & B1.5\,Ia$^{+}$ & 53 & +5.7 & -8.0 & 1.4 $^{(3)}$ \\
HD\,14956 & ~ & B1.5\,Ia & 52 & +7.2 & -7.8 & 3.2 \\
HD\,5551 & ~ & B1.5\,Iab & 57 & +7.8 & -6.3 & 3.1 \\
HD\,193183$^{(1)}$ & ~ & B1.5\,Ib & 50 & +7.0 & -6.2 & 3.6 \\
HD\,13841 & ~ & B1.5\,Ib & 43 & +7.4 & -5.6 & 3.3 \\
HD\,52089 & $\epsilon$~CMa & B1.5\,II & 26 & +1.5 & -4.0 & 6.4 \\
HD\,214993$^{(2)}$ & 12~Lac & B1.5\,III & 25 & +5.2 & -3.0 & 7.3 \\
HD\,215191 & ~ & B1.5\,V & 205 & +6.4 & -2.2 & 10.1 \\
HD\,35299 & ~ & B1.5\,V & 6 & +5.7 & -2.2 & 10.2 \\
HD\,37744 & ~ & B1.5\,V & 37 & +6.2 & -1.9 & 10.3 \\
HD\,41117 & $\chi^2$~Ori & B2\,Ia & 47 & +4.6 & -7.3 & 2.2 $^{(4)}$ \\
HD\,14143 & ~ & B2\,Ia & 55 & +6.7 & -7.0 & 3.3 $^{(4)}$ \\
HD\,15690 & ~ & B2\,Iab--Ib & 46 & +8.0 & -5.8 & 2.9 \\
HD\,206165 & 9~Cep & B2\,Ib & 41 & +4.7 & -6.7 & 3.3 \\
HD\,31327 & ~ & B2\,II & 31 & +6.1 & -5.4 & 4.5 \\
HD\,30836$^{(1)}$ & ~ & B2\,III & 37 & +3.7 & -3.6 & 7.3 \\
HD\,35468 & $\gamma$~Ori & B2\,III & 53 & +1.6 & -2.9 & 8.4 \\
HD\,886 & $\gamma$~Peg & B2\,IV & 8 & +2.8 & -3.0 & 9.5 \\
HD\,3360 & $\zeta$~Cas & B2\,IV & 22 & +3.7 & -2.4 & 9.4 \\
HD\,36285 & ~ & B2\,V & 12 & +6.3 & -1.6 & 11.8 \\
HD\,36629 & ~ & B2\,V & 9 & +7.7 & -1.3 & 12.2 \\
HD\,144218 & beta2 Sco & B2\,V & 57 & +4.9 & -1.2 & 12.3 \\
HD\,208947 & ~ & B2\,V & 25 & +6.4 & -2.5 & 12.3 \\
HD\,198478 & 55~Cyg & B2.5\,Ia & 36 & +4.9 & -8.2 & 2.7 \\
HD\,42087 & 3~Gem & B2.5\,Ib & 39 & +5.8 & -7.3 & 3.4 \\
HD\,207330 & ~ & B2.5\,III & 38 & +4.2 & -4.5 & 8.6 \\
HD\,148605 & 22~Sco & B2.5\,V & 167 & +4.8 & -1.2 & 12.9 \\
HD\,175191 & $\sigma$~Sgr & B2.5\,V & 122 & +2.1 & -2.4 & 12.8 \\
HD\,14134 & ~ & B3\,Ia & 37 & +6.6 & -6.8 & 2.9 \\
HD\,53138 & $o^2$~Cma & B3\,Ia & 36 & +3.0 & -7.7 & 3.0 \\
HD\,51309 & $\iota$~CMa & B3\,Ib & 21 & +4.4 & -6.0 & 5.0 \\
HD\,36212 & ~ & B3\,II & 17 & +7.8 & -4.9 & 5.3 \\
HD\,194779 & ~ & B3\,II & 120 & +7.8 & -3.8 & 6.6 \\
HD\,21483 & ~ & B3\,III & 131 & +7.1 & -2.8 & 9.2 \\
HD\,49567$^{(1)}$ & ~ & B3\,III & 4 & +6.2 & -3.3 & 8.3 \\
HD\,3901 & $\xi$~Cas & B3\,IV & 113 & +4.8 & -3.2 & 11.4 \\
HD\,160762 & $\iota$~Her & B3\,IV & 6 & +3.8 & -2.2 & 10.9 \\
HD\,20365 & 29~Per & B3\,V & 131 & +5.2 & -1.5 & 12.0 \\
HD\,32630 & $\eta$~Aur & B3\,V & 85 & +3.2 & -1.2 & 13.2 \\
HD\,120315 & $\eta$~UMa & B3\,V & 161 & +1.9 & -0.7 & 14.2 \\
HD\,178849 & ~ & B3\,V & 36 & +7.1 & -1.1 & 13.8 \\
HD\,36371$^{(1)}$ & $\chi$~Aur & B4\,Ia & 32 & +4.8 & -7.1 & 2.9 \\
HD\,164353 & 67~Oph & B4\,Ib--II & 22 & +3.9 & -6.0 & 5.4 \\
HD\,41692 & ~ & B4\,III & 31 & +5.4 & -2.7 & 10.2 \\
HD\,180554 & 1~Vul & B4\,IV & 30 & +4.8 & -2.5 & 12.4 \\
HD\,26739 & ~ & B4\,V & 35 & +6.4 & -1.4 & 13.5 \\
HD\,34759$^{(1)}$ & $\rho$~Aur & B4\,V & 62 & +5.2 & -1.3 & 15.6 \\
HD\,58350 & $\eta$~CMa & B5\,Ia & 32 & +2.5 & -6.5 & 3.1 \\
HD\,7902 & ~ & B5\,Iab & 33 & +7.0 & -6.6 & 3.1 \\
HD\,9311 & ~ & B5\,Ib & 24 & +7.3 & -6.0 & 4.6 \\
HD\,191243 & ~ & B5\,II & 26 & +6.1 & -5.0 & 6.1 \\
HD\,170682 & ~ & B5\,III & 120 & +7.9 & -2.4 & 10.0 \\
HD\,211924 & ~ & B5\,III & 27 & +5.4 & -3.2 & 12.0 \\
HD\,147394 & $\tau$~Her & B5\,V & 32 & +3.9 & -1.2 & 14.0 \\
HD\,4142 & ~ & B5\,V & 199 & +5.6 & -1.2 & 12.7 \\
HD\,161572 & ~ & B5\,V & 186 & +7.6 & -0.6 & 14.5 \\
HD\,36936 & ~ & B5\,V & 219 & +7.6 & -0.4 & 14.6 \\
HD\,20809 & ~ & B5\,V & 168 & +5.3 & -1.2 & 12.7 \\
HD\,15497 & ~ & B6\,Ia & 35 & +7.0 & -7.6 & 3.0 \\
HD\,17145 & ~ & B6\,Iab & 33 & +8.2 & -6.3 & 3.7 \\
HD\,170719 & ~ & B6\,III & 28 & +8.1 & -2.3 & 12.1 \\
HD\,34503 & $\tau$~Ori & B6\,III & 35 & +3.6 & -2.6 & 11.8 \\
HD\,49340 & 43 Cam & B6\,III & 184 & +5.1 & -2.2 & 10.7 \\
HD\,23338 & 19~Tau & B6\,IV & 111 & +4.3 & -0.9 & 12.5 \\
HD\,90994 & $\beta$~Sex & B6\,V & 83 & +5.1 & -0.1 & 15.9 \\
HD\,183143 & ~ & B7\,Ia$^{(+)}$ & 42 & +6.9 & -8.7 & 1.0 $^{(3)}$ \\
HD\,199478 & ~ & B7\,Ia & 37 & +5.6 & -7.9 & 2.5 $^{(4)}$ \\
HD\,208501 & 13~Cep & B7\,Iab & 39 & +5.8 & -6.7 & 3.8 \\
HD\,35497 & $\beta$~Tau & B7\,III & 60 & +1.6 & -1.4 & 12.4 \\
HD\,23408 & ~ & B7\,III & 32 & +3.9 & -1.9 & 11.1 \\
HD\,1279 & ~ & B7\,III & 29 & +5.9 & -2.0 & 10.7 \\
HD\,23288 & 16~Tau & B7\,V & 202 & +5.5 & -0.5 & 15.0 \\
HD\,21071 & ~ & B7\,V & 66 & +6.1 & -0.2 & 15.9 \\
HD\,182255 & 3~Vul & B7\,V & 30 & +5.2 & -0.1 & 15.4 \\
HD\,34085 & $\beta$~Ori & B8\,Ia & 32 & +0.1 & -6.9 & 2.8 \\
HD\,14542 & ~ & B8\,Iab & 29 & +7.0 & -6.7 & 3.4 \\
HD\,12301 & 53~Cas & B8\,Ib & 24 & +5.6 & -5.2 & 5.1 \\
HD\,23850 & 27~Tau & B8\,III & 230 & +3.6 & -1.9 & 11.8 \\
HD\,155763 & $\zeta$~Dra & B8\,III & 43 & +3.2 & -2.3 & 12.8 \\
HD\,179761 & 21 Aql & B8\,III & 15 & +5.2 & -1.5 & 12.3 \\
HD\,10205 & $\tau$~And & B8\,III & 89 & +4.9 & -1.7 & 11.9 \\
HD\,17081 & $\pi$~Cet & B8\,IV & 19 & +4.2 & -1.1 & 14.0 \\
HD\,87901 & $\alpha$~Leo & B8\,IVn & 306 & +1.4 & -0.3 & 13.7 \\
HD\,3240 & ~ & B8\,IV & 69 & +5.1 & -0.8 & 14.6 \\
HD\,46075 & ~ & B8\,IV & 53 & +6.7 & -0.7 & 14.9 \\
HD\,23324 & 18~Tau & B8\,V & 212 & +5.6 & -0.2 & 15.6 \\
HD\,23432 & 21~Tau & B8\,V & 160 & +5.8 & -0.1 & 16.7 \\
HD\,21672 & ~ & B8\,V & 250 & +6.6 & +0.3 & 16.4 \\
HD\,171301 & ~ & B8\,V & 33 & +5.5 & +0.3 & 17.3 \\
HD\,224112 & ~ & B8\,V & 40 & +6.8 & -0.1 & 18.0 \\
HD\,184943 & ~ & B8.5\,Iab & 37 & +8.3 & -7.1 & 3.1 \\
HD\,214923 & $\zeta$~Peg & B8.5\,IV & 160 & +3.4 & -0.6 & 15.0 \\
HD\,21279 & ~ & B8.5\,V & 198 & +7.2 & +0.6 & 18.7 \\
HD\,23923 & ~ & B8.5\,V & 308 & +6.2 & +0.3 & 16.8 \\
HD\,21291 & ~ & B9\,Ia & 32 & +4.2 & -7.0 & 3.2 \\
HD\,223960 & ~ & B9\,Ia & 32 & +6.9 & -7.5 & 2.5 \\
HD\,202850 & $\sigma$~Cyg & B9\,Iab & 31 & +4.2 & -5.9 & 4.6 \\
HD\,212593 & 4~Lac & B9\,Iab--Ib & 23 & +4.6 & -5.2 & 5.7 \\
HD\,35600 & ~ & B9\,Ib & 31 & +5.7 & -4.5 & 6.8 \\
HD\,21661 & ~ & B9\,II--III & 5 & +6.4 & -2.4 & 10.9 \\
HD\,57608 & ~ & B9\,II--III & 10 & +6.0 & -2.6 & 11.0 \\
HD\,212097 & 32~Peg & B9\,III & 70 & +4.8 & -2.1 & 12.9 \\
HD\,2011 & 12~Cas & B9\,III & 125 & +5.4 & -1.9 & 14.0 \\
HD\,51688 & ~ & B9\,III & 32 & +6.4 & -0.8 & 14.1 \\
HD\,49606 & 33~Gem & B9\,III & 20 & +5.9 & -0.5 & 12.5 \\
HD\,196867 & $\alpha$~Del & B9\,IV & 108 & +3.8 & -0.7 & 17.0 \\
HD\,21931 & ~ & B9\,V & 161 & +7.4 & +1.0 & 20.1 \\
HD\,38899 & 134~Tau & B9\,V & 25 & +4.9 & +0.2 & 20.0 \\
HD\,176437 & $\gamma$~Lyr & B9.5\,II--III & 74 & +3.2 & -3.2 & 12.1 \\
HD\,181440 & ~ & B9.5\,III & 55 & +5.5 & -0.4 & 15.9 \\
HD\,144206 & ~ & B9.5\,III & 10 & +4.7 & -0.4 & 14.3 \\
HD\,186882$^{(1)}$ & $\delta$~Cyg & B9.5\,IV & 158 & +2.9 & -0.7 & 16.8 \\
HD\,16046 & $\omega$~For & B9.5\,IV & 44 & +5.0 & -0.8 & 18.7 \\
HD\,23873 & ~ & B9.5\,V & 85 & +6.6 & +0.8 & 20.9 \\
HD\,23568 & ~ & B9.5\,V & 181 & +6.8 & +1.0 & 18.4 \\
HD\,222661 & $\omega^2$~Aqr & B9.5\,V & 136 & +4.5 & +1.2 & 22.3 \\
HD\,21389 & ~ & A0\,Ia & 37 & +4.5 & -7.3 & 2.7 $^{(3)}$ \\
HD\,46300 & 13~Mon & A0\,Ib & 7 & +4.5 & -5.1 & 7.0 \\
HD\,87737 & $\eta$~Leo & A0\,Ib & 10 & +3.4 & -5.3 & 6.8 \\
HD\,123299$^{(1)}$ & $\alpha$~Dra & A0\,III & 26 & +3.7 & -0.8 & 18.9 \\
HD\,103287 & $\gamma$~UMa & A0\,V & 177 & +2.4 & +0.3 & 19.2 \\
HD\,172167 & $\alpha$~Lyr & A0\,V & 22 & +0.0 & +0.6 & 21.6 \\
HD\,197345 & $\alpha$~Cyg & A2\,Ia & 26 & +1.2 & -7.1 & 4.5 $^{(3)}$ \\

\end{longtable}

\end{appendix}

\end{document}